\UseRawInputEncoding
\documentclass[aps,pra,reprint,showpacs,superscriptaddress]{revtex4-1}
\usepackage{amsmath}
\usepackage{natbib}
\usepackage{graphicx}
\usepackage{hyperref}
\usepackage[all]{hypcap}
\usepackage{microtype}

\usepackage{bibentry}

\usepackage{array}
\usepackage{supertabular}
\usepackage{hhline}
\usepackage{multirow}
\usepackage{float}
\usepackage{booktabs}
\usepackage{tabularx}
\usepackage{xcolor}

\hypersetup{
    colorlinks,
    linkcolor={blue},
    citecolor={blue},
    urlcolor={blue}
}

\begin{document}
\title{Native defects in monolayer GaS and GaSe: electrical properties and thermodynamic stability}

\author{Daniel Barragan-Yani}
\affiliation{Physics and Materials Science Research Unit, University of Luxembourg, 162a Avenue de la Fa{\"i}ncerie, L-1511 Luxembourg, Luxembourg}
\author{Jonathan M. Polfus}
\affiliation{Department of Chemistry, Centre for Materials Science and Nanotechnology, University of Oslo, PO Box 1033 Blindern, Oslo N-0315, Norway}
\author{Ludger Wirtz}
\affiliation{Physics and Materials Science Research Unit, University of Luxembourg, 162a Avenue de la Fa{\"i}ncerie, L-1511 Luxembourg, Luxembourg}

\date{\today}

\begin{abstract}

{ Structural, electronic and thermodynamic properties of native defects in GaS and GaSe monolayers are investigated by means of accurate ab-initio calculations. Based on their charge transition levels we assess the influence of the studied defects on the electrical properties of the monolayers. Specifically, we show that native defects do not behave as shallow dopants and their presence cannot account for the experimentally observed intrinsic doping. In addition, we predict that native defects are efficient compensation and recombination centers. Besides pointing out their detrimental nature, we also calculate the corresponding finite temperature formation energies and provide a window of growth conditions able to reduce the concentration of all relevant native defects.}

\end{abstract}


\maketitle

\section{\label{intro} Introduction}

Ultrathin two-dimensional (2D) materials exhibit unusual characteristics that show great promise for application in next generation electronic and optoelectronic devices~\cite{Review_Zhu}. The quintessential 2D material is graphene~\cite{Rev_GraphElectro, katsnelson2012graphene}, a monolayer of carbon atoms arranged in a honeycomb lattice, first exfoliated in 2004~\cite{Novoselov1} and intensively studied due to its impressive electrical, optical, magnetic and mechanical properties, mostly associated with its electronic bands exhibiting linear dispersion at the Fermi level~\cite{RevGraphMech, raza2012graphene, Schwierz2010, Han2014, Bonaccorso2010}. Unfortunately, in its pristine semimetallic form, graphene does not posses a band gap and its applicability in semiconductor technology is currently limited. Thus, there is great interest in finding and understanding novel semiconducting 2D materials. 

Beyond graphene, a prototypical example of 2D materials are single-layer transition metal dichalcogenides (TMDs) – MX$_2$ (e.g., M = Mo, W; X = S, Se) – which are direct semiconductors with high absorbance and exhibiting novel optical properties, including valley-selective circular dichroism and coupling of spin and valley degrees of freedom~\cite{annurev_BeyondGraph, Manzeli2017, annurev_OptExiTMDs, Schaibley2016, Mak2018}.  Similar to TMDs but with even more exotic properties, group III-VI metal monochalcogenides – MX (M = Ga, In; X = S, Se, Te) – have recently attarcted interest due to their novel electrical and optical features~\cite{C6NR05976G, Review_Cai}. Specifically, what truly sets group III-VI metal monochalcogenides apart from the greater family of 2D materials, is the direct to indirect band gap transition observed in both experiments and simulations when their structure reach a critical number of layers~\cite{PhysRevB.87.195403, PhysRevB.89.205416, PhysRevB.90.235302, PhysRevB.96.155430, PhysRevB.95.115409, PhysRevB.98.115405, Hamer2019, PhysRevB.100.045404}. Below this critical number of layers, the conduction band minimum (CBM) remains at the brillouin zone center ($\Gamma$-point) while the valence band maximum (VBM) moves away from the $\Gamma$ towards the \textit{K}-point. This process creates a nonparabolic dispersion for holes near the top of the valence band, usually referred to as "Mexican hat" dispersion~\cite{Review_Cai}. Due to the large degeneracy of states associated with it, such electronic dispersion results in a large density of states (DOS) near the band edge, signaling a van Hove singularity~\cite{PhysRevB.87.195403, PhysRevB.89.205416, PhysRevLett.116.206803} which can lead to magnetic or superconducting instabilities~\cite{PhysRevLett.114.236602, Iordanidou2018, SETHULAKSHMI2019107}. Furthermore, the singularity in the DOS and the large number of conducting modes at the band edge can result in enhanced thermoelectric properties of these materials~\cite{doi:10.1063/1.4928559}.   

       
In particular, the ultrathin gallium chalcogenides GaS and GaSe have been found to have exceptionally large photoresponse, which makes them excellent candidates to build efficient flexible photodetectors~\cite{Hu2012, Lei2013, HuWang2013, Li2014, Samani2014}, and strong nonlinear optical properties~\cite{Jie2015, Zhou2015}. Moreover, these materials have been reported to exhibit potential as building blocks of photovoltaic devices~\cite{Cheng2018}, as channel materials in field effect transistors~\cite{Late2012}, as photocalysts for water splitting~\cite{Zhuang2013, Kouser2015, Harvey2015, C8TA08103D} and can transition into a topological insulator state by means of strain and oxygen functionalization~\cite{TopoZhu2012, TopoZhou2018}. From the point of view of their synthesis, single-layers of GaS and GaSe have been obtained by means of micromechanical cleavage technique~\cite{Hu2012, HuWang2013, LateExfo2012, Wu2017}, pulsed laser deposition (PLD)~\cite{Samani2014}, molecular beam epitaxy (MBE)~\cite{BenAziza2016, Choong2017, Chen2018} and vapor phase transport (VPT)~\cite{Lei2013, Li2014, Li2015, Lie1501882, Li2017}. The last two techniques has so far proved to be the most succesful approaches to obtain large-area high-quality monolayers of GaS and GaSe~\cite{Review_Cai}. 
As any other material and regardless of the growth method used to synthesize them, ultrathin gallium monochalcogenides are expected to contain point defects which affect their electrical, magnetic and optical properties~\cite{yu2005fundamentals, mccluskey2012dopants}. However, although the impact of defects in 2D materials is particularly strong due to the reduced dimensionaly, there is currently no clear understanding of their properties in the case of monolayer GaS and GaSe. This despite the fact that some of the most interesting properties and future applications of these materials are closely linked to the presence of defects, e.g., their topological insulator state~\cite{TopoZhou2018} and their potential as material platform for single photon emitters~\cite{Review_Cai, Tonndorf_2017, annurev-2D_Aharono}. In addition, on a more basic level, the origin and nature of intrinsic doping in monolayer GaS and GaSe is still under debate and native defects have been proposed as a possible explanation~\cite{PhysRevB.96.035407, Chen2018, PhysRevB.98.115405}. 

Although some valuable attemps to shed light on these issues have come from theoretical studies~\cite{ChenDefects2015, GuoDefects2017, Hopkinson2019, DabralDefects2019, De_k_2020}, their results are hampered by at least one of the following issues: reduced-size supercells (which causes artificially delocalized defect states), incorrect description of the dielectric properties of the monolayers, severe under- or overestimation of the band gap, incorrect or lack of correction for the electrostatic interaction between charged defects and their periodic images, and incomplete assessment of the possible charge states of the defects. Furthermore, the influence of temperature on defect formation is crucial and closely related to the growth process by which the material is prepared.  Therefore, a complete and accurate inspection of the native defects, as the one available for MoS$_2$~\cite{Komsa2015}, is still lacking in the case of GaS and GaSe. This constitutes the relevance and pertinence of the systematic theoretical study presented in this paper.


By means of accurate density funtional theory (DFT) calculations we obtain physically sound thermodynamic charge transition levels of relevant native defects in monolayer GaS and GaSe. With such information at hand we are able to reveal their electrical properties and to determine whether they are detrimental. In addition, we calculate their formation energies at experimentally relevant conditions, allowing us to predict which point defects are more likely to exist depending on the growth method. More importanly, our results enable us to propose changes in growth conditions such that harmful defects are avoided. 

The paper starts with a detailed description of the methods used in Sec.~\ref{methods}, where we provide a summary of the computational parameters and a precise account of the approach used to calculate defects formation energies. This includes the proper choices of chemical potentials, the correction used to mitigate the finite-size effects and the technique employed to obtain the dielectric constants of the monolayers. Afterwards, our results are reported in Sec.~\ref{results} where we present the basic structural, thermodynamical and electrical properties of all considered defects along with their formation energies at finite temperature. Furthermore, in Sec.~\ref{results}, we discuss the implications of our findings. Finally, in Sec.~\ref{conclusions}, we summarize our key results.

\section{\label{methods} METHODS}

\subsection{\label{setup} Computational setup and basic parameters}

Density functional theory (DFT) calculations were performed using the \textit{VASP}~\cite{PhysRevB.54.11169} simulation package with projector augmented-wave potentials (PAWs) for the effective potential associated to the nucleus and the core electrons. A converged plane-wave energy cutoff of 500 eV was applied and the convergence criterion for forces during ionic relaxation was set to 0.01 eV/\AA . Both the Perdew-Burke-Enzerhof (PBE) functional with van der Waals interactions included via the empirical correction proposed by Grimme (PBE-D) and the Heyd, Scuseria and Enzerhof (HSE06) hybrid functional were used as exchange-correlation functionals. On the one hand, as reported in Refs.~\cite{PhysRevLett.108.235502,Bj_rkman_2012,PhysRevB.101.045428}, the PBE-D approach delivers accurate structural parameters and binding energies for layered materials and it allows us to correctly describe the properties of defects lying outside of the monolayer~\cite{Komsa2015}. On the other hand, although HSE06 calculations do not account for the van der Waals interactions and are more computationally demanding than PBE-D ones, they are useful to accurately assess the electronic properties of defects as they predict band gaps and defect level positions closer to experiments and beyond-DFT methods~\cite{RevModPhys.86.253, alkauskas2011advanced, Chen_2015, Oba_2018}. Throughout this paper, it is clearly pointed out when and why one or the other functional were used. For all results presented in this work spin polarization was taken into account. Since spin-orbit interaction has been found to be mostly relevant for the description of low-lying valence bands~\cite{doi:10.1063/1.4928559, PhysRevB.98.115405}, they are not expected to be crucial for the thermodynamics of defects and they were not included in our calculations. 

\renewcommand{\arraystretch}{1.3}
\setlength{\tabcolsep}{3pt}
\begin{table}[ht]\label{basic}
\normalsize

\caption{Basic properties of monolayer GaS and GaSe as obtained using PBE-D and HSE06 functionals. For comparison, if available, the experimental values are also reported. In-plane lattice constants \textit{a} are given in \AA , band gaps $E_{g}$ are given in eV and finally, we also present the in-plane and out-of-plane components of the dielectric constant. Experimental lattice constants and band gaps are from Refs.~\cite{Jung2015,PhysRevB.96.035407}}

\begin{tabular}{cccccc}

\hhline{======}

\multicolumn{1}{l}{}        &  & \textit{a} & $E_{g}$ & $\varepsilon_{\|}$ & $\varepsilon_{\bot}$ \\ \midrule[0.1mm]
\multirow{3}{*}{GaS} 	& PBE-D & 3.58 & 2.49 & 6.33 & 1.83 \\
                     	& HSE06 & 3.59 & 3.28 & 5.76 & 1.81 \\ 
                     	& Expt. & 3.6 & 3.6 & - & - \\ \midrule[0.1mm]
\multirow{3}{*}{GaSe}      & PBE-D & 3.75 & 2.21 & 7.41 & 1.86 \\ 	
                           & HSE06 & 3.77 & 2.98 & 6.68 & 1.84 \\
                           & Expt. & 3.77 & 3.5 & - & - \\


\hhline{======}

\end{tabular}

\end{table}

Basic parameters of the monolayers were calculated using a converged $16 \times 16 \times 1$ \textit{k}-point grid for the unit cell with 25 \AA ~of vacuum and are reported in Table~\ref{basic}. As can be seen there, both PBE-D and HSE06 deliver similar in-plane lattice parameters, \textit{a}, and slightly underestimate the experimental room-temperature values by $\sim$0.5\%. Regarding their electronic properties, optical band gaps measured via catholuminescence experiments are 3.4 eV and 3.3 eV for single-layer GaS and GaSe, respectively~\cite{Jung2015}. In the case of fundamental gaps, scanning tunneling spectroscopy measurements were conducted and arrived to a value of 3.5 eV in the case of monolayer GaSe~\cite{PhysRevB.96.035407}, which implies an exciton binding energy of 0.2 eV. Such results differ from the predictions of high-level G${_0}$W${_0}$, GW${_0}$ and Koopmans-complaint hybrid calculations, which have been reported for monolayer GaSe and overestimate the band gap and the exciton binding energy~\cite{PhysRevB.84.085314, PhysRevLett.114.236602, PhysRevB.100.235304, De_k_2020}. Using the experimental results and since both GaS and GaSe exhibit similar structural and electronic properties, it is safe to assume that the exciton binding energy of both monolayers would be of the same order and we assume a fundamental band gap of 3.6 eV for the GaS case. When compared with the experimental values, the results presented in this paper using both PBE-D and HSE06 underestimate the band gaps of the monolayers under study. However, HSE06 predictions are never more than 0.5 eV away from the experimental value and, having in mind that the fundamental gap decreases when the dielectric constant of the environment increases, its predictions should be closer to experiments involving substrates. In the case of PBE-D, it considerably underestimates the band gap and in principle cannot be used to draw conclusions on the electrical properties of defects in these monolayers. However, as shown in Sec.~\ref{CTLs}, when aligned with respect to vacuum the levels predicted using PBE-D are in line with those predicted by HSE06. Furthermore, due to the reduction of the band gap as temperature rises, results obtained using PBE-D become more appropiate when trying to draw conclusions for finite temperatures. These two features combined, make it possible to use PBE-D to study the formation energies of defects at relevant growth conditions (see Sec.~\ref{finite-temp}).

Finally, in order to assess the spurious interaction between defects and their periodic images (see Sec.~\ref{Intro_FormationEnergies}), static dielectric constants were calculated using the method presented in Ref.~\cite{Noh2014}, applying either density functional perturbation theory (PBE-D) or self-consistent response to finite electric fields (HSE06) to  $a \times a \times na$ (\textit{n} = 2, 4, 8, 16) supercells, with \textit{a} being the in-plane lattice constants, and analyzing the convergence of the obtained results (see the supplementary material). The obtained values for both in-plane and out-of-plane components of the dielectric constant are also reported in Table~\ref{basic}. Unfortunately, there are no experimental values available for the dielectric properties of the monolayers under study. There are, however, previous calculations for GaSe.~\cite{PhysRevB.100.235304, De_k_2020} Surprinsingly, as the authors of Ref.~\cite{De_k_2020} themselves recognize, their conclusion is that $\varepsilon_{\bot}$ of the monolayer is close to the value of the bulk, which is clearly unphysical. 


\begin{figure*}[t]
\includegraphics[scale=0.65]{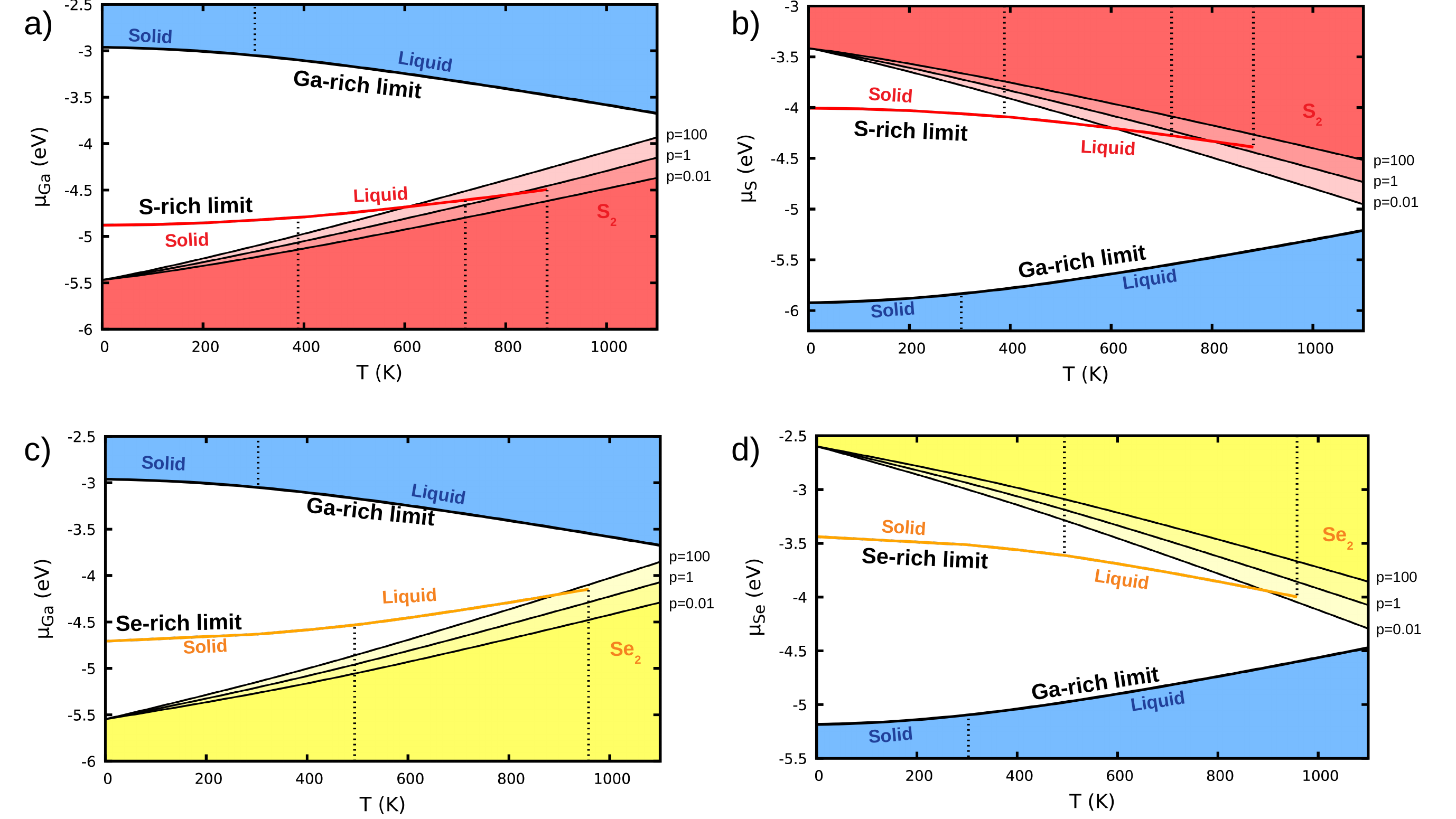} 
\vspace{-0.2cm}
\caption{\label{f:chem_pot} Limiting values for the chemical potentials (a) $\mu_{Ga}$ and (b) $\mu_{S}$ for monolayer GaS, and (c) $\mu_{Ga}$ and (d) $\mu_{Se}$ for monolayer GaSe. The white areas are the regions for which the monolayers are stable. Above or below the lines that limit these areas, species forming the monolayers prefer their elemental phases for the corresponding temperature. Solid red and yellow lines correspond to the reference curves based on thermochemical tables~\cite{Barin1977, 219851, 2005chemical}. Partial pressures are given in bar.}
\end{figure*}

\subsection{\label{Intro_FormationEnergies} Defect formation energies and transition levels}

Defects were studied using $6 \times 6 \times 6$ supercells with a $3 \times 3 \times 1$ \textit{k}-point grid. This supercell size implies a large enough vacuum of $\sim$22\AA ~between monolayers. All other parameters used are the converged ones described in the previous section. In order to avoid local minima and to find accurate relaxed atomic configurations for the defects, the symmetry of the initial defective supercells was broken by introducing small random displacements to all atoms.

As mentioned in the introduction, one of the aims of this work is to produce a complete study of the native defects in which realistic growth conditions are taken into account. To that end, the approach described in Refs.~\cite{PhysRevLett.67.2339, RevModPhys.86.253,alkauskas2011advanced,Oba_2018} and, specifically, the methodology proposed in Ref.~\cite{Komsa2015} for monolayer MoS$_2$ were followed. The key quantity in a defect study where temperature effects are accounted for, is the Gibbs free energy of formation $G^f$. For a given defect $D$ in charge state $q$, $G^f$ is defined as

\begin{multline}\label{e:formation_energy}
G^{f}[D^{q}](T,P) = F[D^{q}] - F[\rm bulk] + \textit{P} V^{\textit{f}} \\
 - \sum_i n_i \mu_i (T,P)  + q[E_{\rm VBM} + E_{\rm F}] + E_{\rm corr} ,
\end{multline}

where $P$ is the pressure, $T$ is the temperature, $F[D^{q}]$ and $F[\rm bulk]$ are the Helmholtz free energies of the defective and pristine supercells, respectively, $V^{\textit{f}}$ is the formation volume of the defect, $n_i$ is an integer that indicates the number of atoms of type $i$ added or removed to create defect $D$ and $\mu_i$ is the chemical potential of the same added or removed atoms. The fifth term in Eq.~\ref{e:formation_energy} accounts for the energy needed to charge the system. Consequently, it contains the valence band maximum (VBM) energy, $E_{\rm VBM}$, and the Fermi energy referenced to the VBM, $E_{\rm F}$. The final term, $E_{\rm corr}$, accounts for the correction needed to eliminate the spurious electrostatic interactions between periodic images of the defects under study. At $T = 0$ K, the Gibbs free energy of formation reduces to the formation energy, $G^f \rightarrow E^f$.  

Following Ref.~\cite{Komsa2015}, here it is assumed that the $F[D^{q}] - F[\rm bulk]$ difference in Eq.~\ref{e:formation_energy} cancels out most of the vibrational contributions to the Helmholtz free of the system. Furthermore, due to the large gap of monolayer GaS and GaSe, it is also assumed that finite temperature effects on the electronic contribution to the free energies are negligible. This means that $F[D^{q}]$ and $F[\rm bulk]$ can be replaced by $E_0[D^{q}]$ and $E_0[\rm bulk]$, i.e., the corresponding total energies as obtained from DFT calculations at 0 K. Regarding the defect formation volume, it is assumed to be zero as the concept is not valid for an atomically-thin material.

As it is clear from the statement expressed in Eq.~\ref{e:formation_energy}, it is assumed that the pressure and temperature dependence of $G^f$ enters via the chemical potentials, which define the environmental conditions. Naturally, in order to produce results comparable with experiments, one is interested in realistic limits for the chemical potentials. Thus, it is assumed that one limiting condition for the chemical potentials of Ga, S and Se is that they are in equilibrium with the corresponding monolayer for all temperatures and pressures considered in our study,

\begin{equation}\label{e:eq_mu_Se}
\mu_{GaS} = \mu_{Ga} + \mu_{S} , \; \mu_{GaSe} = \mu_{Ga} + \mu_{Se}.
\end{equation}        

In addition, a further limiting condition is that $\mu_{i}$ cannot be lower than the corresponding values for the stable elemental phases, $\mu^{\rm ref}_{i}$. In other words, this second limiting condition means that the differences $\Delta \mu_i = \mu_i - \mu^{\rm ref}_i$ must add up to the formation enthalpy, $\Delta H^{GaS}_{\rm f}$, of the monolayers: 

 \begin{align}\label{e:eq_delta_mus}
 \Delta \mu_{Ga} + \Delta \mu_{S} &= \Delta H^{GaS}_{\rm f} \\
 \Delta \mu_{Ga} + \Delta \mu_{Se} &= \Delta H^{GaSe}_{\rm f}.
\end{align}      

For the actual calculation of $\mu_{i}$, it should be kept in mind that for elemental phases the chemical potentials are proportional to the corresponding total Gibbs free energy, $\mu^{\rm ref} = G^{\rm ref}/N^{\rm ref}$, where $N^{\rm ref}$ is the number of particles in the unit cell of interest. At temperature $T$ and reference pressure $P^{\circ}$, the free energy is given by
\begin{multline}\label{e:chem_pot_Gibbs}
G^{\rm ref}(T,P) = E_0 + E^{\rm vib} + \Delta H (T,P^{\circ}) - T \Delta S (T,P^{\circ}),
\end{multline}

where $E^{\rm vib}$ is the energy of zero-point vibrations, directly obtained from DFT results, and $\Delta H$ and $\Delta S$ account for the changes in enthalpy and entropy, respectively, associated to finite temperatures. Values for $\Delta H$ and $\Delta S$ at the standard pressure of $P^{\circ} = 100$ kPa are readily available from thermochemical tables.~\cite{Barin1977, 219851, 2005chemical}

$E_0$ for Ga, S and Se, was calculated using the fact that at $T=0$ all these elements prefer solid phases. Specifically, Ga stabilizes in an orthorhombic structure, and both S and Se crystallize in monoclinic allotropes formed by S$_8$ and Se$_8$ rings. After obtaining all the needed values, it is possible to use Eq.~\ref{e:chem_pot_Gibbs} to focus on realistic growth conditions. As mentioned in the Sec.~\ref{intro}, the most succesful approaches to obtain large-area high-quality monolayers of GaS and GaSe are MBE~\cite{BenAziza2016, Choong2017, Chen2018}, and VPT~\cite{Lei2013, Li2014, Li2015, Lie1501882, Li2017}, which usually imply growth temperatures of about 700 K for the former and 1000 K for the latter. Therefore, when discussing Gibbs formation energies of defects in monolayer GaS and GaSe, it will always refer to these two temperatures. 

On the one hand, since Ga melts at 302.9 K, it is expected to be liquid at the relevant growth temperatures and we can directly apply Eq.~\ref{e:chem_pot_Gibbs} using thermochemical tables for $\Delta H$ and $\Delta S$.~\cite{219851} On the other hand, at these same temperatures both S and Se are expected to be found in their diatomic gas phase~\cite{Barin1977, 219851, 2005chemical}. As a result, Eq.~\ref{e:chem_pot_Gibbs} cannot be directly used. Following Ref.~\citep{Komsa2015}, assuming ideal gas behavior, the chemical potential of S$_2$ and Se$_2$ can be written as

\begin{multline}\label{e:chem_pot_Gibbs_2}
\mu_{X}(T,P) = \dfrac{1}{2} \left[  E_{X_{2}} + E^{\rm vib}_{X_{2}}(0 \rm K) + \Delta H (T,\textit{P}^0) \right. \\
\left. - T \Delta S (T,\textit{P}^0) + k_B T \rm ln \left( \dfrac{\textit{P}_{\textit{X}_{2}}}{\textit{P}^0}  \right)   \right]  ,
\end{multline}

where $X = \rm S, Se$ and $\textit{P}_{\textit{X}_{2}}$ is the partial pressure of the corresponding diatomic gas. $E_{X_{2}}$ is the total energy of the corresponding isolated diatomic molecules and $E^{\rm vib}_{X_{2}}(0 \rm K)$ is their energy of zero-point vibrations. Both $E_{X_{2}}$ and $E^{\rm vib}_{X_{2}}(0 \rm K)$ are directly obtained from DFT calculations. In the case of $\Delta H$ and $\Delta S$, as done for Ga, a $P^{\circ} = 100$ kPa is assumed and thermochemical tables are used~\cite{Barin1977, 219851, 2005chemical}.  

\begin{figure}[t]
\includegraphics[width=\columnwidth]{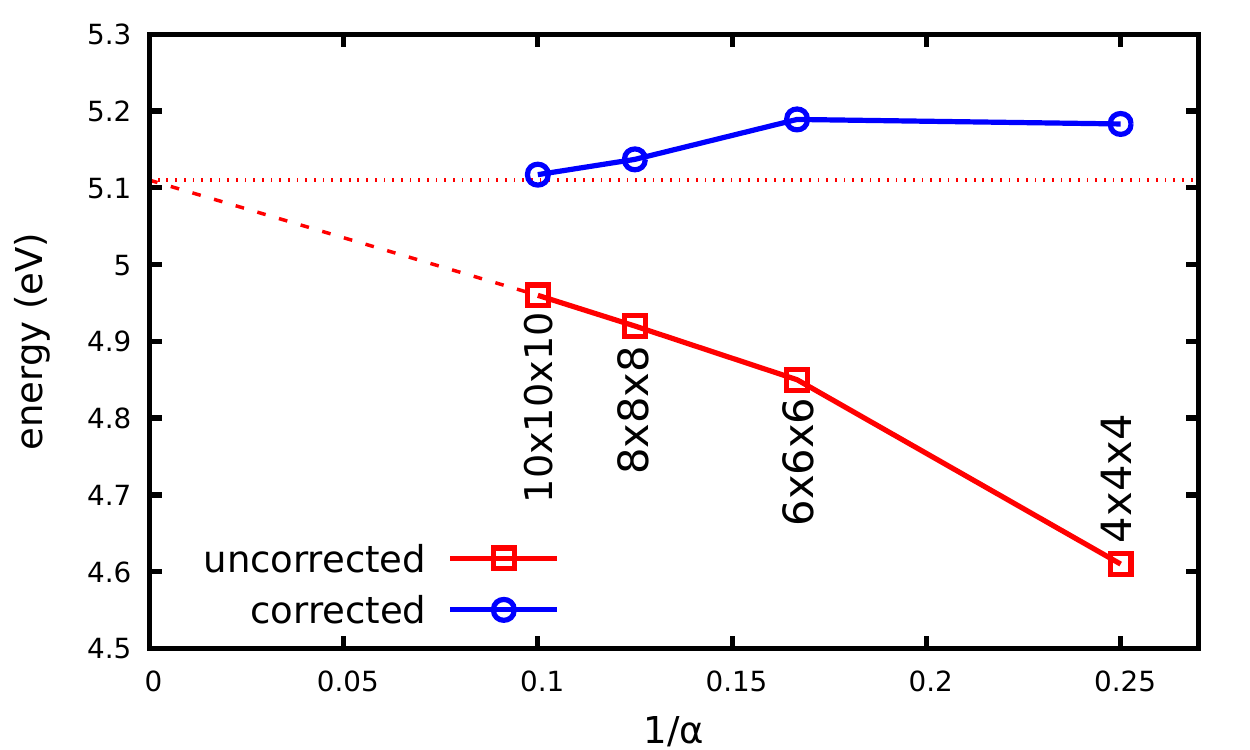} 
\vspace{-0.4cm}
\caption{\label{f:conv_formation} Formation energies of the negatively charged S vacancy, $\rm V^{-1}_S$, in monolayer GaS at different supercell sizes. The red and blue lines shows the uncorrected and corrected values, respectively. Corrections are carried out using the FNV-2D scheme. The used supercells have a $\alpha \times \alpha \times \alpha $ repetition pattern, with $\alpha$ being the number of times the in-plane lattice constant \textit{a} is repeated in each direction. The dilute limit is calculated by extrapolating the uncorrected formation energies obtained using the three largest supercells.}
\end{figure}

The resulting limits for the chemical potentials in the monolayers under study and using PBE-D are shown in Fig.\ref{f:chem_pot}, where the white areas are the regions for which the monolayers are stable. Above or below the lines that limit these areas Ga, S and Se prefer their elemental phases at the corresponding temperature. Specifically, if the Ga-rich limit is exceeded, Ga will form precipitates and if S- or Se-rich limits are the ones exceeded, there will be S and Se desorption from the monolayer. In all cases, a rich limit means that the condition $\mu_{i} = \mu^{\rm ref}_i$ has been achieved for one of the species in the monolayer. For example, in Fig.\ref{f:chem_pot}(a), $\mu_{Ga} = \mu^{\rm ref}_{Ga}$ in the Ga-rich limit and, using Eq.~\ref{e:eq_mu_Se}, $\mu_{Ga} = \mu_{GaS}-\mu^{\rm ref}_{S}$ in the S-rich limit. A further detail shown in Fig.\ref{f:chem_pot}, are solid red and yellow lines which correspond to the tabulated values for the lowest energy phases~\cite{Barin1977, 219851, 2005chemical} and serve as references. At this point it is important to remark that an often overlooked issue when dealing with small molecules, e.g. diatomic S or Se, is that depending on the used computational parameters one can have relatively large errors on their calculated energies. This naturally affects the obtained chemical potentials shown in Fig.\ref{f:chem_pot}. However, what is crucial for defect calculations is to be consistent with respect to the choice of computational methods used to obtain all the needed chemical potentials in a given study~\cite{RevModPhys.86.253}. 

Being aware that realistic growth temperatures for the monolayers of GaS and GaSe are between 700 K and 1000 K, it is possible to see in Fig.\ref{f:chem_pot} that the reduction induced in the chemical potentials due to temperature effects is considerable. Thus, it is not correct to use $\mu_{i}(\rm T = 0~K)$ to obtain reliable Gibbs formation energies of defects at relevant finite temperature. Moreover, Fig.\ref{f:chem_pot} also shows that the actual values of the chemical potentials are also dependent on the S and Se partial pressures. Throughout this paper, Eq.~\ref{e:formation_energy} and PBE-D are applied to calculate defect Gibbs formation energies at 700 K and 1000 K, using the values for $\mu_{i}$ shown in Fig.\ref{f:chem_pot} for S and Se partial pressures of 1 bar. The reason why PBE-D is used for finite temperature calculations is explained in Sec.~\ref{finite-temp}. 


\begin{figure}[b]
\includegraphics[width=1\columnwidth]{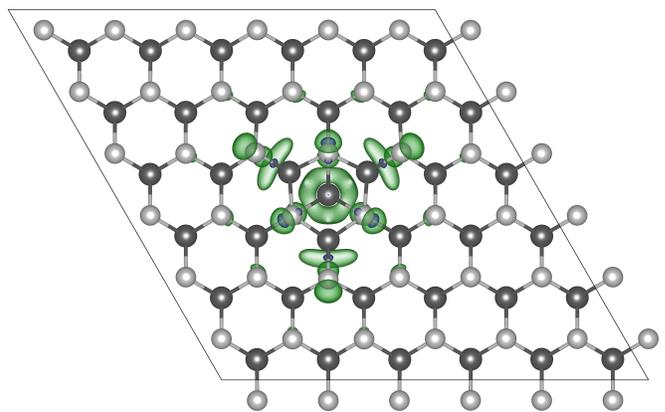} 
\vspace{-0.4cm}
\caption{\label{f:charge_density} Charge density difference between the neutral and positively charged Ga interstitial in GaS. The isosurface level is set to 0.003 $e$/\AA$^3$. Ga and S atoms are shown in dark and light gray, respectively.}
\end{figure}

\begin{figure*}[ht]
\includegraphics[scale=0.65]{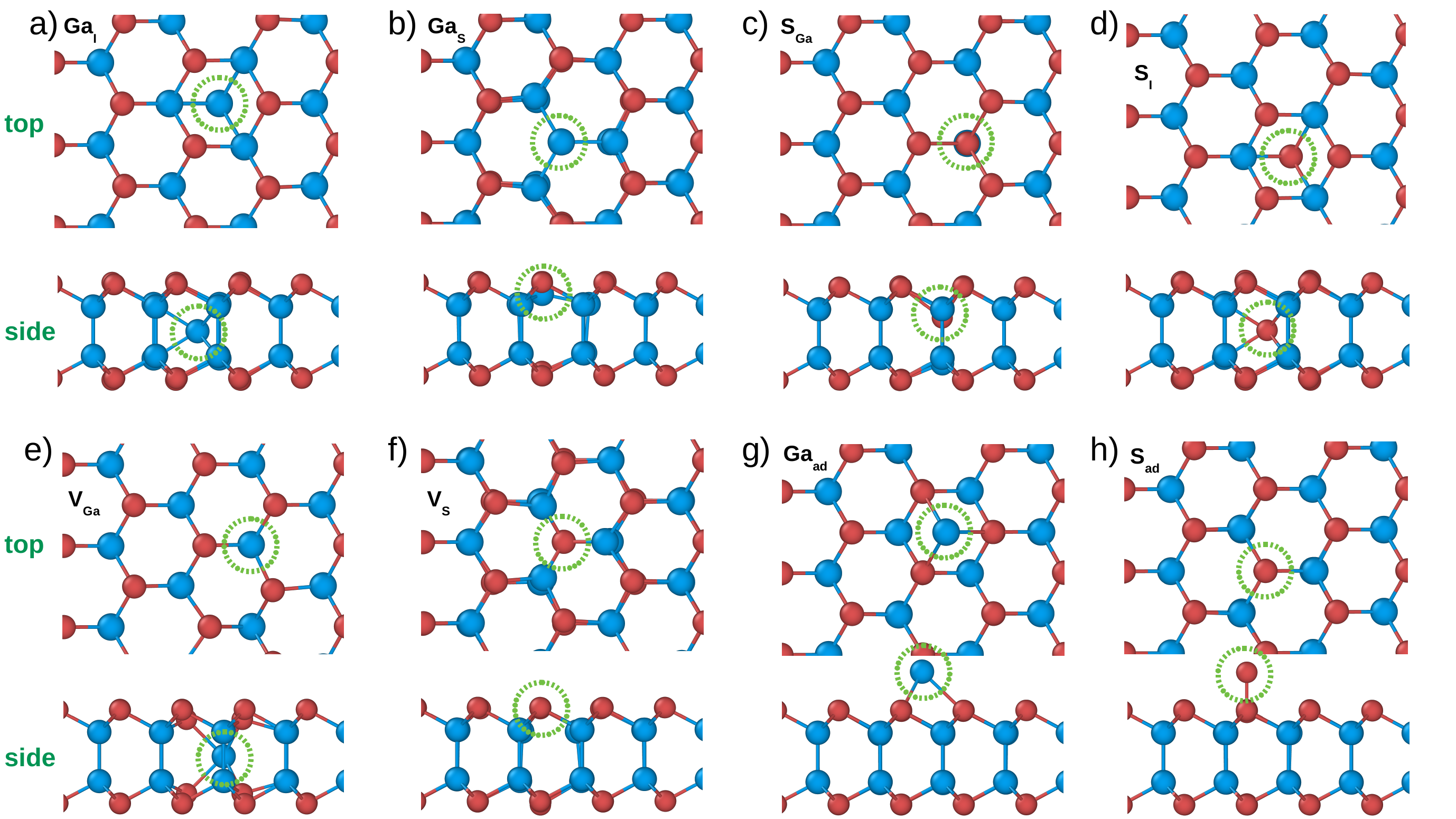} 
\vspace{-0.4cm}
\caption{\label{f:structures} Relaxed structures of native defects in monolayer GaS in their neutral charge state, as obtained from PBE-D calculations. (a) Ga$_{\rm i}$, (b) Ga$_{\rm S}$, (c) S$_{\rm Ga}$, (d) S$_{\rm i}$, (e) V$_{\rm Ga}$, (f) V$_{\rm S}$, (g) Ga$_{\rm ad}$, (h) S$_{\rm ad}$. For each case top and side views are shown, and green circles mark the position of the defects. Ga and S atoms are shown in red and blue, respectively. Results using the HSE06 functional only show negligible differences with respect to the structures shown here. Visualization of the structures is done with OVITO~\cite{0965-0393-18-1-015012}. }
\end{figure*}

After having determined accurate values for $\mu_{i}$, only the $E_{\rm corr}$ term has to yet be calculated in order to take full advantage of Eq.~\ref{e:formation_energy}. As mentioned before, this term is needed when using the supercell approach and takes care of correcting the spurious electrostatic interactions due to defect periodic images immersed in a jellium background. The existence and relevance of these spurious effect is known since the 1985 seminal work of Leslie and Gillan~\cite{Leslie_1985}. After a great deal of effort in this direction, currently there are several reliable schemes to calculate $E_{\rm corr}$ for bulk materials~\cite{PhysRevB.51.4014, PhysRevB.73.035215, PhysRevB.78.235104, PhysRevLett.102.016402}. Among those, the one proposed by Freysoldt, Neugebauer and Van de Walle (FNV) in Ref.~\citenum{PhysRevLett.102.016402} is rigorous and has been proven to be the most accurate option~\cite{PhysRevB.86.045112,  PhysRevB.89.195205}. Since there is less screening in the case of a supercell containing a defective monolayer, the spurious electrostatic interactions between charged defects in such systems are expected to have a greater impact. This is the reason why during the last years there has been interest in finding accurate schemes that include the inhomogeneity of the dielectric environment natural to monolayers~\cite{PhysRevLett.110.095505, PhysRevX.4.031044, PhysRevLett.114.196801, PhysRevB.96.155424, doi:10.1063/1.4978238, PhysRevMaterials.1.071001, PhysRevB.97.205425}. Unfortunately, contrary to the bulk case, there is still no detailed study of how these methods compare to each other. However, based on the success of the FNV method for bulk systems and the rigorous method behind it, its extension for two-dimensional materials (FNV-2D) is used in this paper~\cite{PhysRevB.97.205425}. The formation energy of the negatively charged S vacancy, $\rm V^{-1}_S$, in monolayer GaS at different supercell sizes is shown in Fig.~\ref{f:conv_formation}. For each supercell size used, both the uncorrected and corrected values are presented. As can be seen there, for this very localized defect the FNV-2D is able to deliver accurate predictions even for the smallest case included in our study, i.e, the $4 \times 4 \times 4$ supercell. Although using such small simulation box would reduce the computational cost of our study, it is important not to forget that we must also use a supercell size that guarantees the correct localization of the defects states. This is the reason why in this paper $6 \times 6 \times 6$ supercells were used: they allow the calculation of formation energies values that are less than $0.1$ eV away from dilute limit and are large enough to contain even the most delocalized defects under study, i.e., the interstitials. As an example, in Fig.~\ref{f:charge_density} we present the charge density difference between the neutral and positively charged Ga interstitial in GaS in a $6 \times 6 \times 6$ supercell, where it is clear that the hole is correctly localized within the boundaries of the used supercell.

A further defect characterization tool are charge transition levels (CTLs), defined as the Fermi level for which the formation energies of charge states $q$ and $q'$ are equal. They must not be confused with defect Kohn-Sham states, which are called \textit{states} throughout this paper. Charge transition levels for defect $D$ at $T = 0$ K can be obtained as follows:

\begin{equation}\label{e:charge_transition_level}
\epsilon(q/q') = \frac{E^{\rm f}[D^{q};E_{\rm F}=0]-E^{\rm f}[D^{q'};E_{\rm F}=0]}{q'-q}  ,
\end{equation} 

The transition levels are relevant because they can be directly related to experiments in which the defects are able to fully relax after the charge transition~\cite{RevModPhys.86.253}.

\begin{figure}[t]
\includegraphics[width=1.1\columnwidth]{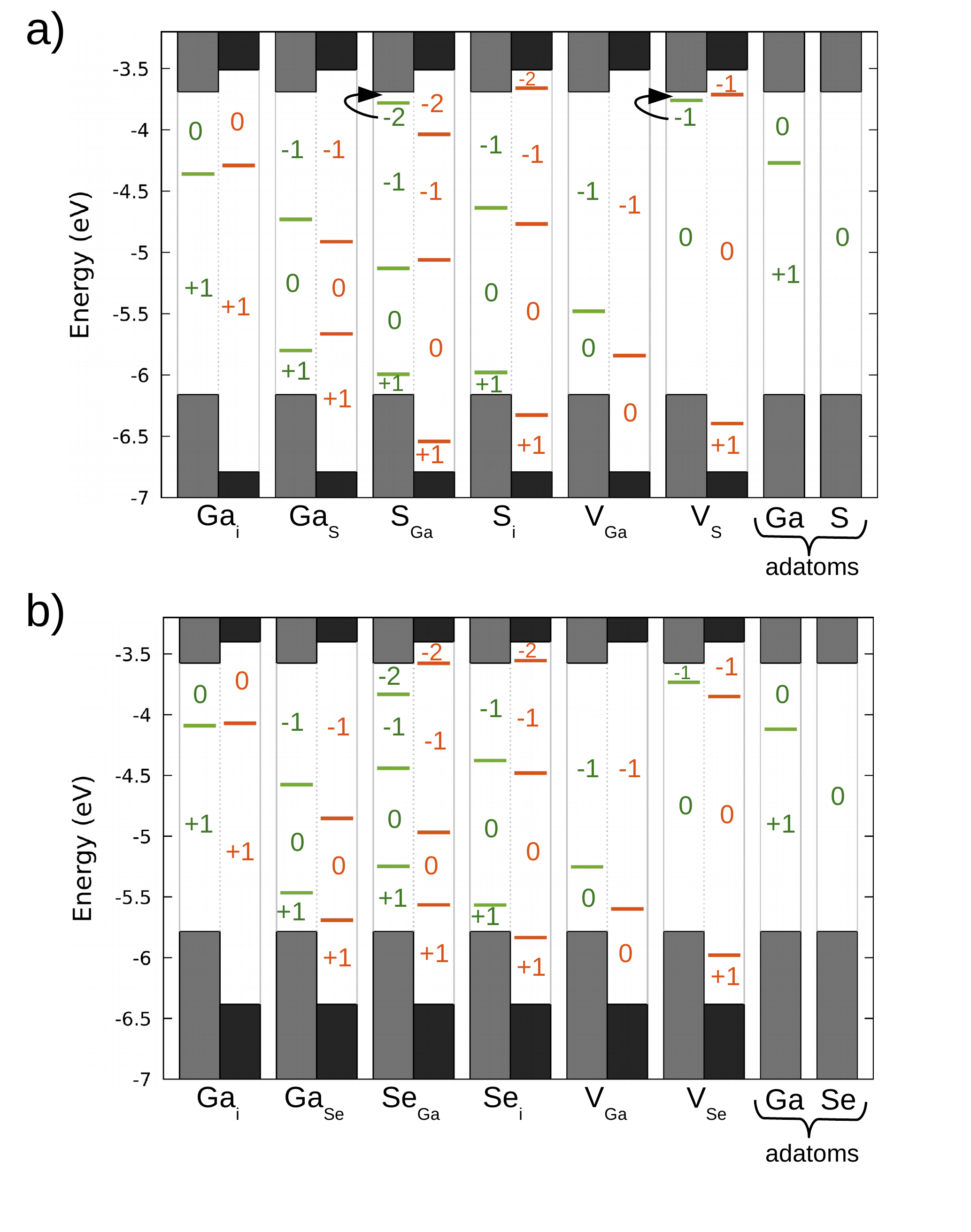} 
\vspace{-1.0cm}
\caption{\label{f:transition_levels} Charge transition levels, CTLs, of the native defects in monolayer (a) GaS and (b) GaSe. Results are shown wit respect to vacuum level. For each defect we present the CTLs obtained with both PBE-D and HSE06. Valence and conduction bands are show in light and dark gray for PBE-D and HSE06, respectively. Green (red) horizontal lines and numbers mark the CTLs and charge state, respectively, of each defect as obtained with PBE-D (HSE06). Black arrows are used for clarity and do not imply any dynamical process. They are used to point out the charge state of small stability regions for which the numbers in the figure cannot fit.}
\end{figure}

\section{\label{results} RESULTS}

In order to understand the properties of native defects in the monolayers under study, we start the presentation of our results by discussing their CLTs together with their basic geometries. With this first part we aim at understanding whether native defects are a possible explanation to the experimentally observed intrinsic doping on the monolayers. Since it is not central to our goal, we only discuss the electronic structure of the studied defects in a brief manner. Based on such results, in a second part we present the formation energies of the same relevant native defects at realistic growth temperatures. Our aim there is to reveal whether harmful defects are likely to form when growing monolayers using state-of-the-art methods. More importantly, in this second section we propose growth conditions that would limit the formation of the most detrimental native defects.

\begin{figure*}[t]
\includegraphics[scale=0.62]{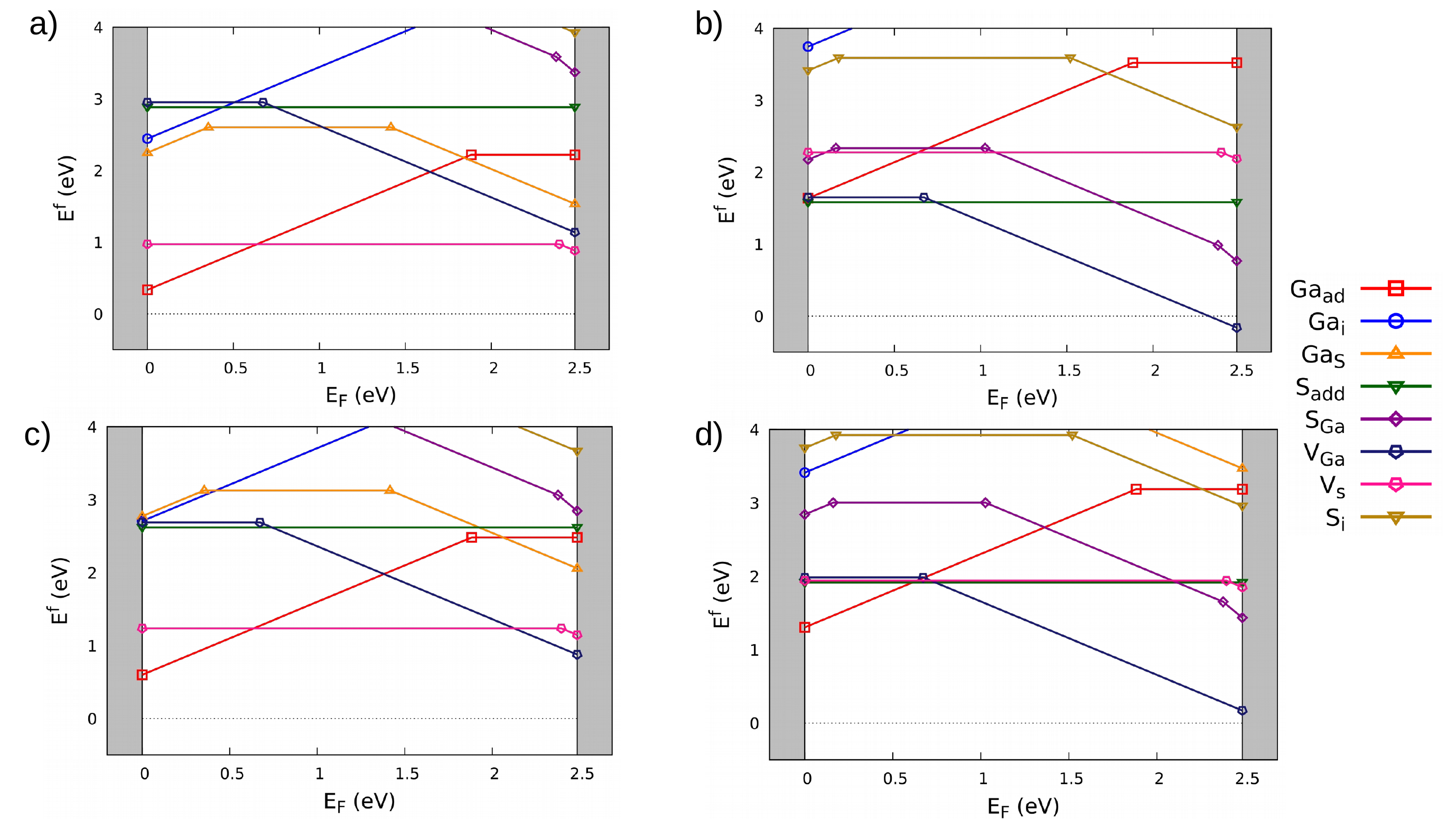} 
\vspace{-0.4cm}
\caption{\label{f:Formation_FinTemp_GaS} Formation energies of native defects in monolayer GaS at 700 K for the (a) Ga-rich limit and the (b) S-rich limit, and at 1000 K for the (c) Ga-rich limit and the (d) S-rich limit. Results obtained with PBE-D and not given with respect to vacuum.}
\end{figure*}

\subsection{\label{CTLs} Transition levels and intrinsic doping}

It may be tempting to directly compare calculated Kohn-Sham (KS) states for a given system with experiments. But it is important to keep in mind that KS states, even the ones obatined with hybrid functionals, cannot be directly associated to any experimentally significant levels~\cite{RevModPhys.86.253}. The relevance of CTLs to characterize defects in semiconductors lies on the fact that they are, as mentioned in Sec.~\ref{Intro_FormationEnergies}, one of the quantities which can be predicted via ab-initio methods and are amenable to comparison with experiment. This interesting feature of CTLs is rooted on the fact that they are calculated using Eq.~\ref{e:charge_transition_level} and total energies, which are well described by DFT. Furthermore, although defects can go through excitations in which their charge state does not change, it is more common for them to trap charge carriers. This is the process described by the CTLs.

In Fig.~\ref{f:structures} we present the relaxed geometries of the studied native defects in monolayer GaS and in their charge neutral state. We show optimized geometries as obtained with PBE-D, which delivers a better description for out-of-plane defects, i.e. adatoms, and we do not find considerable differences with HSE results for defects inside the monolayer. Since relevant features of the optimized geometries of the native defects are common to both monolayer GaS and GaSe, we only show the corresponding GaSe configurations in the Supplementary Material.  

In addition, we present the vacuum-aligned transition levels of all relevant defects for both monolayers in Fig.~\ref{f:transition_levels}. For most defects we present the obtained CTLs using both PBE-D and HSE06. However, due to the previously discussed fact that HSE06 does not account for van der Waals interactions, for adatoms we only show the PBE-D results. Naturally, due to its more accurate description of the band gap, CTLs predicted using HSE06 are expected to be more accurate at $T = 0$ K and are the ones we use to draw conclusions on the electrical properties of the studied defects. Nevertheless, by showing both PBE-D and HSE06 results in the same figure, we are able to observe that both functionals provide a consistent picture of the electrical properties of the defects and predict CTLs that agree well when given with respect to vacuum. This first general conclusion is fundamental to explain why we use PBE-D for finite-temperature calculations, as we argue in Sec.~\ref{finite-temp}.  

\subsubsection{\label{GAi} Gallium interstitial}    

The optimized atomic configuration of the neutral Ga interstitial, Ga$_{\rm i}$, is shown in Fig.~\ref{f:structures}(a). As we can see there, at this charge state such interstitial has only a marginal effect on the positions of its surrounding atoms. Since Ga$_{\rm i}$ is located in the middle of the monolayer, its optimized geometry indicates that the $D_{3h}$ symmetry of the pristine monolayer is retained. The electronic structure of the neutral Ga$_{\rm i}$ features three occupied defect states within the band gap, deep in the case of the GaS monolayer and close to the conduction band in the GaSe system. These findings suggest that the Ga$_{\rm i}$ may be prone to donate electrons, which is actually the case. As we can see in Figs.~\ref{f:transition_levels}(a) and ~\ref{f:transition_levels}(b), with respect to the VBM this defect has $\epsilon(+1/0) = 2.49$ eV and $\epsilon(+1/0) = 2.31$ eV for GaS and GaSe, respectively. This means that for both materials the Ga$_{\rm i}$ is a deep donor, with ionization energies larger than 0.65 eV. Furthermore, also for both materials, as the defect donates the electron it exhibits a Jahn-Teller distortion and the symmetry is broken to $C_s$.

\subsubsection{\label{antisites} Antisites} 

Contrary to the case of the Ga$_{\rm i}$, the antisites in monolayer GaS and GaSe induce significant changes in the positions of their surrounding atoms. Their relaxed configurations in the neutral state and for the GaS system are shown in Fig.~\ref{f:structures}(b) and Fig.~\ref{f:structures}(c). For the corresponding relaxed structures in monolayer GaSe please refer to the supplementary material. For both monolayers and for all relevant charge states the antisites reduce the $D_{3h}$ symmetry to a $C_{3v}$ case. We do not find extra symmetry breaking when these defects become charged. As they trap or donate electrons, the antisites in the monolayers under study induce localized, but symmetric, displacements in their surrounding atoms. 

In the specific cases of the Ga$_{\rm S}$ and Ga$_{\rm Se}$, their electronic structure is characterized by three defects states in the gap, one occupied and two unoccupied. Such features suggest that these antisites can, in principle, be amphoteric. The obtained CTLs for Ga$_{\rm S}$ and Ga$_{\rm Se}$ are presented in Figs.~\ref{f:transition_levels}(a) and ~\ref{f:transition_levels}(b), and detailed values of the CTLs are summarized in Table~\ref{antisites_table}. As we can see in both the figures and the table, the amphoteric character of the Ga$_{\rm S}$ and Ga$_{\rm Se}$ is confirmed by the calculated CTLs. In addition, based in the position of the CTLs, we conclude that these antisites can behave as both deep acceptors or deep donors.
       
Regarding the S$_{\rm Ga}$ and Se$_{\rm Ga}$ antisites, their electronic structure is dominated by three defect states in the gap, one occupied and very close to the VBM and two unoccupied and deep in the gap. Due to these features, as in the case of Ga$_{\rm S}$ and Ga$_{\rm Se}$, one would expect that both S$_{\rm Ga}$ and Se$_{\rm Ga}$ antisites have an amphoteric behavior. The calculated CTLs, shown in Figs.~\ref{f:transition_levels}(a) and ~\ref{f:transition_levels}(b) and summarized in Table~\ref{antisites_table}, confirm their amphoteric character. The main difference with respect to the case of Ga$_{\rm S}$ and Ga$_{\rm Se}$, is that both S$_{\rm Ga}$ and Se$_{\rm Ga}$ exhibit also an extremely deep $\epsilon(-1/-2)$, that would be stable only under strong $n$-type doping. Based on the position of the CTLs, we can predict that both S$_{\rm Ga}$ and Se$_{\rm Ga}$ behave as deep acceptors or deep donors.

\renewcommand{\arraystretch}{1.3}
\setlength{\tabcolsep}{3pt}
\begin{table}[ht]\label{antisites_table}
\normalsize

\caption{Charge transition levels, CTLs, for the antisites in monolayer GaS and GaSe. Values obtained using the HSE06 functional and given in eV and with respect to the corresponding VBM.}

\begin{tabular}{cccccc}

\hhline{======}

\multicolumn{1}{l}{}        &  & $\epsilon(+1/0)$ & $\epsilon(0/-1)$ & $\epsilon(-1/-2)$ \\ \midrule[0.1mm]
\multirow{2}{*}{GaS} 	& Ga$_{\rm S}$ & 0.83 & 1.88 & - \\
                     	& S$_{\rm Ga}$ & 0.25 & 1.73 & 2.75  \\  \midrule[0.1mm]
\multirow{2}{*}{GaSe}      & Ga$_{\rm Se}$ & 0.69 & 1.53 & - \\ 	
                           & Se$_{\rm Ga}$ & 0.82 & 1.41 & 2.81 \\


\hhline{======}

\end{tabular}

\end{table}

\begin{figure*}[t]
\includegraphics[scale=0.62]{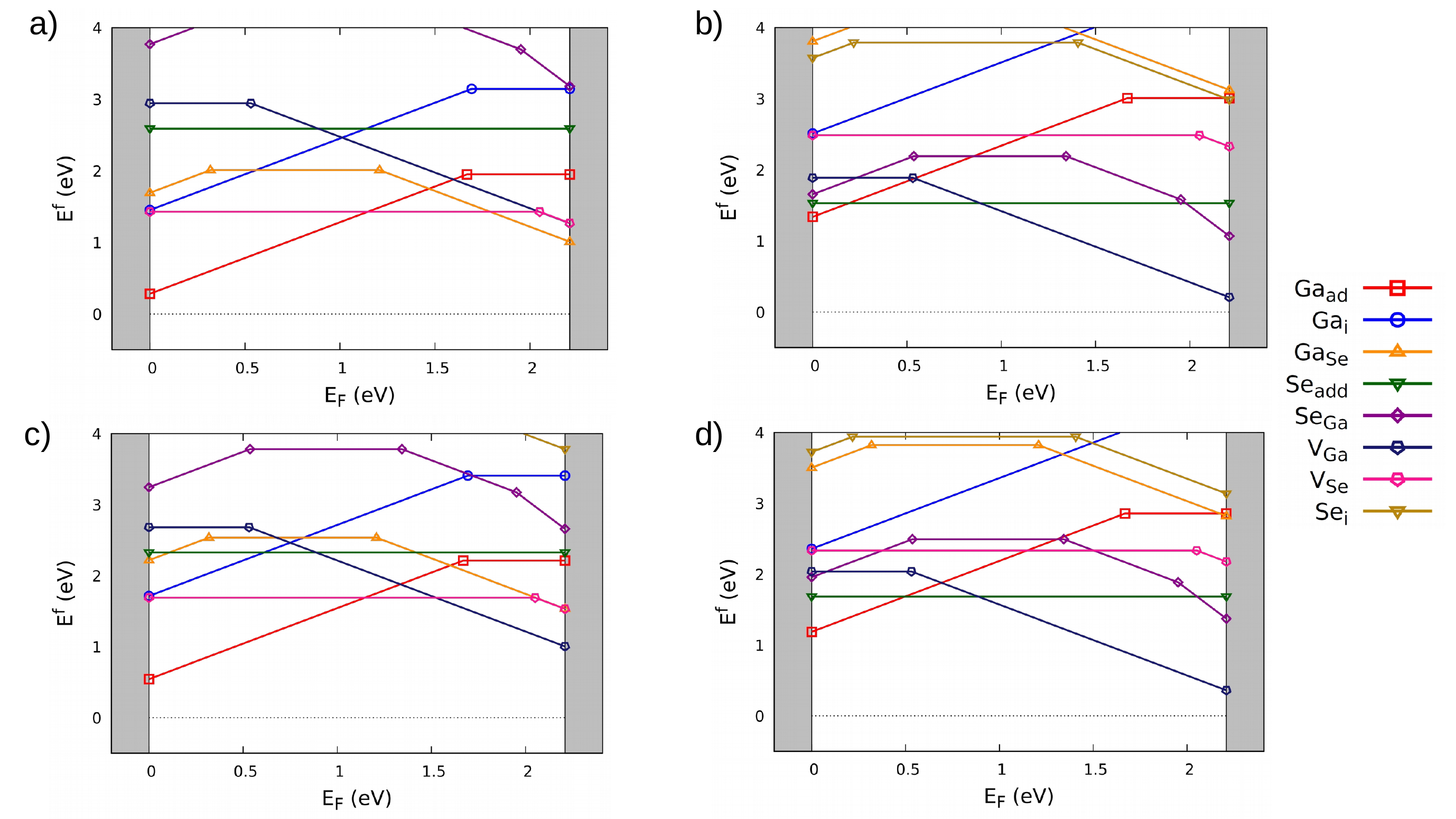} 
\vspace{-0.4cm}
\caption{\label{f:Formation_FinTemp_GaSe} Formation energies of native defects in monolayer GaSe at 700 K for the (a) Ga-rich limit and the (b) Se-rich limit, and at 1000 K for the (c) Ga-rich limit and the (d) Se-rich limit.}
\end{figure*}

\subsubsection{\label{anion_inter} Chalcogenide interstitials} 

Analogous to the case of the Ga$_{\rm i}$ and as can be seen in Fig.~\ref{f:structures}(d), the neutral chalcogenide interstitials S$_{\rm i}$ and Se$_{\rm i}$ do not induce large changes in the positions of their surrounding atoms and retain the $D_{3h}$ symmetry of the pristine monolayers. However, contrary to what we found for Ga$_{\rm i}$, associated to the charging process of both S and Se interstitials, we only observe localized but symmetric distortions, i.e., the $D_{3h}$ symmetry is kept even in their charged states. Regarding their electronic structure, the chalcogenide interstitials feature three defect states in the band gap, two of which are occupied. Among these defects states, only one, induced by Se$_{\rm i}$, is located close to the VBM and the rest are deep within the band gap. These results suggest again an amphoteric tendency, which is confirmed by the calculated CTLs shown in Figs.~\ref{f:transition_levels}(a) and ~\ref{f:transition_levels}(b) and summarized in Table~\ref{anion_inter_table}. Based on the position of the CTLs, we can predict that both S$_{\rm i}$ and Se$_{\rm i}$ behave as deep acceptors or deep donors.

\renewcommand{\arraystretch}{1.3}
\setlength{\tabcolsep}{3pt}
\begin{table}[ht]\label{anion_inter_table}
\normalsize

\caption{Charge transition levels, CTLs, for the chalcogenide interstitials in monolayer GaS and GaSe. Values obtained using the HSE06 functional and given in eV and with respect to the corresponding VBM.}

\begin{tabular}{cccccc}

\hhline{======}

\multicolumn{1}{l}{}        &  & $\epsilon(+1/0)$ & $\epsilon(0/-1)$ & $\epsilon(-1/-2)$ \\ \midrule[0.1mm]
\multirow{1}{*}{GaS}       & S$_{\rm i}$ & 0.46 & 2.02 & 3.14  \\  \midrule[0.1mm]
\multirow{1}{*}{GaSe}      & Se$_{\rm i}$ & 0.55 & 1.9 & 2.83 \\


\hhline{======}

\end{tabular}

\end{table}

\subsubsection{\label{vacancies} Vacancies}

In the case of the gallium vacancy $\rm V_{Ga}$, we do observe large relaxations associated to the presence of the defect, Fig.~\ref{f:structures}(e). Specifically, the neighbouring gallium atom relaxes to the middle of the monolayer. Due to this change, the neutral $\rm V_{Ga}$ retains the $D_{3h}$ symmetry of the non-defective monolayers. Such symmetric configuration is also observed when the $\rm V_{Ga}$  traps an extra electron, a process that only induces a symmetric outward "breathing" distortion affecting the atoms surrounding the vacancy. The electronic structure of this defect is characterized by the appearance of three defect states in the band gap: one just above the VBM, occupied (unoccupied) in its spin-up (spin-down) channel, and two unoccupied and deep in the gap. In principle, with such features one would expect a tendency towards trapping extra electrons for the $\rm V_{Ga}$. As we can see in Figs.~\ref{f:transition_levels}(a) and ~\ref{f:transition_levels}(b), and in Table~\ref{vacancies_table} this defect is a deep acceptor with ionization energies larger than 0.7 eV for both monolayers.

Chalcogenide vacancies, $\rm V_{S}$ and $\rm V_{Se}$, break the symmetry of the system to $C_{3v}$ as can be seen in Fig.~\ref{f:structures}(f). Although relaxations are observed when these vacancies change their charge state, the symmetry is not lowered further. For S and Se vacancies the electronic structure is dominated by three defect states, all of which are deep in the gap and only one is occupied. In principle, these antisites could then be either positively or negatively charged. However, the obtained CTLs presented Figs.~\ref{f:transition_levels}(a) and ~\ref{f:transition_levels}(b), and in Table~\ref{vacancies_table} show that such charging processes would happen only under extreme conditions. Our assesment is that $\rm V_{S}$ and $\rm V_{S}$ are extremely deep amphoteric defects.

\renewcommand{\arraystretch}{1.3}
\setlength{\tabcolsep}{3pt}
\begin{table}[ht]\label{vacancies_table}
\normalsize

\caption{Charge transition levels, CTLs, for the single vacancies in monolayer GaS and GaSe. Values obtained using the HSE06 functional and given in eV and with respect to the corresponding VBM.}

\begin{tabular}{cccccc}

\hhline{======}

\multicolumn{1}{l}{}        &  & $\epsilon(+1/0)$ & $\epsilon(0/-1)$ \\ \midrule[0.1mm]
\multirow{2}{*}{GaS} 	& V$_{\rm Ga}$ & - & 0.95 \\
                     	& V$_{\rm S}$ & 0.39 & 3.08  \\  \midrule[0.1mm]
\multirow{2}{*}{GaSe}      & V$_{\rm Ga}$ & - & 0.79 \\ 	
                           & V$_{\rm Se}$ & 0.4 & 2.53 \\


\hhline{======}

\end{tabular}

\end{table}


\subsubsection{\label{adatoms} Adatoms}

As mentioned at the beginning of this section, conclusions regarding the electrical properties of intrinsic defects in monolayer GaS and GaSe should be obtained from CTLs calculated using hybrids. Unfortunately, since HSE06 does not account for van der Waals interactions, we cannot use it to correctly predict the relaxed configuration of out-of-plane defects. Thus, it cannot deliver reliable results for adatoms. This is the reason why for these defects, Figs.~\ref{f:transition_levels}(a) and ~\ref{f:transition_levels}(b) only show PBE-D results. In the case of gallium adatoms, whose relaxed configuration is directly on top of the center of the hexagons as shown in Fig.~\ref{f:structures}(g), PBE-D predicts $\epsilon(+1/0) = 1.88$ eV and $\epsilon(+1/0) = 1.67$ eV for monolayer GaS and GaSe, respectively. In addition, PBE-D predicts that S and Se adatoms prefer to be located on top of chalcogenide sites, see Fig.~\ref{f:structures}(h), and are always neutral. The key to draw relevant conclusions regarding the electrical activity of adatoms from these PBE-D results is that, despite suffering from the band gap problem, CTLs obtained using PBE-D agree well with HSE06 ones when aligned with respect to vacuum. This means that opening the gap would only result in gallium adatoms being deep donors, even deeper compared to what PBE-D predicts, and the chalcogenide adatoms are expected to be electrically inactive as they are neutral for all doping conditions.

\subsubsection{\label{intrinsic_doping} Native defects and intrinsic doping}

The central goal of our research is to shed light on the currently unclear origin and nature of intrinsic doping in monolayer GaS and GaSe. Experimental studies have measured shifts in the Fermi level and have proposed defects as a possible explanation for such behavior~\cite{PhysRevB.96.035407, Chen2018, PhysRevB.98.115405}. However, our results do not agree with this hypothesis. As shown in Figs.~\ref{f:transition_levels}(a) and ~\ref{f:transition_levels}(b), no native defect exhibits a shallow nature. Specifically, in the monolayers under study all ionization energies are above 0.6 eV. This means that native defects in these systems cannot be easily ionized and are not efficient dopants. Therefore, we conclude that the intrinsic doping observed in experiments should be caused by charge transfer with the substrates or by the presence of shallow extrinsic dopants. 

The fact that native defects cannot explain intrinsic doping in monolayer GaS and GaSe does not mean that they are irrelevant. Actually, it is quite the opposite situation. Being deep defects, they can act as carrier traps and degrade the efficiency of any device which uses these monolayers. In addition, being stable in different charge states, the studied native defects can act as compensating centers and, as a consequence, make it difficult to engineer the doping conditions of the monolayers. We dedicate the next section to elucidate which growth conditions can lower the concentrations of harmful defects.

\subsection{\label{finite-temp} Formation energies at realistic growth conditions}

We now use the methodology described in Sec.~\ref{Intro_FormationEnergies} and tackle the calculation of formation energies at finite temperatures. To that end, we keep in mind that high-quality monolayers of GaS and GaSe are grown at about 700 K via MBE ~\cite{BenAziza2016, Choong2017, Chen2018}, and at about 1000 K via VPT~\cite{Lei2013, Li2014, Li2015, Lie1501882, Li2017}. Thus, in this section we present our results for these two experimentally relevant temperatures. In Sec.~\ref{Intro_FormationEnergies} we clearly explained that in our approach, based on Ref.~\cite{Komsa2015}, the temperature dependence entered via the chemical potentials. However, to be closer to reality, the temperature dependence of the formation energies should also enter via changes in the band gap, which reduces as temperature rises. For bulk GaS and GaSe it has been reported that the band gap decreases by about 0.45 meV/K and 0.51 meV/K, respectively~\cite{temp_GaS,temp_GaSe}. Assuming a rate of the same magnitude for the monolayers means that at 1000 K the experimental band gap would be reduced to $E_g = 3.15$ eV for monolayer GaS and to $E_g = 2.99$ eV for monolayer GaSe. Actually, assuming an analogous behavior to the one observed for MoS$_2$, for which the rate of change of the band gap increases from 0.4 meV/K in bulk~\cite{Ho_1998,doi:10.1063/1.4882301} to 0.7 meV/K in the monolayers~\cite{Choi2017}, one would expect that the experimental band gaps of monolayer GaS and GaSe could be reduced further to $E_g \simeq 2.8$ eV for monolayer GaS and to $E_g \simeq 2.6$ eV, respectively. Based on the values reported in Table~\ref{basic} and keeping in mind that the fundamental gap decreases due to the presence of substrates, we can safely say that at high temperatures PBE-D should deliver band gaps, defect formation energies and CTLs more in line with experiments. In addition, since it takes van der Waals interactions into account, PBE-D allows us to correctly describe the physics of adatoms. Due to these two relevant advantages, we adopted PBE-D for our finite-temperature calculations. The resulting formation energies for both monolayers at 700 K and 1000 K are presented in Figs.~\ref{f:Formation_FinTemp_GaS} and~\ref{f:Formation_FinTemp_GaSe}. In there, chemical potentials are obtained using the methodology described in Sec.~\ref{Intro_FormationEnergies} and from Fig.~\ref{f:chem_pot}. Ideally, one would use the finite temperature formation energies and translate them to concentrations~\cite{RevModPhys.86.253}. The usual procedure to do so implies the self-consistent calculation of the intrinsic Fermi level~\cite{RevModPhys.86.253,Komsa2015}. Although simple and clear, this method cannot be directly applied to monolayers due to the strong influence of substrates, i.e., the Fermi level in the monolayer is likely to be far from the intrinsic one. Therefore, in the following, we use the calculated formation energies and differences between them to qualitatively draw conclusions about the relative abundance of specific defects. 

For Ga-rich conditions at 700 K, Figs.~\ref{f:Formation_FinTemp_GaS}(a) and~\ref{f:Formation_FinTemp_GaSe}(a), the native defects that are most likely to occur are chalcogenide vacancies and gallium adatoms.  A little bit less common, but expected to occur in considerable quantities, are gallium vacancies and the Ga$_{\rm S}$/Ga$_{\rm Se}$ antisites. For the same Ga-rich conditions but at 1000 K, Figs.~\ref{f:Formation_FinTemp_GaS}(c) and~\ref{f:Formation_FinTemp_GaSe}(c), we predict a more or less equal defect landscape, with the only difference of slightly larger formation energies for chalcogenide vacancies, gallium adatoms and Ga$_{\rm S}$/Ga$_{\rm Se}$ antisites, and slightly lower formation energies for gallium vacancies. The consequence of these findings can be understood based on the discussion presented in Sec.~\ref{CTLs}. On the one hand, S and Se vacancies are expected to be electrically harmless. On the other hand, gallium adatoms, gallium vacancies and the Ga$_{\rm S}$/Ga$_{\rm Se}$ antisites exhibit deep CTLs and are expected to be harmful. Specifically, Ga$_{\rm S}$/Ga$_{\rm Se}$ have relatively high formation energies and are not expected to play a relevant role as compensation center. However, they do feature a deep CTLs very close to the center of the band gap in both monolayers and are expected to be very efficient charge carrier traps. In the case of gallium adatoms (deep donors) and gallium vacancies (deep acceptors), our results indicate that both are expected to be efficient compensation centers. Naturally, under Ga-rich conditions, gallium adatoms are more likely to occur than gallium vacancies, causing an asymmetric behavior with respect to doping. In other words, when grown under Ga-rich conditions, monolayer GaS and GaSe are easier to dope \textit{n}-type than \textit{p}-type.    
  
In the case of chalcogenide-rich conditions at 700 K, Figs.~\ref{f:Formation_FinTemp_GaS}(b) and~\ref{f:Formation_FinTemp_GaSe}(b), the native defects that are most likely to occur are S and Se adatoms, gallium vacancies and gallium adatoms.  Due to their slightly higher, but still low, formation energies we predict considerable amounts of chalcogenide vacancies and S$_{\rm Ga}$/Se$_{\rm Ga}$ antisites. As expected, at 1000 K and the same chalcogenide-rich conditions, Figs.~\ref{f:Formation_FinTemp_GaS}(d) and~\ref{f:Formation_FinTemp_GaSe}(d), we predict an essentially identical defect landscape. At such temperature chalcogenide adatoms, gallium vacancies and S$_{\rm Ga}$/Se$_{\rm Ga}$ antisites exhibit slightly larger formation energies, and gallium adatoms have slightly lower formation energies. Based on the discussion presented in Sec.~\ref{CTLs} we know that S and Se adatoms are expected to be electrically harmless and that gallium adatoms, gallium vacancies and S$_{\rm Ga}$/Se$_{\rm Ga}$ antisites induce deep CTLs and are expected to be harmful. Analogous to what we found for Ga$_{\rm S}$/Ga$_{\rm Se}$ antisites in Ga-rich conditions, when samples are grown under chalcogenide-rich conditions, S$_{\rm Ga}$/Se$_{\rm Ga}$ antisites have relatively high formation energies and are not expected to be efficient compensation centers but very efficient charge carrier traps. Regarding compensation in samples grown under chalcogenide-rich conditions, we predict a strongly asymmetric process driven by the presence of gallium adatoms (deep donors) and gallium vacancies (deep acceptors), and heavily dominated by the latter. Thus, our results show that monolayers GaS and GaSe grown under chalcogenide-rich conditions will be far easier to dope \textit{p}-type than \textit{n}-type. The reason for the strong asymmetry is caused by the large difference in formation energies of the two defect types.    

\begin{figure}[t]
\includegraphics[width=1\columnwidth]{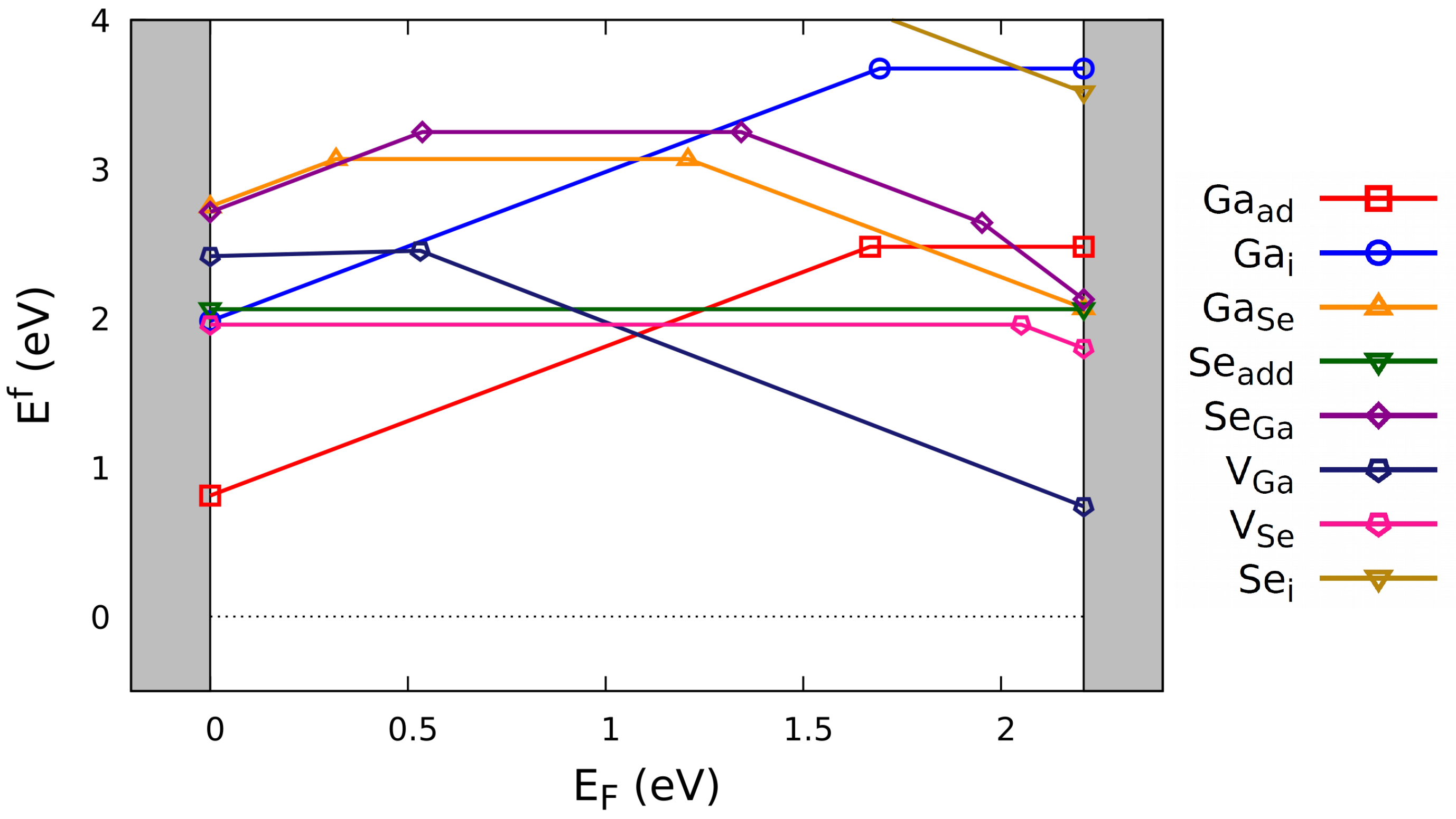} 
\vspace{-0.4cm}
\caption{\label{f:intermediate} Formation energies of native defects in monolayer GaSe at 700 k for intermediate growth conditions, $\mu_{\rm Ga} = -3.84$ eV and $\mu_{\rm Se} = -4.3$ eV. Results obtained with PBE-D and not given with respect to vacuum.}
\end{figure} 

At this point we are certain that, although some native defects are harmless, most of them are harmful and should be avoided. To this end we can use a strong tendency hidden in Figs.~\ref{f:Formation_FinTemp_GaS} and~\ref{f:Formation_FinTemp_GaSe}: harmful defects observed at Ga-rich conditions have higher formation energies in chalcogenide-rich conditions, and vice versa. Therefore, we propose that growing monolayer GaS and GaSe at intermediate conditions should increase (decrease) the formation energies (concentration) of all harmful defects at the same time. As an example of the approach, in Fig.~\ref{f:intermediate}  we show the formation energies of native defects in monolayer GaSe at 700 k for intermediate growth conditions, $\mu_{\rm Ga} = -3.84$ eV and $\mu_{\rm Se} = -4.3$ eV (which are certainly within the stability region of the monolayer, see Fig.~\ref{f:chem_pot}). In there we can see how such small change in the chemical potentials leads to a situation where the formation energies of all harmful defects (gallium/chalcogenide vacancies and antisites) are simultaneously increased above $\sim 1$ eV. Since most high-quality monolayer GaS and GaSe grown via MBE or VPT methods are actually grown in chalcogenide-rich conditions, our proposal to experimentalists is to slightly increase the gallium concentration or decrease the S/Se partial pressures.

\section{\label{conclusions} CONCLUSIONS}

We have performed first-principles calculations to assess the electrical and thermodynamic properties of native defects in GaS and GaSe monolayers. To that end we calculated their charge transition levels and formation energies, using state-of-the-art schemes to account for finite-size effects and the temperature dependence of the chemical potentials. Based on the position of the obtained charge transition levels, we conclude that among the studied native defects there are no shallow dopants and the experimentally observed intrinsic doping should then be caused by charge transfer to/from the substrates or by the presence of shallow extrinsic dopants. In addition, the position of the charge transition levels also allowed us to predict that most native defects are efficient compensation and recombination centers. Thus, native defects should be avoided in order to grow technologically reliable GaS and GaSe monolayers. By studying relevant formation energies at realistic growth temperatures, we are able to prove that by growing GaS and GaSe monolayers at intermediate conditions, far from the chalcogenide- or Ga-rich regimes, it should be possible to simultaneously decrease the concentration of all relevant harmful native defects. In other words, when growing monolayer GaS and GaSe under S/Se-rich conditions, our findings translate to a simple experimental directive: to slightly increase the gallium concentration or to slightly decrease the S/Se partial pressures.

\bibliography{DefectsTHERMO_GA}{}

\begin{thebibliography}{99}%
\makeatletter
\providecommand \@ifxundefined [1]{%
 \@ifx{#1\undefined}
}%
\providecommand \@ifnum [1]{%
 \ifnum #1\expandafter \@firstoftwo
 \else \expandafter \@secondoftwo
 \fi
}%
\providecommand \@ifx [1]{%
 \ifx #1\expandafter \@firstoftwo
 \else \expandafter \@secondoftwo
 \fi
}%
\providecommand \natexlab [1]{#1}%
\providecommand \enquote  [1]{``#1''}%
\providecommand \bibnamefont  [1]{#1}%
\providecommand \bibfnamefont [1]{#1}%
\providecommand \citenamefont [1]{#1}%
\providecommand \href@noop [0]{\@secondoftwo}%
\providecommand \href [0]{\begingroup \@sanitize@url \@href}%
\providecommand \@href[1]{\@@startlink{#1}\@@href}%
\providecommand \@@href[1]{\endgroup#1\@@endlink}%
\providecommand \@sanitize@url [0]{\catcode `\\12\catcode `\$12\catcode
  `\&12\catcode `\#12\catcode `\^12\catcode `\_12\catcode `\%12\relax}%
\providecommand \@@startlink[1]{}%
\providecommand \@@endlink[0]{}%
\providecommand \url  [0]{\begingroup\@sanitize@url \@url }%
\providecommand \@url [1]{\endgroup\@href {#1}{\urlprefix }}%
\providecommand \urlprefix  [0]{URL }%
\providecommand \Eprint [0]{\href }%
\providecommand \doibase [0]{http://dx.doi.org/}%
\providecommand \selectlanguage [0]{\@gobble}%
\providecommand \bibinfo  [0]{\@secondoftwo}%
\providecommand \bibfield  [0]{\@secondoftwo}%
\providecommand \translation [1]{[#1]}%
\providecommand \BibitemOpen [0]{}%
\providecommand \bibitemStop [0]{}%
\providecommand \bibitemNoStop [0]{.\EOS\space}%
\providecommand \EOS [0]{\spacefactor3000\relax}%
\providecommand \BibitemShut  [1]{\csname bibitem#1\endcsname}%
\let\auto@bib@innerbib\@empty
\bibitem [{\citenamefont {Li}\ \emph {et~al.}(2017{\natexlab{a}})\citenamefont
  {Li}, \citenamefont {Tao}, \citenamefont {Chen}, \citenamefont {Fang},
  \citenamefont {Li}, \citenamefont {Wang}, \citenamefont {Xu},\ and\
  \citenamefont {Zhu}}]{Review_Zhu}%
  \BibitemOpen
  \bibfield  {author} {\bibinfo {author} {\bibfnamefont {X.}~\bibnamefont
  {Li}}, \bibinfo {author} {\bibfnamefont {L.}~\bibnamefont {Tao}}, \bibinfo
  {author} {\bibfnamefont {Z.}~\bibnamefont {Chen}}, \bibinfo {author}
  {\bibfnamefont {H.}~\bibnamefont {Fang}}, \bibinfo {author} {\bibfnamefont
  {X.}~\bibnamefont {Li}}, \bibinfo {author} {\bibfnamefont {X.}~\bibnamefont
  {Wang}}, \bibinfo {author} {\bibfnamefont {J.-B.}\ \bibnamefont {Xu}}, \ and\
  \bibinfo {author} {\bibfnamefont {H.}~\bibnamefont {Zhu}},\ }\href {\doibase
  10.1063/1.4983646} {\bibfield  {journal} {\bibinfo  {journal} {Applied
  Physics Reviews}\ }\textbf {\bibinfo {volume} {4}},\ \bibinfo {pages}
  {021306} (\bibinfo {year} {2017}{\natexlab{a}})}\BibitemShut {NoStop}%
\bibitem [{\citenamefont {Castro~Neto}\ \emph {et~al.}(2009)\citenamefont
  {Castro~Neto}, \citenamefont {Guinea}, \citenamefont {Peres}, \citenamefont
  {Novoselov},\ and\ \citenamefont {Geim}}]{Rev_GraphElectro}%
  \BibitemOpen
  \bibfield  {author} {\bibinfo {author} {\bibfnamefont {A.~H.}\ \bibnamefont
  {Castro~Neto}}, \bibinfo {author} {\bibfnamefont {F.}~\bibnamefont {Guinea}},
  \bibinfo {author} {\bibfnamefont {N.~M.~R.}\ \bibnamefont {Peres}}, \bibinfo
  {author} {\bibfnamefont {K.~S.}\ \bibnamefont {Novoselov}}, \ and\ \bibinfo
  {author} {\bibfnamefont {A.~K.}\ \bibnamefont {Geim}},\ }\href {\doibase
  10.1103/RevModPhys.81.109} {\bibfield  {journal} {\bibinfo  {journal} {Rev.
  Mod. Phys.}\ }\textbf {\bibinfo {volume} {81}},\ \bibinfo {pages} {109}
  (\bibinfo {year} {2009})}\BibitemShut {NoStop}%
\bibitem [{\citenamefont {Katsnelson}\ and\ \citenamefont
  {Kat︠s︡nelʹson}(2012)}]{katsnelson2012graphene}%
  \BibitemOpen
  \bibfield  {author} {\bibinfo {author} {\bibfnamefont {M.}~\bibnamefont
  {Katsnelson}}\ and\ \bibinfo {author} {\bibfnamefont {M.}~\bibnamefont
  {Kat︠s︡nelʹson}},\ }\href
  {https://books.google.co.uk/books?id=FgwW3aVRBT0C} {\emph {\bibinfo {title}
  {Graphene: Carbon in Two Dimensions}}}\ (\bibinfo  {publisher} {Cambridge
  University Press},\ \bibinfo {year} {2012})\BibitemShut {NoStop}%
\bibitem [{\citenamefont {Novoselov}\ \emph {et~al.}(2004)\citenamefont
  {Novoselov}, \citenamefont {Geim}, \citenamefont {Morozov}, \citenamefont
  {Jiang}, \citenamefont {Zhang}, \citenamefont {Dubonos}, \citenamefont
  {Grigorieva},\ and\ \citenamefont {Firsov}}]{Novoselov1}%
  \BibitemOpen
  \bibfield  {author} {\bibinfo {author} {\bibfnamefont {K.~S.}\ \bibnamefont
  {Novoselov}}, \bibinfo {author} {\bibfnamefont {A.~K.}\ \bibnamefont {Geim}},
  \bibinfo {author} {\bibfnamefont {S.~V.}\ \bibnamefont {Morozov}}, \bibinfo
  {author} {\bibfnamefont {D.}~\bibnamefont {Jiang}}, \bibinfo {author}
  {\bibfnamefont {Y.}~\bibnamefont {Zhang}}, \bibinfo {author} {\bibfnamefont
  {S.~V.}\ \bibnamefont {Dubonos}}, \bibinfo {author} {\bibfnamefont {I.~V.}\
  \bibnamefont {Grigorieva}}, \ and\ \bibinfo {author} {\bibfnamefont {A.~A.}\
  \bibnamefont {Firsov}},\ }\href {\doibase 10.1126/science.1102896} {\bibfield
   {journal} {\bibinfo  {journal} {Science}\ }\textbf {\bibinfo {volume}
  {306}},\ \bibinfo {pages} {666} (\bibinfo {year} {2004})}\BibitemShut
  {NoStop}%
\bibitem [{\citenamefont {Galiotis}\ \emph {et~al.}(2015)\citenamefont
  {Galiotis}, \citenamefont {Frank}, \citenamefont {Koukaras},\ and\
  \citenamefont {Sfyris}}]{RevGraphMech}%
  \BibitemOpen
  \bibfield  {author} {\bibinfo {author} {\bibfnamefont {C.}~\bibnamefont
  {Galiotis}}, \bibinfo {author} {\bibfnamefont {O.}~\bibnamefont {Frank}},
  \bibinfo {author} {\bibfnamefont {E.~N.}\ \bibnamefont {Koukaras}}, \ and\
  \bibinfo {author} {\bibfnamefont {D.}~\bibnamefont {Sfyris}},\ }\href
  {\doibase 10.1146/annurev-chembioeng-061114-123216} {\bibfield  {journal}
  {\bibinfo  {journal} {Annual Review of Chemical and Biomolecular
  Engineering}\ }\textbf {\bibinfo {volume} {6}},\ \bibinfo {pages} {121}
  (\bibinfo {year} {2015})}\BibitemShut {NoStop}%
\bibitem [{\citenamefont {Raza}(2012)}]{raza2012graphene}%
  \BibitemOpen
  \bibfield  {author} {\bibinfo {author} {\bibfnamefont {H.}~\bibnamefont
  {Raza}},\ }\href {https://books.google.lu/books?id=XBjXpLwWCEsC} {\emph
  {\bibinfo {title} {Graphene Nanoelectronics: Metrology, Synthesis, Properties
  and Applications}}},\ NanoScience and Technology\ (\bibinfo  {publisher}
  {Springer Berlin Heidelberg},\ \bibinfo {year} {2012})\BibitemShut {NoStop}%
\bibitem [{\citenamefont {Schwierz}(2010)}]{Schwierz2010}%
  \BibitemOpen
  \bibfield  {author} {\bibinfo {author} {\bibfnamefont {F.}~\bibnamefont
  {Schwierz}},\ }\href {\doibase 10.1038/nnano.2010.89} {\bibfield  {journal}
  {\bibinfo  {journal} {Nature Nanotechnology}\ }\textbf {\bibinfo {volume}
  {5}},\ \bibinfo {pages} {487} (\bibinfo {year} {2010})}\BibitemShut {NoStop}%
\bibitem [{\citenamefont {Han}\ \emph {et~al.}(2014)\citenamefont {Han},
  \citenamefont {Kawakami}, \citenamefont {Gmitra},\ and\ \citenamefont
  {Fabian}}]{Han2014}%
  \BibitemOpen
  \bibfield  {author} {\bibinfo {author} {\bibfnamefont {W.}~\bibnamefont
  {Han}}, \bibinfo {author} {\bibfnamefont {R.~K.}\ \bibnamefont {Kawakami}},
  \bibinfo {author} {\bibfnamefont {M.}~\bibnamefont {Gmitra}}, \ and\ \bibinfo
  {author} {\bibfnamefont {J.}~\bibnamefont {Fabian}},\ }\href {\doibase
  10.1038/nnano.2014.214} {\bibfield  {journal} {\bibinfo  {journal} {Nature
  Nanotechnology}\ }\textbf {\bibinfo {volume} {9}},\ \bibinfo {pages} {794}
  (\bibinfo {year} {2014})}\BibitemShut {NoStop}%
\bibitem [{\citenamefont {Bonaccorso}\ \emph {et~al.}(2010)\citenamefont
  {Bonaccorso}, \citenamefont {Sun}, \citenamefont {Hasan},\ and\ \citenamefont
  {Ferrari}}]{Bonaccorso2010}%
  \BibitemOpen
  \bibfield  {author} {\bibinfo {author} {\bibfnamefont {F.}~\bibnamefont
  {Bonaccorso}}, \bibinfo {author} {\bibfnamefont {Z.}~\bibnamefont {Sun}},
  \bibinfo {author} {\bibfnamefont {T.}~\bibnamefont {Hasan}}, \ and\ \bibinfo
  {author} {\bibfnamefont {A.~C.}\ \bibnamefont {Ferrari}},\ }\href {\doibase
  10.1038/nphoton.2010.186} {\bibfield  {journal} {\bibinfo  {journal} {Nature
  Photonics}\ }\textbf {\bibinfo {volume} {4}},\ \bibinfo {pages} {611}
  (\bibinfo {year} {2010})}\BibitemShut {NoStop}%
\bibitem [{\citenamefont {Das}\ \emph {et~al.}(2015)\citenamefont {Das},
  \citenamefont {Robinson}, \citenamefont {Dubey}, \citenamefont {Terrones},\
  and\ \citenamefont {Terrones}}]{annurev_BeyondGraph}%
  \BibitemOpen
  \bibfield  {author} {\bibinfo {author} {\bibfnamefont {S.}~\bibnamefont
  {Das}}, \bibinfo {author} {\bibfnamefont {J.~A.}\ \bibnamefont {Robinson}},
  \bibinfo {author} {\bibfnamefont {M.}~\bibnamefont {Dubey}}, \bibinfo
  {author} {\bibfnamefont {H.}~\bibnamefont {Terrones}}, \ and\ \bibinfo
  {author} {\bibfnamefont {M.}~\bibnamefont {Terrones}},\ }\href {\doibase
  10.1146/annurev-matsci-070214-021034} {\bibfield  {journal} {\bibinfo
  {journal} {Annual Review of Materials Research}\ }\textbf {\bibinfo {volume}
  {45}},\ \bibinfo {pages} {1} (\bibinfo {year} {2015})}\BibitemShut {NoStop}%
\bibitem [{\citenamefont {Manzeli}\ \emph {et~al.}(2017)\citenamefont
  {Manzeli}, \citenamefont {Ovchinnikov}, \citenamefont {Pasquier},
  \citenamefont {Yazyev},\ and\ \citenamefont {Kis}}]{Manzeli2017}%
  \BibitemOpen
  \bibfield  {author} {\bibinfo {author} {\bibfnamefont {S.}~\bibnamefont
  {Manzeli}}, \bibinfo {author} {\bibfnamefont {D.}~\bibnamefont
  {Ovchinnikov}}, \bibinfo {author} {\bibfnamefont {D.}~\bibnamefont
  {Pasquier}}, \bibinfo {author} {\bibfnamefont {O.~V.}\ \bibnamefont
  {Yazyev}}, \ and\ \bibinfo {author} {\bibfnamefont {A.}~\bibnamefont {Kis}},\
  }\href {\doibase 10.1038/natrevmats.2017.33} {\bibfield  {journal} {\bibinfo
  {journal} {Nature Reviews Materials}\ }\textbf {\bibinfo {volume} {2}},\
  \bibinfo {pages} {17033} (\bibinfo {year} {2017})}\BibitemShut {NoStop}%
\bibitem [{\citenamefont {Berkelbach}\ and\ \citenamefont
  {Reichman}(2018)}]{annurev_OptExiTMDs}%
  \BibitemOpen
  \bibfield  {author} {\bibinfo {author} {\bibfnamefont {T.~C.}\ \bibnamefont
  {Berkelbach}}\ and\ \bibinfo {author} {\bibfnamefont {D.~R.}\ \bibnamefont
  {Reichman}},\ }\href {\doibase 10.1146/annurev-conmatphys-033117-054009}
  {\bibfield  {journal} {\bibinfo  {journal} {Annual Review of Condensed Matter
  Physics}\ }\textbf {\bibinfo {volume} {9}},\ \bibinfo {pages} {379} (\bibinfo
  {year} {2018})}\BibitemShut {NoStop}%
\bibitem [{\citenamefont {Schaibley}\ \emph {et~al.}(2016)\citenamefont
  {Schaibley}, \citenamefont {Yu}, \citenamefont {Clark}, \citenamefont
  {Rivera}, \citenamefont {Ross}, \citenamefont {Seyler}, \citenamefont {Yao},\
  and\ \citenamefont {Xu}}]{Schaibley2016}%
  \BibitemOpen
  \bibfield  {author} {\bibinfo {author} {\bibfnamefont {J.~R.}\ \bibnamefont
  {Schaibley}}, \bibinfo {author} {\bibfnamefont {H.}~\bibnamefont {Yu}},
  \bibinfo {author} {\bibfnamefont {G.}~\bibnamefont {Clark}}, \bibinfo
  {author} {\bibfnamefont {P.}~\bibnamefont {Rivera}}, \bibinfo {author}
  {\bibfnamefont {J.~S.}\ \bibnamefont {Ross}}, \bibinfo {author}
  {\bibfnamefont {K.~L.}\ \bibnamefont {Seyler}}, \bibinfo {author}
  {\bibfnamefont {W.}~\bibnamefont {Yao}}, \ and\ \bibinfo {author}
  {\bibfnamefont {X.}~\bibnamefont {Xu}},\ }\href {\doibase
  10.1038/natrevmats.2016.55} {\bibfield  {journal} {\bibinfo  {journal}
  {Nature Reviews Materials}\ }\textbf {\bibinfo {volume} {1}},\ \bibinfo
  {pages} {16055} (\bibinfo {year} {2016})}\BibitemShut {NoStop}%
\bibitem [{\citenamefont {Mak}\ \emph {et~al.}(2018)\citenamefont {Mak},
  \citenamefont {Xiao},\ and\ \citenamefont {Shan}}]{Mak2018}%
  \BibitemOpen
  \bibfield  {author} {\bibinfo {author} {\bibfnamefont {K.~F.}\ \bibnamefont
  {Mak}}, \bibinfo {author} {\bibfnamefont {D.}~\bibnamefont {Xiao}}, \ and\
  \bibinfo {author} {\bibfnamefont {J.}~\bibnamefont {Shan}},\ }\href {\doibase
  10.1038/s41566-018-0204-6} {\bibfield  {journal} {\bibinfo  {journal} {Nature
  Photonics}\ }\textbf {\bibinfo {volume} {12}},\ \bibinfo {pages} {451}
  (\bibinfo {year} {2018})}\BibitemShut {NoStop}%
\bibitem [{\citenamefont {Xu}\ \emph {et~al.}(2016)\citenamefont {Xu},
  \citenamefont {Yin}, \citenamefont {Huang}, \citenamefont {Shifa},
  \citenamefont {Chu}, \citenamefont {Wang}, \citenamefont {Cheng},
  \citenamefont {Wang},\ and\ \citenamefont {He}}]{C6NR05976G}%
  \BibitemOpen
  \bibfield  {author} {\bibinfo {author} {\bibfnamefont {K.}~\bibnamefont
  {Xu}}, \bibinfo {author} {\bibfnamefont {L.}~\bibnamefont {Yin}}, \bibinfo
  {author} {\bibfnamefont {Y.}~\bibnamefont {Huang}}, \bibinfo {author}
  {\bibfnamefont {T.~A.}\ \bibnamefont {Shifa}}, \bibinfo {author}
  {\bibfnamefont {J.}~\bibnamefont {Chu}}, \bibinfo {author} {\bibfnamefont
  {F.}~\bibnamefont {Wang}}, \bibinfo {author} {\bibfnamefont {R.}~\bibnamefont
  {Cheng}}, \bibinfo {author} {\bibfnamefont {Z.}~\bibnamefont {Wang}}, \ and\
  \bibinfo {author} {\bibfnamefont {J.}~\bibnamefont {He}},\ }\href {\doibase
  10.1039/C6NR05976G} {\bibfield  {journal} {\bibinfo  {journal} {Nanoscale}\
  }\textbf {\bibinfo {volume} {8}},\ \bibinfo {pages} {16802} (\bibinfo {year}
  {2016})}\BibitemShut {NoStop}%
\bibitem [{\citenamefont {Cai}\ \emph {et~al.}(2019)\citenamefont {Cai},
  \citenamefont {Gu}, \citenamefont {Lin}, \citenamefont {Yu}, \citenamefont
  {Geohegan},\ and\ \citenamefont {Xiao}}]{Review_Cai}%
  \BibitemOpen
  \bibfield  {author} {\bibinfo {author} {\bibfnamefont {H.}~\bibnamefont
  {Cai}}, \bibinfo {author} {\bibfnamefont {Y.}~\bibnamefont {Gu}}, \bibinfo
  {author} {\bibfnamefont {Y.-C.}\ \bibnamefont {Lin}}, \bibinfo {author}
  {\bibfnamefont {Y.}~\bibnamefont {Yu}}, \bibinfo {author} {\bibfnamefont
  {D.~B.}\ \bibnamefont {Geohegan}}, \ and\ \bibinfo {author} {\bibfnamefont
  {K.}~\bibnamefont {Xiao}},\ }\href {\doibase 10.1063/1.5123487} {\bibfield
  {journal} {\bibinfo  {journal} {Applied Physics Reviews}\ }\textbf {\bibinfo
  {volume} {6}},\ \bibinfo {pages} {041312} (\bibinfo {year}
  {2019})}\BibitemShut {NoStop}%
\bibitem [{\citenamefont {Z\'olyomi}\ \emph {et~al.}(2013)\citenamefont
  {Z\'olyomi}, \citenamefont {Drummond},\ and\ \citenamefont
  {Fal'ko}}]{PhysRevB.87.195403}%
  \BibitemOpen
  \bibfield  {author} {\bibinfo {author} {\bibfnamefont {V.}~\bibnamefont
  {Z\'olyomi}}, \bibinfo {author} {\bibfnamefont {N.~D.}\ \bibnamefont
  {Drummond}}, \ and\ \bibinfo {author} {\bibfnamefont {V.~I.}\ \bibnamefont
  {Fal'ko}},\ }\href {\doibase 10.1103/PhysRevB.87.195403} {\bibfield
  {journal} {\bibinfo  {journal} {Phys. Rev. B}\ }\textbf {\bibinfo {volume}
  {87}},\ \bibinfo {pages} {195403} (\bibinfo {year} {2013})}\BibitemShut
  {NoStop}%
\bibitem [{\citenamefont {Z\'olyomi}\ \emph {et~al.}(2014)\citenamefont
  {Z\'olyomi}, \citenamefont {Drummond},\ and\ \citenamefont
  {Fal'ko}}]{PhysRevB.89.205416}%
  \BibitemOpen
  \bibfield  {author} {\bibinfo {author} {\bibfnamefont {V.}~\bibnamefont
  {Z\'olyomi}}, \bibinfo {author} {\bibfnamefont {N.~D.}\ \bibnamefont
  {Drummond}}, \ and\ \bibinfo {author} {\bibfnamefont {V.~I.}\ \bibnamefont
  {Fal'ko}},\ }\href {\doibase 10.1103/PhysRevB.89.205416} {\bibfield
  {journal} {\bibinfo  {journal} {Phys. Rev. B}\ }\textbf {\bibinfo {volume}
  {89}},\ \bibinfo {pages} {205416} (\bibinfo {year} {2014})}\BibitemShut
  {NoStop}%
\bibitem [{\citenamefont {Rybkovskiy}\ \emph {et~al.}(2014)\citenamefont
  {Rybkovskiy}, \citenamefont {Osadchy},\ and\ \citenamefont
  {Obraztsova}}]{PhysRevB.90.235302}%
  \BibitemOpen
  \bibfield  {author} {\bibinfo {author} {\bibfnamefont {D.~V.}\ \bibnamefont
  {Rybkovskiy}}, \bibinfo {author} {\bibfnamefont {A.~V.}\ \bibnamefont
  {Osadchy}}, \ and\ \bibinfo {author} {\bibfnamefont {E.~D.}\ \bibnamefont
  {Obraztsova}},\ }\href {\doibase 10.1103/PhysRevB.90.235302} {\bibfield
  {journal} {\bibinfo  {journal} {Phys. Rev. B}\ }\textbf {\bibinfo {volume}
  {90}},\ \bibinfo {pages} {235302} (\bibinfo {year} {2014})}\BibitemShut
  {NoStop}%
\bibitem [{\citenamefont {Zhou}\ \emph {et~al.}(2017)\citenamefont {Zhou},
  \citenamefont {Zhang}, \citenamefont {Sun}, \citenamefont {Lou},
  \citenamefont {Zhang}, \citenamefont {Yang},\ and\ \citenamefont
  {Chang}}]{PhysRevB.96.155430}%
  \BibitemOpen
  \bibfield  {author} {\bibinfo {author} {\bibfnamefont {M.}~\bibnamefont
  {Zhou}}, \bibinfo {author} {\bibfnamefont {R.}~\bibnamefont {Zhang}},
  \bibinfo {author} {\bibfnamefont {J.}~\bibnamefont {Sun}}, \bibinfo {author}
  {\bibfnamefont {W.-K.}\ \bibnamefont {Lou}}, \bibinfo {author} {\bibfnamefont
  {D.}~\bibnamefont {Zhang}}, \bibinfo {author} {\bibfnamefont
  {W.}~\bibnamefont {Yang}}, \ and\ \bibinfo {author} {\bibfnamefont
  {K.}~\bibnamefont {Chang}},\ }\href {\doibase 10.1103/PhysRevB.96.155430}
  {\bibfield  {journal} {\bibinfo  {journal} {Phys. Rev. B}\ }\textbf {\bibinfo
  {volume} {96}},\ \bibinfo {pages} {155430} (\bibinfo {year}
  {2017})}\BibitemShut {NoStop}%
\bibitem [{\citenamefont {Demirci}\ \emph {et~al.}(2017)\citenamefont
  {Demirci}, \citenamefont {Avazl}, \citenamefont {Durgun},\ and\ \citenamefont
  {Cahangirov}}]{PhysRevB.95.115409}%
  \BibitemOpen
  \bibfield  {author} {\bibinfo {author} {\bibfnamefont {S.}~\bibnamefont
  {Demirci}}, \bibinfo {author} {\bibfnamefont {N.}~\bibnamefont {Avazl}},
  \bibinfo {author} {\bibfnamefont {E.}~\bibnamefont {Durgun}}, \ and\ \bibinfo
  {author} {\bibfnamefont {S.}~\bibnamefont {Cahangirov}},\ }\href {\doibase
  10.1103/PhysRevB.95.115409} {\bibfield  {journal} {\bibinfo  {journal} {Phys.
  Rev. B}\ }\textbf {\bibinfo {volume} {95}},\ \bibinfo {pages} {115409}
  (\bibinfo {year} {2017})}\BibitemShut {NoStop}%
\bibitem [{\citenamefont {Ben~Aziza}\ \emph {et~al.}(2018)\citenamefont
  {Ben~Aziza}, \citenamefont {Z\'olyomi}, \citenamefont {Henck}, \citenamefont
  {Pierucci}, \citenamefont {Silly}, \citenamefont {Avila}, \citenamefont
  {Magorrian}, \citenamefont {Chaste}, \citenamefont {Chen}, \citenamefont
  {Yoon}, \citenamefont {Xiao}, \citenamefont {Sirotti}, \citenamefont
  {Asensio}, \citenamefont {Lhuillier}, \citenamefont {Eddrief}, \citenamefont
  {Fal'ko},\ and\ \citenamefont {Ouerghi}}]{PhysRevB.98.115405}%
  \BibitemOpen
  \bibfield  {author} {\bibinfo {author} {\bibfnamefont {Z.}~\bibnamefont
  {Ben~Aziza}}, \bibinfo {author} {\bibfnamefont {V.}~\bibnamefont
  {Z\'olyomi}}, \bibinfo {author} {\bibfnamefont {H.}~\bibnamefont {Henck}},
  \bibinfo {author} {\bibfnamefont {D.}~\bibnamefont {Pierucci}}, \bibinfo
  {author} {\bibfnamefont {M.~G.}\ \bibnamefont {Silly}}, \bibinfo {author}
  {\bibfnamefont {J.}~\bibnamefont {Avila}}, \bibinfo {author} {\bibfnamefont
  {S.~J.}\ \bibnamefont {Magorrian}}, \bibinfo {author} {\bibfnamefont
  {J.}~\bibnamefont {Chaste}}, \bibinfo {author} {\bibfnamefont
  {C.}~\bibnamefont {Chen}}, \bibinfo {author} {\bibfnamefont {M.}~\bibnamefont
  {Yoon}}, \bibinfo {author} {\bibfnamefont {K.}~\bibnamefont {Xiao}}, \bibinfo
  {author} {\bibfnamefont {F.}~\bibnamefont {Sirotti}}, \bibinfo {author}
  {\bibfnamefont {M.~C.}\ \bibnamefont {Asensio}}, \bibinfo {author}
  {\bibfnamefont {E.}~\bibnamefont {Lhuillier}}, \bibinfo {author}
  {\bibfnamefont {M.}~\bibnamefont {Eddrief}}, \bibinfo {author} {\bibfnamefont
  {V.~I.}\ \bibnamefont {Fal'ko}}, \ and\ \bibinfo {author} {\bibfnamefont
  {A.}~\bibnamefont {Ouerghi}},\ }\href {\doibase 10.1103/PhysRevB.98.115405}
  {\bibfield  {journal} {\bibinfo  {journal} {Phys. Rev. B}\ }\textbf {\bibinfo
  {volume} {98}},\ \bibinfo {pages} {115405} (\bibinfo {year}
  {2018})}\BibitemShut {NoStop}%
\bibitem [{\citenamefont {Hamer}\ \emph {et~al.}(2019)\citenamefont {Hamer},
  \citenamefont {Zultak}, \citenamefont {Tyurnina}, \citenamefont
  {Z{\'o}lyomi}, \citenamefont {Terry}, \citenamefont {Barinov}, \citenamefont
  {Garner}, \citenamefont {Donoghue}, \citenamefont {Rooney}, \citenamefont
  {Kandyba}, \citenamefont {Giampietri}, \citenamefont {Graham}, \citenamefont
  {Teutsch}, \citenamefont {Xia}, \citenamefont {Koperski}, \citenamefont
  {Haigh}, \citenamefont {Fal'ko}, \citenamefont {Gorbachev},\ and\
  \citenamefont {Wilson}}]{Hamer2019}%
  \BibitemOpen
  \bibfield  {author} {\bibinfo {author} {\bibfnamefont {M.~J.}\ \bibnamefont
  {Hamer}}, \bibinfo {author} {\bibfnamefont {J.}~\bibnamefont {Zultak}},
  \bibinfo {author} {\bibfnamefont {A.~V.}\ \bibnamefont {Tyurnina}}, \bibinfo
  {author} {\bibfnamefont {V.}~\bibnamefont {Z{\'o}lyomi}}, \bibinfo {author}
  {\bibfnamefont {D.}~\bibnamefont {Terry}}, \bibinfo {author} {\bibfnamefont
  {A.}~\bibnamefont {Barinov}}, \bibinfo {author} {\bibfnamefont
  {A.}~\bibnamefont {Garner}}, \bibinfo {author} {\bibfnamefont
  {J.}~\bibnamefont {Donoghue}}, \bibinfo {author} {\bibfnamefont {A.~P.}\
  \bibnamefont {Rooney}}, \bibinfo {author} {\bibfnamefont {V.}~\bibnamefont
  {Kandyba}}, \bibinfo {author} {\bibfnamefont {A.}~\bibnamefont {Giampietri}},
  \bibinfo {author} {\bibfnamefont {A.}~\bibnamefont {Graham}}, \bibinfo
  {author} {\bibfnamefont {N.}~\bibnamefont {Teutsch}}, \bibinfo {author}
  {\bibfnamefont {X.}~\bibnamefont {Xia}}, \bibinfo {author} {\bibfnamefont
  {M.}~\bibnamefont {Koperski}}, \bibinfo {author} {\bibfnamefont {S.~J.}\
  \bibnamefont {Haigh}}, \bibinfo {author} {\bibfnamefont {V.~I.}\ \bibnamefont
  {Fal'ko}}, \bibinfo {author} {\bibfnamefont {R.~V.}\ \bibnamefont
  {Gorbachev}}, \ and\ \bibinfo {author} {\bibfnamefont {N.~R.}\ \bibnamefont
  {Wilson}},\ }\href {\doibase 10.1021/acsnano.8b08726} {\bibfield  {journal}
  {\bibinfo  {journal} {ACS Nano}\ }\textbf {\bibinfo {volume} {13}},\ \bibinfo
  {pages} {2136} (\bibinfo {year} {2019})}\BibitemShut {NoStop}%
\bibitem [{\citenamefont {Budweg}\ \emph {et~al.}(2019)\citenamefont {Budweg},
  \citenamefont {Yadav}, \citenamefont {Grupp}, \citenamefont {Leitenstorfer},
  \citenamefont {Trushin}, \citenamefont {Pauly},\ and\ \citenamefont
  {Brida}}]{PhysRevB.100.045404}%
  \BibitemOpen
  \bibfield  {author} {\bibinfo {author} {\bibfnamefont {A.}~\bibnamefont
  {Budweg}}, \bibinfo {author} {\bibfnamefont {D.}~\bibnamefont {Yadav}},
  \bibinfo {author} {\bibfnamefont {A.}~\bibnamefont {Grupp}}, \bibinfo
  {author} {\bibfnamefont {A.}~\bibnamefont {Leitenstorfer}}, \bibinfo {author}
  {\bibfnamefont {M.}~\bibnamefont {Trushin}}, \bibinfo {author} {\bibfnamefont
  {F.}~\bibnamefont {Pauly}}, \ and\ \bibinfo {author} {\bibfnamefont
  {D.}~\bibnamefont {Brida}},\ }\href {\doibase 10.1103/PhysRevB.100.045404}
  {\bibfield  {journal} {\bibinfo  {journal} {Phys. Rev. B}\ }\textbf {\bibinfo
  {volume} {100}},\ \bibinfo {pages} {045404} (\bibinfo {year}
  {2019})}\BibitemShut {NoStop}%
\bibitem [{\citenamefont {Seixas}\ \emph {et~al.}(2016)\citenamefont {Seixas},
  \citenamefont {Rodin}, \citenamefont {Carvalho},\ and\ \citenamefont
  {Castro~Neto}}]{PhysRevLett.116.206803}%
  \BibitemOpen
  \bibfield  {author} {\bibinfo {author} {\bibfnamefont {L.}~\bibnamefont
  {Seixas}}, \bibinfo {author} {\bibfnamefont {A.~S.}\ \bibnamefont {Rodin}},
  \bibinfo {author} {\bibfnamefont {A.}~\bibnamefont {Carvalho}}, \ and\
  \bibinfo {author} {\bibfnamefont {A.~H.}\ \bibnamefont {Castro~Neto}},\
  }\href {\doibase 10.1103/PhysRevLett.116.206803} {\bibfield  {journal}
  {\bibinfo  {journal} {Phys. Rev. Lett.}\ }\textbf {\bibinfo {volume} {116}},\
  \bibinfo {pages} {206803} (\bibinfo {year} {2016})}\BibitemShut {NoStop}%
\bibitem [{\citenamefont {Cao}\ \emph {et~al.}(2015)\citenamefont {Cao},
  \citenamefont {Li},\ and\ \citenamefont {Louie}}]{PhysRevLett.114.236602}%
  \BibitemOpen
  \bibfield  {author} {\bibinfo {author} {\bibfnamefont {T.}~\bibnamefont
  {Cao}}, \bibinfo {author} {\bibfnamefont {Z.}~\bibnamefont {Li}}, \ and\
  \bibinfo {author} {\bibfnamefont {S.~G.}\ \bibnamefont {Louie}},\ }\href
  {\doibase 10.1103/PhysRevLett.114.236602} {\bibfield  {journal} {\bibinfo
  {journal} {Phys. Rev. Lett.}\ }\textbf {\bibinfo {volume} {114}},\ \bibinfo
  {pages} {236602} (\bibinfo {year} {2015})}\BibitemShut {NoStop}%
\bibitem [{\citenamefont {Iordanidou}\ \emph {et~al.}(2018)\citenamefont
  {Iordanidou}, \citenamefont {Houssa}, \citenamefont {Kioseoglou},
  \citenamefont {Afanas'ev}, \citenamefont {Stesmans},\ and\ \citenamefont
  {Persson}}]{Iordanidou2018}%
  \BibitemOpen
  \bibfield  {author} {\bibinfo {author} {\bibfnamefont {K.}~\bibnamefont
  {Iordanidou}}, \bibinfo {author} {\bibfnamefont {M.}~\bibnamefont {Houssa}},
  \bibinfo {author} {\bibfnamefont {J.}~\bibnamefont {Kioseoglou}}, \bibinfo
  {author} {\bibfnamefont {V.~V.}\ \bibnamefont {Afanas'ev}}, \bibinfo {author}
  {\bibfnamefont {A.}~\bibnamefont {Stesmans}}, \ and\ \bibinfo {author}
  {\bibfnamefont {C.}~\bibnamefont {Persson}},\ }\href {\doibase
  10.1021/acsanm.8b01476} {\bibfield  {journal} {\bibinfo  {journal} {ACS
  Applied Nano Materials}\ }\textbf {\bibinfo {volume} {1}},\ \bibinfo {pages}
  {6656} (\bibinfo {year} {2018})}\BibitemShut {NoStop}%
\bibitem [{\citenamefont {Sethulakshmi}\ \emph {et~al.}(2019)\citenamefont
  {Sethulakshmi}, \citenamefont {Mishra}, \citenamefont {Ajayan}, \citenamefont
  {Kawazoe}, \citenamefont {Roy}, \citenamefont {Singh},\ and\ \citenamefont
  {Tiwary}}]{SETHULAKSHMI2019107}%
  \BibitemOpen
  \bibfield  {author} {\bibinfo {author} {\bibfnamefont {N.}~\bibnamefont
  {Sethulakshmi}}, \bibinfo {author} {\bibfnamefont {A.}~\bibnamefont
  {Mishra}}, \bibinfo {author} {\bibfnamefont {P.}~\bibnamefont {Ajayan}},
  \bibinfo {author} {\bibfnamefont {Y.}~\bibnamefont {Kawazoe}}, \bibinfo
  {author} {\bibfnamefont {A.~K.}\ \bibnamefont {Roy}}, \bibinfo {author}
  {\bibfnamefont {A.~K.}\ \bibnamefont {Singh}}, \ and\ \bibinfo {author}
  {\bibfnamefont {C.~S.}\ \bibnamefont {Tiwary}},\ }\href {\doibase
  https://doi.org/10.1016/j.mattod.2019.03.015} {\bibfield  {journal} {\bibinfo
   {journal} {Materials Today}\ }\textbf {\bibinfo {volume} {27}},\ \bibinfo
  {pages} {107 } (\bibinfo {year} {2019})}\BibitemShut {NoStop}%
\bibitem [{\citenamefont {Wickramaratne}\ \emph {et~al.}(2015)\citenamefont
  {Wickramaratne}, \citenamefont {Zahid},\ and\ \citenamefont
  {Lake}}]{doi:10.1063/1.4928559}%
  \BibitemOpen
  \bibfield  {author} {\bibinfo {author} {\bibfnamefont {D.}~\bibnamefont
  {Wickramaratne}}, \bibinfo {author} {\bibfnamefont {F.}~\bibnamefont
  {Zahid}}, \ and\ \bibinfo {author} {\bibfnamefont {R.~K.}\ \bibnamefont
  {Lake}},\ }\href {\doibase 10.1063/1.4928559} {\bibfield  {journal} {\bibinfo
   {journal} {Journal of Applied Physics}\ }\textbf {\bibinfo {volume} {118}},\
  \bibinfo {pages} {075101} (\bibinfo {year} {2015})}\BibitemShut {NoStop}%
\bibitem [{\citenamefont {Hu}\ \emph {et~al.}(2012)\citenamefont {Hu},
  \citenamefont {Wen}, \citenamefont {Wang}, \citenamefont {Tan},\ and\
  \citenamefont {Xiao}}]{Hu2012}%
  \BibitemOpen
  \bibfield  {author} {\bibinfo {author} {\bibfnamefont {P.}~\bibnamefont
  {Hu}}, \bibinfo {author} {\bibfnamefont {Z.}~\bibnamefont {Wen}}, \bibinfo
  {author} {\bibfnamefont {L.}~\bibnamefont {Wang}}, \bibinfo {author}
  {\bibfnamefont {P.}~\bibnamefont {Tan}}, \ and\ \bibinfo {author}
  {\bibfnamefont {K.}~\bibnamefont {Xiao}},\ }\href {\doibase
  10.1021/nn300889c} {\bibfield  {journal} {\bibinfo  {journal} {ACS Nano}\
  }\textbf {\bibinfo {volume} {6}},\ \bibinfo {pages} {5988} (\bibinfo {year}
  {2012})}\BibitemShut {NoStop}%
\bibitem [{\citenamefont {Lei}\ \emph {et~al.}(2013)\citenamefont {Lei},
  \citenamefont {Ge}, \citenamefont {Liu}, \citenamefont {Najmaei},
  \citenamefont {Shi}, \citenamefont {You}, \citenamefont {Lou}, \citenamefont
  {Vajtai},\ and\ \citenamefont {Ajayan}}]{Lei2013}%
  \BibitemOpen
  \bibfield  {author} {\bibinfo {author} {\bibfnamefont {S.}~\bibnamefont
  {Lei}}, \bibinfo {author} {\bibfnamefont {L.}~\bibnamefont {Ge}}, \bibinfo
  {author} {\bibfnamefont {Z.}~\bibnamefont {Liu}}, \bibinfo {author}
  {\bibfnamefont {S.}~\bibnamefont {Najmaei}}, \bibinfo {author} {\bibfnamefont
  {G.}~\bibnamefont {Shi}}, \bibinfo {author} {\bibfnamefont {G.}~\bibnamefont
  {You}}, \bibinfo {author} {\bibfnamefont {J.}~\bibnamefont {Lou}}, \bibinfo
  {author} {\bibfnamefont {R.}~\bibnamefont {Vajtai}}, \ and\ \bibinfo {author}
  {\bibfnamefont {P.~M.}\ \bibnamefont {Ajayan}},\ }\href {\doibase
  10.1021/nl4010089} {\bibfield  {journal} {\bibinfo  {journal} {Nano Letters}\
  }\textbf {\bibinfo {volume} {13}},\ \bibinfo {pages} {2777} (\bibinfo {year}
  {2013})}\BibitemShut {NoStop}%
\bibitem [{\citenamefont {Hu}\ \emph {et~al.}(2013)\citenamefont {Hu},
  \citenamefont {Wang}, \citenamefont {Yoon}, \citenamefont {Zhang},
  \citenamefont {Feng}, \citenamefont {Wang}, \citenamefont {Wen},
  \citenamefont {Idrobo}, \citenamefont {Miyamoto}, \citenamefont {Geohegan},\
  and\ \citenamefont {Xiao}}]{HuWang2013}%
  \BibitemOpen
  \bibfield  {author} {\bibinfo {author} {\bibfnamefont {P.}~\bibnamefont
  {Hu}}, \bibinfo {author} {\bibfnamefont {L.}~\bibnamefont {Wang}}, \bibinfo
  {author} {\bibfnamefont {M.}~\bibnamefont {Yoon}}, \bibinfo {author}
  {\bibfnamefont {J.}~\bibnamefont {Zhang}}, \bibinfo {author} {\bibfnamefont
  {W.}~\bibnamefont {Feng}}, \bibinfo {author} {\bibfnamefont {X.}~\bibnamefont
  {Wang}}, \bibinfo {author} {\bibfnamefont {Z.}~\bibnamefont {Wen}}, \bibinfo
  {author} {\bibfnamefont {J.~C.}\ \bibnamefont {Idrobo}}, \bibinfo {author}
  {\bibfnamefont {Y.}~\bibnamefont {Miyamoto}}, \bibinfo {author}
  {\bibfnamefont {D.~B.}\ \bibnamefont {Geohegan}}, \ and\ \bibinfo {author}
  {\bibfnamefont {K.}~\bibnamefont {Xiao}},\ }\href {\doibase
  10.1021/nl400107k} {\bibfield  {journal} {\bibinfo  {journal} {Nano Letters}\
  }\textbf {\bibinfo {volume} {13}},\ \bibinfo {pages} {1649} (\bibinfo {year}
  {2013})}\BibitemShut {NoStop}%
\bibitem [{\citenamefont {Li}\ \emph {et~al.}(2014)\citenamefont {Li},
  \citenamefont {Lin}, \citenamefont {Puretzky}, \citenamefont {Idrobo},
  \citenamefont {Ma}, \citenamefont {Chi}, \citenamefont {Yoon}, \citenamefont
  {Rouleau}, \citenamefont {Kravchenko}, \citenamefont {Geohegan},\ and\
  \citenamefont {Xiao}}]{Li2014}%
  \BibitemOpen
  \bibfield  {author} {\bibinfo {author} {\bibfnamefont {X.}~\bibnamefont
  {Li}}, \bibinfo {author} {\bibfnamefont {M.-W.}\ \bibnamefont {Lin}},
  \bibinfo {author} {\bibfnamefont {A.~A.}\ \bibnamefont {Puretzky}}, \bibinfo
  {author} {\bibfnamefont {J.~C.}\ \bibnamefont {Idrobo}}, \bibinfo {author}
  {\bibfnamefont {C.}~\bibnamefont {Ma}}, \bibinfo {author} {\bibfnamefont
  {M.}~\bibnamefont {Chi}}, \bibinfo {author} {\bibfnamefont {M.}~\bibnamefont
  {Yoon}}, \bibinfo {author} {\bibfnamefont {C.~M.}\ \bibnamefont {Rouleau}},
  \bibinfo {author} {\bibfnamefont {I.~I.}\ \bibnamefont {Kravchenko}},
  \bibinfo {author} {\bibfnamefont {D.~B.}\ \bibnamefont {Geohegan}}, \ and\
  \bibinfo {author} {\bibfnamefont {K.}~\bibnamefont {Xiao}},\ }\href {\doibase
  10.1038/srep05497} {\bibfield  {journal} {\bibinfo  {journal} {Scientific
  Reports}\ }\textbf {\bibinfo {volume} {4}},\ \bibinfo {pages} {5497}
  (\bibinfo {year} {2014})}\BibitemShut {NoStop}%
\bibitem [{\citenamefont {Mahjouri-Samani}\ \emph {et~al.}(2014)\citenamefont
  {Mahjouri-Samani}, \citenamefont {Gresback}, \citenamefont {Tian},
  \citenamefont {Wang}, \citenamefont {Puretzky}, \citenamefont {Rouleau},
  \citenamefont {Eres}, \citenamefont {Ivanov}, \citenamefont {Xiao},
  \citenamefont {McGuire}, \citenamefont {Duscher},\ and\ \citenamefont
  {Geohegan}}]{Samani2014}%
  \BibitemOpen
  \bibfield  {author} {\bibinfo {author} {\bibfnamefont {M.}~\bibnamefont
  {Mahjouri-Samani}}, \bibinfo {author} {\bibfnamefont {R.}~\bibnamefont
  {Gresback}}, \bibinfo {author} {\bibfnamefont {M.}~\bibnamefont {Tian}},
  \bibinfo {author} {\bibfnamefont {K.}~\bibnamefont {Wang}}, \bibinfo {author}
  {\bibfnamefont {A.~A.}\ \bibnamefont {Puretzky}}, \bibinfo {author}
  {\bibfnamefont {C.~M.}\ \bibnamefont {Rouleau}}, \bibinfo {author}
  {\bibfnamefont {G.}~\bibnamefont {Eres}}, \bibinfo {author} {\bibfnamefont
  {I.~N.}\ \bibnamefont {Ivanov}}, \bibinfo {author} {\bibfnamefont
  {K.}~\bibnamefont {Xiao}}, \bibinfo {author} {\bibfnamefont {M.~A.}\
  \bibnamefont {McGuire}}, \bibinfo {author} {\bibfnamefont {G.}~\bibnamefont
  {Duscher}}, \ and\ \bibinfo {author} {\bibfnamefont {D.~B.}\ \bibnamefont
  {Geohegan}},\ }\href {\doibase 10.1002/adfm.201401440} {\bibfield  {journal}
  {\bibinfo  {journal} {Advanced Functional Materials}\ }\textbf {\bibinfo
  {volume} {24}},\ \bibinfo {pages} {6365} (\bibinfo {year}
  {2014})}\BibitemShut {NoStop}%
\bibitem [{\citenamefont {Jie}\ \emph {et~al.}(2015)\citenamefont {Jie},
  \citenamefont {Chen}, \citenamefont {Li}, \citenamefont {Xie}, \citenamefont
  {Hui}, \citenamefont {Lau}, \citenamefont {Cui},\ and\ \citenamefont
  {Hao}}]{Jie2015}%
  \BibitemOpen
  \bibfield  {author} {\bibinfo {author} {\bibfnamefont {W.}~\bibnamefont
  {Jie}}, \bibinfo {author} {\bibfnamefont {X.}~\bibnamefont {Chen}}, \bibinfo
  {author} {\bibfnamefont {D.}~\bibnamefont {Li}}, \bibinfo {author}
  {\bibfnamefont {L.}~\bibnamefont {Xie}}, \bibinfo {author} {\bibfnamefont
  {Y.~Y.}\ \bibnamefont {Hui}}, \bibinfo {author} {\bibfnamefont {S.~P.}\
  \bibnamefont {Lau}}, \bibinfo {author} {\bibfnamefont {X.}~\bibnamefont
  {Cui}}, \ and\ \bibinfo {author} {\bibfnamefont {J.}~\bibnamefont {Hao}},\
  }\href {\doibase 10.1002/anie.201409837} {\bibfield  {journal} {\bibinfo
  {journal} {Angewandte Chemie International Edition}\ }\textbf {\bibinfo
  {volume} {54}},\ \bibinfo {pages} {1185} (\bibinfo {year}
  {2015})}\BibitemShut {NoStop}%
\bibitem [{\citenamefont {Zhou}\ \emph {et~al.}(2015)\citenamefont {Zhou},
  \citenamefont {Cheng}, \citenamefont {Zhou}, \citenamefont {Cao},
  \citenamefont {Hong}, \citenamefont {Liao}, \citenamefont {Wu}, \citenamefont
  {Peng}, \citenamefont {Liu},\ and\ \citenamefont {Yu}}]{Zhou2015}%
  \BibitemOpen
  \bibfield  {author} {\bibinfo {author} {\bibfnamefont {X.}~\bibnamefont
  {Zhou}}, \bibinfo {author} {\bibfnamefont {J.}~\bibnamefont {Cheng}},
  \bibinfo {author} {\bibfnamefont {Y.}~\bibnamefont {Zhou}}, \bibinfo {author}
  {\bibfnamefont {T.}~\bibnamefont {Cao}}, \bibinfo {author} {\bibfnamefont
  {H.}~\bibnamefont {Hong}}, \bibinfo {author} {\bibfnamefont {Z.}~\bibnamefont
  {Liao}}, \bibinfo {author} {\bibfnamefont {S.}~\bibnamefont {Wu}}, \bibinfo
  {author} {\bibfnamefont {H.}~\bibnamefont {Peng}}, \bibinfo {author}
  {\bibfnamefont {K.}~\bibnamefont {Liu}}, \ and\ \bibinfo {author}
  {\bibfnamefont {D.}~\bibnamefont {Yu}},\ }\href {\doibase
  10.1021/jacs.5b04305} {\bibfield  {journal} {\bibinfo  {journal} {Journal of
  the American Chemical Society}\ }\textbf {\bibinfo {volume} {137}},\ \bibinfo
  {pages} {7994} (\bibinfo {year} {2015})}\BibitemShut {NoStop}%
\bibitem [{\citenamefont {Cheng}\ \emph {et~al.}(2018)\citenamefont {Cheng},
  \citenamefont {Guo}, \citenamefont {Han}, \citenamefont {Jiang},
  \citenamefont {Zhang}, \citenamefont {Ahuja}, \citenamefont {Su},\ and\
  \citenamefont {Zhao}}]{Cheng2018}%
  \BibitemOpen
  \bibfield  {author} {\bibinfo {author} {\bibfnamefont {K.}~\bibnamefont
  {Cheng}}, \bibinfo {author} {\bibfnamefont {Y.}~\bibnamefont {Guo}}, \bibinfo
  {author} {\bibfnamefont {N.}~\bibnamefont {Han}}, \bibinfo {author}
  {\bibfnamefont {X.}~\bibnamefont {Jiang}}, \bibinfo {author} {\bibfnamefont
  {J.}~\bibnamefont {Zhang}}, \bibinfo {author} {\bibfnamefont
  {R.}~\bibnamefont {Ahuja}}, \bibinfo {author} {\bibfnamefont
  {Y.}~\bibnamefont {Su}}, \ and\ \bibinfo {author} {\bibfnamefont
  {J.}~\bibnamefont {Zhao}},\ }\href {\doibase 10.1063/1.5020618} {\bibfield
  {journal} {\bibinfo  {journal} {Applied Physics Letters}\ }\textbf {\bibinfo
  {volume} {112}},\ \bibinfo {pages} {143902} (\bibinfo {year}
  {2018})}\BibitemShut {NoStop}%
\bibitem [{\citenamefont {Late}\ \emph
  {et~al.}(2012{\natexlab{a}})\citenamefont {Late}, \citenamefont {Liu},
  \citenamefont {Luo}, \citenamefont {Yan}, \citenamefont {Matte},
  \citenamefont {Grayson}, \citenamefont {Rao},\ and\ \citenamefont
  {Dravid}}]{Late2012}%
  \BibitemOpen
  \bibfield  {author} {\bibinfo {author} {\bibfnamefont {D.~J.}\ \bibnamefont
  {Late}}, \bibinfo {author} {\bibfnamefont {B.}~\bibnamefont {Liu}}, \bibinfo
  {author} {\bibfnamefont {J.}~\bibnamefont {Luo}}, \bibinfo {author}
  {\bibfnamefont {A.}~\bibnamefont {Yan}}, \bibinfo {author} {\bibfnamefont
  {H.~S. S.~R.}\ \bibnamefont {Matte}}, \bibinfo {author} {\bibfnamefont
  {M.}~\bibnamefont {Grayson}}, \bibinfo {author} {\bibfnamefont {C.~N.~R.}\
  \bibnamefont {Rao}}, \ and\ \bibinfo {author} {\bibfnamefont {V.~P.}\
  \bibnamefont {Dravid}},\ }\href {\doibase 10.1002/adma.201201361} {\bibfield
  {journal} {\bibinfo  {journal} {Advanced Materials}\ }\textbf {\bibinfo
  {volume} {24}},\ \bibinfo {pages} {3549} (\bibinfo {year}
  {2012}{\natexlab{a}})}\BibitemShut {NoStop}%
\bibitem [{\citenamefont {Zhuang}\ and\ \citenamefont
  {Hennig}(2013)}]{Zhuang2013}%
  \BibitemOpen
  \bibfield  {author} {\bibinfo {author} {\bibfnamefont {H.~L.}\ \bibnamefont
  {Zhuang}}\ and\ \bibinfo {author} {\bibfnamefont {R.~G.}\ \bibnamefont
  {Hennig}},\ }\href {\doibase 10.1021/cm401661x} {\bibfield  {journal}
  {\bibinfo  {journal} {Chemistry of Materials}\ }\textbf {\bibinfo {volume}
  {25}},\ \bibinfo {pages} {3232} (\bibinfo {year} {2013})}\BibitemShut
  {NoStop}%
\bibitem [{\citenamefont {Kouser}\ \emph {et~al.}(2015)\citenamefont {Kouser},
  \citenamefont {Thannikoth}, \citenamefont {Gupta}, \citenamefont {Waghmare},\
  and\ \citenamefont {Rao}}]{Kouser2015}%
  \BibitemOpen
  \bibfield  {author} {\bibinfo {author} {\bibfnamefont {S.}~\bibnamefont
  {Kouser}}, \bibinfo {author} {\bibfnamefont {A.}~\bibnamefont {Thannikoth}},
  \bibinfo {author} {\bibfnamefont {U.}~\bibnamefont {Gupta}}, \bibinfo
  {author} {\bibfnamefont {U.~V.}\ \bibnamefont {Waghmare}}, \ and\ \bibinfo
  {author} {\bibfnamefont {C.~N.~R.}\ \bibnamefont {Rao}},\ }\href {\doibase
  10.1002/smll.201501077} {\bibfield  {journal} {\bibinfo  {journal} {Small}\
  }\textbf {\bibinfo {volume} {11}},\ \bibinfo {pages} {4723} (\bibinfo {year}
  {2015})}\BibitemShut {NoStop}%
\bibitem [{\citenamefont {Harvey}\ \emph {et~al.}(2015)\citenamefont {Harvey},
  \citenamefont {Backes}, \citenamefont {Gholamvand}, \citenamefont {Hanlon},
  \citenamefont {McAteer}, \citenamefont {Nerl}, \citenamefont {McGuire},
  \citenamefont {Seral-Ascaso}, \citenamefont {Ramasse}, \citenamefont
  {McEvoy}, \citenamefont {Winters}, \citenamefont {Berner}, \citenamefont
  {McCloskey}, \citenamefont {Donegan}, \citenamefont {Duesberg}, \citenamefont
  {Nicolosi},\ and\ \citenamefont {Coleman}}]{Harvey2015}%
  \BibitemOpen
  \bibfield  {author} {\bibinfo {author} {\bibfnamefont {A.}~\bibnamefont
  {Harvey}}, \bibinfo {author} {\bibfnamefont {C.}~\bibnamefont {Backes}},
  \bibinfo {author} {\bibfnamefont {Z.}~\bibnamefont {Gholamvand}}, \bibinfo
  {author} {\bibfnamefont {D.}~\bibnamefont {Hanlon}}, \bibinfo {author}
  {\bibfnamefont {D.}~\bibnamefont {McAteer}}, \bibinfo {author} {\bibfnamefont
  {H.~C.}\ \bibnamefont {Nerl}}, \bibinfo {author} {\bibfnamefont
  {E.}~\bibnamefont {McGuire}}, \bibinfo {author} {\bibfnamefont
  {A.}~\bibnamefont {Seral-Ascaso}}, \bibinfo {author} {\bibfnamefont {Q.~M.}\
  \bibnamefont {Ramasse}}, \bibinfo {author} {\bibfnamefont {N.}~\bibnamefont
  {McEvoy}}, \bibinfo {author} {\bibfnamefont {S.}~\bibnamefont {Winters}},
  \bibinfo {author} {\bibfnamefont {N.~C.}\ \bibnamefont {Berner}}, \bibinfo
  {author} {\bibfnamefont {D.}~\bibnamefont {McCloskey}}, \bibinfo {author}
  {\bibfnamefont {J.~F.}\ \bibnamefont {Donegan}}, \bibinfo {author}
  {\bibfnamefont {G.~S.}\ \bibnamefont {Duesberg}}, \bibinfo {author}
  {\bibfnamefont {V.}~\bibnamefont {Nicolosi}}, \ and\ \bibinfo {author}
  {\bibfnamefont {J.~N.}\ \bibnamefont {Coleman}},\ }\href {\doibase
  10.1021/acs.chemmater.5b00910} {\bibfield  {journal} {\bibinfo  {journal}
  {Chemistry of Materials}\ }\textbf {\bibinfo {volume} {27}},\ \bibinfo
  {pages} {3483} (\bibinfo {year} {2015})}\BibitemShut {NoStop}%
\bibitem [{\citenamefont {Cui}\ \emph {et~al.}(2018)\citenamefont {Cui},
  \citenamefont {Peng}, \citenamefont {Sun}, \citenamefont {Qian},\ and\
  \citenamefont {Huang}}]{C8TA08103D}%
  \BibitemOpen
  \bibfield  {author} {\bibinfo {author} {\bibfnamefont {Y.}~\bibnamefont
  {Cui}}, \bibinfo {author} {\bibfnamefont {L.}~\bibnamefont {Peng}}, \bibinfo
  {author} {\bibfnamefont {L.}~\bibnamefont {Sun}}, \bibinfo {author}
  {\bibfnamefont {Q.}~\bibnamefont {Qian}}, \ and\ \bibinfo {author}
  {\bibfnamefont {Y.}~\bibnamefont {Huang}},\ }\href {\doibase
  10.1039/C8TA08103D} {\bibfield  {journal} {\bibinfo  {journal} {J. Mater.
  Chem. A}\ }\textbf {\bibinfo {volume} {6}},\ \bibinfo {pages} {22768}
  (\bibinfo {year} {2018})}\BibitemShut {NoStop}%
\bibitem [{\citenamefont {Zhu}\ \emph {et~al.}(2012)\citenamefont {Zhu},
  \citenamefont {Cheng},\ and\ \citenamefont
  {Schwingenschl\"ogl}}]{TopoZhu2012}%
  \BibitemOpen
  \bibfield  {author} {\bibinfo {author} {\bibfnamefont {Z.}~\bibnamefont
  {Zhu}}, \bibinfo {author} {\bibfnamefont {Y.}~\bibnamefont {Cheng}}, \ and\
  \bibinfo {author} {\bibfnamefont {U.}~\bibnamefont {Schwingenschl\"ogl}},\
  }\href {\doibase 10.1103/PhysRevLett.108.266805} {\bibfield  {journal}
  {\bibinfo  {journal} {Phys. Rev. Lett.}\ }\textbf {\bibinfo {volume} {108}},\
  \bibinfo {pages} {266805} (\bibinfo {year} {2012})}\BibitemShut {NoStop}%
\bibitem [{\citenamefont {Zhou}\ \emph {et~al.}(2018)\citenamefont {Zhou},
  \citenamefont {Liu}, \citenamefont {Zhao},\ and\ \citenamefont
  {Yao}}]{TopoZhou2018}%
  \BibitemOpen
  \bibfield  {author} {\bibinfo {author} {\bibfnamefont {S.}~\bibnamefont
  {Zhou}}, \bibinfo {author} {\bibfnamefont {C.-C.}\ \bibnamefont {Liu}},
  \bibinfo {author} {\bibfnamefont {J.}~\bibnamefont {Zhao}}, \ and\ \bibinfo
  {author} {\bibfnamefont {Y.}~\bibnamefont {Yao}},\ }\href {\doibase
  10.1038/s41535-018-0089-0} {\bibfield  {journal} {\bibinfo  {journal} {npj
  Quantum Materials}\ }\textbf {\bibinfo {volume} {3}},\ \bibinfo {pages} {16}
  (\bibinfo {year} {2018})}\BibitemShut {NoStop}%
\bibitem [{\citenamefont {Late}\ \emph
  {et~al.}(2012{\natexlab{b}})\citenamefont {Late}, \citenamefont {Liu},
  \citenamefont {Matte}, \citenamefont {Rao},\ and\ \citenamefont
  {Dravid}}]{LateExfo2012}%
  \BibitemOpen
  \bibfield  {author} {\bibinfo {author} {\bibfnamefont {D.~J.}\ \bibnamefont
  {Late}}, \bibinfo {author} {\bibfnamefont {B.}~\bibnamefont {Liu}}, \bibinfo
  {author} {\bibfnamefont {H.~S. S.~R.}\ \bibnamefont {Matte}}, \bibinfo
  {author} {\bibfnamefont {C.~N.~R.}\ \bibnamefont {Rao}}, \ and\ \bibinfo
  {author} {\bibfnamefont {V.~P.}\ \bibnamefont {Dravid}},\ }\href {\doibase
  10.1002/adfm.201102913} {\bibfield  {journal} {\bibinfo  {journal} {Advanced
  Functional Materials}\ }\textbf {\bibinfo {volume} {22}},\ \bibinfo {pages}
  {1894} (\bibinfo {year} {2012}{\natexlab{b}})}\BibitemShut {NoStop}%
\bibitem [{\citenamefont {Wu}\ \emph {et~al.}()\citenamefont {Wu},
  \citenamefont {Zhang}, \citenamefont {Lee}, \citenamefont {Duesberg},
  \citenamefont {Syrlybekov}, \citenamefont {Liu}, \citenamefont {Abid},
  \citenamefont {Abid}, \citenamefont {Liu}, \citenamefont {Zhang},
  \citenamefont {Coileáin}, \citenamefont {Xu}, \citenamefont {Cho},
  \citenamefont {Choi}, \citenamefont {Chun}, \citenamefont {Wang},
  \citenamefont {Liu},\ and\ \citenamefont {Wu}}]{Wu2017}%
  \BibitemOpen
  \bibfield  {author} {\bibinfo {author} {\bibfnamefont {Y.}~\bibnamefont
  {Wu}}, \bibinfo {author} {\bibfnamefont {D.}~\bibnamefont {Zhang}}, \bibinfo
  {author} {\bibfnamefont {K.}~\bibnamefont {Lee}}, \bibinfo {author}
  {\bibfnamefont {G.~S.}\ \bibnamefont {Duesberg}}, \bibinfo {author}
  {\bibfnamefont {A.}~\bibnamefont {Syrlybekov}}, \bibinfo {author}
  {\bibfnamefont {X.}~\bibnamefont {Liu}}, \bibinfo {author} {\bibfnamefont
  {M.}~\bibnamefont {Abid}}, \bibinfo {author} {\bibfnamefont {M.}~\bibnamefont
  {Abid}}, \bibinfo {author} {\bibfnamefont {Y.}~\bibnamefont {Liu}}, \bibinfo
  {author} {\bibfnamefont {L.}~\bibnamefont {Zhang}}, \bibinfo {author}
  {\bibfnamefont {C.~Ã.}\ \bibnamefont {Coileáin}}, \bibinfo {author}
  {\bibfnamefont {H.}~\bibnamefont {Xu}}, \bibinfo {author} {\bibfnamefont
  {J.}~\bibnamefont {Cho}}, \bibinfo {author} {\bibfnamefont {M.}~\bibnamefont
  {Choi}}, \bibinfo {author} {\bibfnamefont {B.~S.}\ \bibnamefont {Chun}},
  \bibinfo {author} {\bibfnamefont {H.}~\bibnamefont {Wang}}, \bibinfo {author}
  {\bibfnamefont {H.}~\bibnamefont {Liu}}, \ and\ \bibinfo {author}
  {\bibfnamefont {H.-C.}\ \bibnamefont {Wu}},\ }\href {\doibase
  10.1002/admt.201600197} {\bibfield  {journal} {\bibinfo  {journal} {Advanced
  Materials Technologies}\ }\textbf {\bibinfo {volume} {2}},\ \bibinfo {pages}
  {1600197}}\BibitemShut {NoStop}%
\bibitem [{\citenamefont {Ben~Aziza}\ \emph {et~al.}(2016)\citenamefont
  {Ben~Aziza}, \citenamefont {Henck}, \citenamefont {Pierucci}, \citenamefont
  {Silly}, \citenamefont {Lhuillier}, \citenamefont {Patriarche}, \citenamefont
  {Sirotti}, \citenamefont {Eddrief},\ and\ \citenamefont
  {Ouerghi}}]{BenAziza2016}%
  \BibitemOpen
  \bibfield  {author} {\bibinfo {author} {\bibfnamefont {Z.}~\bibnamefont
  {Ben~Aziza}}, \bibinfo {author} {\bibfnamefont {H.}~\bibnamefont {Henck}},
  \bibinfo {author} {\bibfnamefont {D.}~\bibnamefont {Pierucci}}, \bibinfo
  {author} {\bibfnamefont {M.~G.}\ \bibnamefont {Silly}}, \bibinfo {author}
  {\bibfnamefont {E.}~\bibnamefont {Lhuillier}}, \bibinfo {author}
  {\bibfnamefont {G.}~\bibnamefont {Patriarche}}, \bibinfo {author}
  {\bibfnamefont {F.}~\bibnamefont {Sirotti}}, \bibinfo {author} {\bibfnamefont
  {M.}~\bibnamefont {Eddrief}}, \ and\ \bibinfo {author} {\bibfnamefont
  {A.}~\bibnamefont {Ouerghi}},\ }\href {\doibase 10.1021/acsnano.6b05521}
  {\bibfield  {journal} {\bibinfo  {journal} {ACS Nano}\ }\textbf {\bibinfo
  {volume} {10}},\ \bibinfo {pages} {9679} (\bibinfo {year}
  {2016})}\BibitemShut {NoStop}%
\bibitem [{\citenamefont {Lee}\ \emph {et~al.}(2017)\citenamefont {Lee},
  \citenamefont {Krishnamoorthy}, \citenamefont {O'Hara}, \citenamefont
  {Brenner}, \citenamefont {Johnson}, \citenamefont {Jamison}, \citenamefont
  {Myers}, \citenamefont {Kawakami}, \citenamefont {Hwang},\ and\ \citenamefont
  {Rajan}}]{Choong2017}%
  \BibitemOpen
  \bibfield  {author} {\bibinfo {author} {\bibfnamefont {C.~H.}\ \bibnamefont
  {Lee}}, \bibinfo {author} {\bibfnamefont {S.}~\bibnamefont {Krishnamoorthy}},
  \bibinfo {author} {\bibfnamefont {D.~J.}\ \bibnamefont {O'Hara}}, \bibinfo
  {author} {\bibfnamefont {M.~R.}\ \bibnamefont {Brenner}}, \bibinfo {author}
  {\bibfnamefont {J.~M.}\ \bibnamefont {Johnson}}, \bibinfo {author}
  {\bibfnamefont {J.~S.}\ \bibnamefont {Jamison}}, \bibinfo {author}
  {\bibfnamefont {R.~C.}\ \bibnamefont {Myers}}, \bibinfo {author}
  {\bibfnamefont {R.~K.}\ \bibnamefont {Kawakami}}, \bibinfo {author}
  {\bibfnamefont {J.}~\bibnamefont {Hwang}}, \ and\ \bibinfo {author}
  {\bibfnamefont {S.}~\bibnamefont {Rajan}},\ }\href {\doibase
  10.1063/1.4977697} {\bibfield  {journal} {\bibinfo  {journal} {Journal of
  Applied Physics}\ }\textbf {\bibinfo {volume} {121}},\ \bibinfo {pages}
  {094302} (\bibinfo {year} {2017})}\BibitemShut {NoStop}%
\bibitem [{\citenamefont {Chen}\ \emph {et~al.}(2018)\citenamefont {Chen},
  \citenamefont {Kim}, \citenamefont {Ovchinnikov}, \citenamefont {Kuc},
  \citenamefont {Heine}, \citenamefont {Renault},\ and\ \citenamefont
  {Kis}}]{Chen2018}%
  \BibitemOpen
  \bibfield  {author} {\bibinfo {author} {\bibfnamefont {M.-W.}\ \bibnamefont
  {Chen}}, \bibinfo {author} {\bibfnamefont {H.}~\bibnamefont {Kim}}, \bibinfo
  {author} {\bibfnamefont {D.}~\bibnamefont {Ovchinnikov}}, \bibinfo {author}
  {\bibfnamefont {A.}~\bibnamefont {Kuc}}, \bibinfo {author} {\bibfnamefont
  {T.}~\bibnamefont {Heine}}, \bibinfo {author} {\bibfnamefont
  {O.}~\bibnamefont {Renault}}, \ and\ \bibinfo {author} {\bibfnamefont
  {A.}~\bibnamefont {Kis}},\ }\href {\doibase 10.1038/s41699-017-0047-x}
  {\bibfield  {journal} {\bibinfo  {journal} {npj 2D Materials and
  Applications}\ }\textbf {\bibinfo {volume} {2}},\ \bibinfo {pages} {2}
  (\bibinfo {year} {2018})}\BibitemShut {NoStop}%
\bibitem [{\citenamefont {Li}\ \emph {et~al.}(2015)\citenamefont {Li},
  \citenamefont {Basile}, \citenamefont {Huang}, \citenamefont {Ma},
  \citenamefont {Lee}, \citenamefont {Vlassiouk}, \citenamefont {Puretzky},
  \citenamefont {Lin}, \citenamefont {Yoon}, \citenamefont {Chi}, \citenamefont
  {Idrobo}, \citenamefont {Rouleau}, \citenamefont {Sumpter}, \citenamefont
  {Geohegan},\ and\ \citenamefont {Xiao}}]{Li2015}%
  \BibitemOpen
  \bibfield  {author} {\bibinfo {author} {\bibfnamefont {X.}~\bibnamefont
  {Li}}, \bibinfo {author} {\bibfnamefont {L.}~\bibnamefont {Basile}}, \bibinfo
  {author} {\bibfnamefont {B.}~\bibnamefont {Huang}}, \bibinfo {author}
  {\bibfnamefont {C.}~\bibnamefont {Ma}}, \bibinfo {author} {\bibfnamefont
  {J.}~\bibnamefont {Lee}}, \bibinfo {author} {\bibfnamefont {I.~V.}\
  \bibnamefont {Vlassiouk}}, \bibinfo {author} {\bibfnamefont {A.~A.}\
  \bibnamefont {Puretzky}}, \bibinfo {author} {\bibfnamefont {M.-W.}\
  \bibnamefont {Lin}}, \bibinfo {author} {\bibfnamefont {M.}~\bibnamefont
  {Yoon}}, \bibinfo {author} {\bibfnamefont {M.}~\bibnamefont {Chi}}, \bibinfo
  {author} {\bibfnamefont {J.~C.}\ \bibnamefont {Idrobo}}, \bibinfo {author}
  {\bibfnamefont {C.~M.}\ \bibnamefont {Rouleau}}, \bibinfo {author}
  {\bibfnamefont {B.~G.}\ \bibnamefont {Sumpter}}, \bibinfo {author}
  {\bibfnamefont {D.~B.}\ \bibnamefont {Geohegan}}, \ and\ \bibinfo {author}
  {\bibfnamefont {K.}~\bibnamefont {Xiao}},\ }\href {\doibase
  10.1021/acsnano.5b01943} {\bibfield  {journal} {\bibinfo  {journal} {ACS
  Nano}\ }\textbf {\bibinfo {volume} {9}},\ \bibinfo {pages} {8078} (\bibinfo
  {year} {2015})}\BibitemShut {NoStop}%
\bibitem [{\citenamefont {Li}\ \emph {et~al.}(2016)\citenamefont {Li},
  \citenamefont {Lin}, \citenamefont {Lin}, \citenamefont {Huang},
  \citenamefont {Puretzky}, \citenamefont {Ma}, \citenamefont {Wang},
  \citenamefont {Zhou}, \citenamefont {Pantelides}, \citenamefont {Chi},
  \citenamefont {Kravchenko}, \citenamefont {Fowlkes}, \citenamefont {Rouleau},
  \citenamefont {Geohegan},\ and\ \citenamefont {Xiao}}]{Lie1501882}%
  \BibitemOpen
  \bibfield  {author} {\bibinfo {author} {\bibfnamefont {X.}~\bibnamefont
  {Li}}, \bibinfo {author} {\bibfnamefont {M.-W.}\ \bibnamefont {Lin}},
  \bibinfo {author} {\bibfnamefont {J.}~\bibnamefont {Lin}}, \bibinfo {author}
  {\bibfnamefont {B.}~\bibnamefont {Huang}}, \bibinfo {author} {\bibfnamefont
  {A.~A.}\ \bibnamefont {Puretzky}}, \bibinfo {author} {\bibfnamefont
  {C.}~\bibnamefont {Ma}}, \bibinfo {author} {\bibfnamefont {K.}~\bibnamefont
  {Wang}}, \bibinfo {author} {\bibfnamefont {W.}~\bibnamefont {Zhou}}, \bibinfo
  {author} {\bibfnamefont {S.~T.}\ \bibnamefont {Pantelides}}, \bibinfo
  {author} {\bibfnamefont {M.}~\bibnamefont {Chi}}, \bibinfo {author}
  {\bibfnamefont {I.}~\bibnamefont {Kravchenko}}, \bibinfo {author}
  {\bibfnamefont {J.}~\bibnamefont {Fowlkes}}, \bibinfo {author} {\bibfnamefont
  {C.~M.}\ \bibnamefont {Rouleau}}, \bibinfo {author} {\bibfnamefont {D.~B.}\
  \bibnamefont {Geohegan}}, \ and\ \bibinfo {author} {\bibfnamefont
  {K.}~\bibnamefont {Xiao}},\ }\href {\doibase 10.1126/sciadv.1501882}
  {\bibfield  {journal} {\bibinfo  {journal} {Science Advances}\ }\textbf
  {\bibinfo {volume} {2}} (\bibinfo {year} {2016}),\
  10.1126/sciadv.1501882}\BibitemShut {NoStop}%
\bibitem [{\citenamefont {Li}\ \emph {et~al.}(2017{\natexlab{b}})\citenamefont
  {Li}, \citenamefont {Dong}, \citenamefont {Idrobo}, \citenamefont {Puretzky},
  \citenamefont {Rouleau}, \citenamefont {Geohegan}, \citenamefont {Ding},\
  and\ \citenamefont {Xiao}}]{Li2017}%
  \BibitemOpen
  \bibfield  {author} {\bibinfo {author} {\bibfnamefont {X.}~\bibnamefont
  {Li}}, \bibinfo {author} {\bibfnamefont {J.}~\bibnamefont {Dong}}, \bibinfo
  {author} {\bibfnamefont {J.~C.}\ \bibnamefont {Idrobo}}, \bibinfo {author}
  {\bibfnamefont {A.~A.}\ \bibnamefont {Puretzky}}, \bibinfo {author}
  {\bibfnamefont {C.~M.}\ \bibnamefont {Rouleau}}, \bibinfo {author}
  {\bibfnamefont {D.~B.}\ \bibnamefont {Geohegan}}, \bibinfo {author}
  {\bibfnamefont {F.}~\bibnamefont {Ding}}, \ and\ \bibinfo {author}
  {\bibfnamefont {K.}~\bibnamefont {Xiao}},\ }\href {\doibase
  10.1021/jacs.6b11076} {\bibfield  {journal} {\bibinfo  {journal} {Journal of
  the American Chemical Society}\ }\textbf {\bibinfo {volume} {139}},\ \bibinfo
  {pages} {482} (\bibinfo {year} {2017}{\natexlab{b}})}\BibitemShut {NoStop}%
\bibitem [{\citenamefont {YU}\ and\ \citenamefont
  {Cardona}(2005)}]{yu2005fundamentals}%
  \BibitemOpen
  \bibfield  {author} {\bibinfo {author} {\bibfnamefont {P.}~\bibnamefont
  {YU}}\ and\ \bibinfo {author} {\bibfnamefont {M.}~\bibnamefont {Cardona}},\
  }\href {https://books.google.lu/books?id=W9pdJZoAeyEC} {\emph {\bibinfo
  {title} {Fundamentals of Semiconductors: Physics and Materials
  Properties}}},\ \bibinfo {series} {Advanced texts in physics}\ No.\ \bibinfo
  {number} {Bd. 3}\ (\bibinfo  {publisher} {Springer Berlin Heidelberg},\
  \bibinfo {year} {2005})\BibitemShut {NoStop}%
\bibitem [{\citenamefont {McCluskey}\ and\ \citenamefont
  {Haller}(2012)}]{mccluskey2012dopants}%
  \BibitemOpen
  \bibfield  {author} {\bibinfo {author} {\bibfnamefont {M.}~\bibnamefont
  {McCluskey}}\ and\ \bibinfo {author} {\bibfnamefont {E.}~\bibnamefont
  {Haller}},\ }\href {https://books.google.lu/books?id=fV8k6YMxf-MC} {\emph
  {\bibinfo {title} {Dopants and Defects in Semiconductors}}}\ (\bibinfo
  {publisher} {Taylor \& Francis},\ \bibinfo {year} {2012})\BibitemShut
  {NoStop}%
\bibitem [{\citenamefont {Tonndorf}\ \emph {et~al.}(2017)\citenamefont
  {Tonndorf}, \citenamefont {Schwarz}, \citenamefont {Kern}, \citenamefont
  {Niehues}, \citenamefont {Pozo-Zamudio}, \citenamefont {Dmitriev},
  \citenamefont {Bakhtinov}, \citenamefont {Borisenko}, \citenamefont
  {Kolesnikov}, \citenamefont {Tartakovskii}, \citenamefont {de~Vasconcellos},\
  and\ \citenamefont {Bratschitsch}}]{Tonndorf_2017}%
  \BibitemOpen
  \bibfield  {author} {\bibinfo {author} {\bibfnamefont {P.}~\bibnamefont
  {Tonndorf}}, \bibinfo {author} {\bibfnamefont {S.}~\bibnamefont {Schwarz}},
  \bibinfo {author} {\bibfnamefont {J.}~\bibnamefont {Kern}}, \bibinfo {author}
  {\bibfnamefont {I.}~\bibnamefont {Niehues}}, \bibinfo {author} {\bibfnamefont
  {O.~D.}\ \bibnamefont {Pozo-Zamudio}}, \bibinfo {author} {\bibfnamefont
  {A.~I.}\ \bibnamefont {Dmitriev}}, \bibinfo {author} {\bibfnamefont {A.~P.}\
  \bibnamefont {Bakhtinov}}, \bibinfo {author} {\bibfnamefont {D.~N.}\
  \bibnamefont {Borisenko}}, \bibinfo {author} {\bibfnamefont {N.~N.}\
  \bibnamefont {Kolesnikov}}, \bibinfo {author} {\bibfnamefont {A.~I.}\
  \bibnamefont {Tartakovskii}}, \bibinfo {author} {\bibfnamefont {S.~M.}\
  \bibnamefont {de~Vasconcellos}}, \ and\ \bibinfo {author} {\bibfnamefont
  {R.}~\bibnamefont {Bratschitsch}},\ }\href {\doibase
  10.1088/2053-1583/aa525b} {\bibfield  {journal} {\bibinfo  {journal} {2D
  Materials}\ }\textbf {\bibinfo {volume} {4}},\ \bibinfo {pages} {021010}
  (\bibinfo {year} {2017})}\BibitemShut {NoStop}%
\bibitem [{\citenamefont {Toth}\ and\ \citenamefont
  {Aharonovich}(2019)}]{annurev-2D_Aharono}%
  \BibitemOpen
  \bibfield  {author} {\bibinfo {author} {\bibfnamefont {M.}~\bibnamefont
  {Toth}}\ and\ \bibinfo {author} {\bibfnamefont {I.}~\bibnamefont
  {Aharonovich}},\ }\href {\doibase 10.1146/annurev-physchem-042018-052628}
  {\bibfield  {journal} {\bibinfo  {journal} {Annual Review of Physical
  Chemistry}\ }\textbf {\bibinfo {volume} {70}},\ \bibinfo {pages} {123}
  (\bibinfo {year} {2019})},\ \bibinfo {note} {pMID: 30735459}\BibitemShut
  {NoStop}%
\bibitem [{\citenamefont {Ben~Aziza}\ \emph {et~al.}(2017)\citenamefont
  {Ben~Aziza}, \citenamefont {Pierucci}, \citenamefont {Henck}, \citenamefont
  {Silly}, \citenamefont {David}, \citenamefont {Yoon}, \citenamefont
  {Sirotti}, \citenamefont {Xiao}, \citenamefont {Eddrief}, \citenamefont
  {Girard},\ and\ \citenamefont {Ouerghi}}]{PhysRevB.96.035407}%
  \BibitemOpen
  \bibfield  {author} {\bibinfo {author} {\bibfnamefont {Z.}~\bibnamefont
  {Ben~Aziza}}, \bibinfo {author} {\bibfnamefont {D.}~\bibnamefont {Pierucci}},
  \bibinfo {author} {\bibfnamefont {H.}~\bibnamefont {Henck}}, \bibinfo
  {author} {\bibfnamefont {M.~G.}\ \bibnamefont {Silly}}, \bibinfo {author}
  {\bibfnamefont {C.}~\bibnamefont {David}}, \bibinfo {author} {\bibfnamefont
  {M.}~\bibnamefont {Yoon}}, \bibinfo {author} {\bibfnamefont {F.}~\bibnamefont
  {Sirotti}}, \bibinfo {author} {\bibfnamefont {K.}~\bibnamefont {Xiao}},
  \bibinfo {author} {\bibfnamefont {M.}~\bibnamefont {Eddrief}}, \bibinfo
  {author} {\bibfnamefont {J.-C.}\ \bibnamefont {Girard}}, \ and\ \bibinfo
  {author} {\bibfnamefont {A.}~\bibnamefont {Ouerghi}},\ }\href {\doibase
  10.1103/PhysRevB.96.035407} {\bibfield  {journal} {\bibinfo  {journal} {Phys.
  Rev. B}\ }\textbf {\bibinfo {volume} {96}},\ \bibinfo {pages} {035407}
  (\bibinfo {year} {2017})}\BibitemShut {NoStop}%
\bibitem [{\citenamefont {Chen}\ \emph {et~al.}(2015)\citenamefont {Chen},
  \citenamefont {Li}, \citenamefont {Huang},\ and\ \citenamefont
  {Li}}]{ChenDefects2015}%
  \BibitemOpen
  \bibfield  {author} {\bibinfo {author} {\bibfnamefont {H.}~\bibnamefont
  {Chen}}, \bibinfo {author} {\bibfnamefont {Y.}~\bibnamefont {Li}}, \bibinfo
  {author} {\bibfnamefont {L.}~\bibnamefont {Huang}}, \ and\ \bibinfo {author}
  {\bibfnamefont {J.}~\bibnamefont {Li}},\ }\href {\doibase 10.1039/C5RA08329J}
  {\bibfield  {journal} {\bibinfo  {journal} {RSC Adv.}\ }\textbf {\bibinfo
  {volume} {5}},\ \bibinfo {pages} {50883} (\bibinfo {year}
  {2015})}\BibitemShut {NoStop}%
\bibitem [{\citenamefont {Guo}\ \emph {et~al.}(2017)\citenamefont {Guo},
  \citenamefont {Zhou}, \citenamefont {Bai},\ and\ \citenamefont
  {Zhao}}]{GuoDefects2017}%
  \BibitemOpen
  \bibfield  {author} {\bibinfo {author} {\bibfnamefont {Y.}~\bibnamefont
  {Guo}}, \bibinfo {author} {\bibfnamefont {S.}~\bibnamefont {Zhou}}, \bibinfo
  {author} {\bibfnamefont {Y.}~\bibnamefont {Bai}}, \ and\ \bibinfo {author}
  {\bibfnamefont {J.}~\bibnamefont {Zhao}},\ }\href {\doibase
  10.1063/1.4993639} {\bibfield  {journal} {\bibinfo  {journal} {The Journal of
  Chemical Physics}\ }\textbf {\bibinfo {volume} {147}},\ \bibinfo {pages}
  {104709} (\bibinfo {year} {2017})}\BibitemShut {NoStop}%
\bibitem [{\citenamefont {Hopkinson}\ \emph {et~al.}(2019)\citenamefont
  {Hopkinson}, \citenamefont {Z{\'o}lyomi}, \citenamefont {Rooney},
  \citenamefont {Clark}, \citenamefont {Terry}, \citenamefont {Hamer},
  \citenamefont {Lewis}, \citenamefont {Allen}, \citenamefont {Kirkland},
  \citenamefont {Andreev}, \citenamefont {Kudrynskyi}, \citenamefont
  {Kovalyuk}, \citenamefont {Patan{\`e}}, \citenamefont {Fal'ko}, \citenamefont
  {Gorbachev},\ and\ \citenamefont {Haigh}}]{Hopkinson2019}%
  \BibitemOpen
  \bibfield  {author} {\bibinfo {author} {\bibfnamefont {D.~G.}\ \bibnamefont
  {Hopkinson}}, \bibinfo {author} {\bibfnamefont {V.}~\bibnamefont
  {Z{\'o}lyomi}}, \bibinfo {author} {\bibfnamefont {A.~P.}\ \bibnamefont
  {Rooney}}, \bibinfo {author} {\bibfnamefont {N.}~\bibnamefont {Clark}},
  \bibinfo {author} {\bibfnamefont {D.~J.}\ \bibnamefont {Terry}}, \bibinfo
  {author} {\bibfnamefont {M.}~\bibnamefont {Hamer}}, \bibinfo {author}
  {\bibfnamefont {D.~J.}\ \bibnamefont {Lewis}}, \bibinfo {author}
  {\bibfnamefont {C.~S.}\ \bibnamefont {Allen}}, \bibinfo {author}
  {\bibfnamefont {A.~I.}\ \bibnamefont {Kirkland}}, \bibinfo {author}
  {\bibfnamefont {Y.}~\bibnamefont {Andreev}}, \bibinfo {author} {\bibfnamefont
  {Z.}~\bibnamefont {Kudrynskyi}}, \bibinfo {author} {\bibfnamefont
  {Z.}~\bibnamefont {Kovalyuk}}, \bibinfo {author} {\bibfnamefont
  {A.}~\bibnamefont {Patan{\`e}}}, \bibinfo {author} {\bibfnamefont {V.~I.}\
  \bibnamefont {Fal'ko}}, \bibinfo {author} {\bibfnamefont {R.}~\bibnamefont
  {Gorbachev}}, \ and\ \bibinfo {author} {\bibfnamefont {S.~J.}\ \bibnamefont
  {Haigh}},\ }\href {\doibase 10.1021/acsnano.8b08253} {\bibfield  {journal}
  {\bibinfo  {journal} {ACS Nano}\ }\textbf {\bibinfo {volume} {13}},\ \bibinfo
  {pages} {5112} (\bibinfo {year} {2019})}\BibitemShut {NoStop}%
\bibitem [{\citenamefont {Dabral}\ \emph {et~al.}(2019)\citenamefont {Dabral},
  \citenamefont {Lu}, \citenamefont {Chiappe}, \citenamefont {Houssa},\ and\
  \citenamefont {Pourtois}}]{DabralDefects2019}%
  \BibitemOpen
  \bibfield  {author} {\bibinfo {author} {\bibfnamefont {A.}~\bibnamefont
  {Dabral}}, \bibinfo {author} {\bibfnamefont {A.~K.~A.}\ \bibnamefont {Lu}},
  \bibinfo {author} {\bibfnamefont {D.}~\bibnamefont {Chiappe}}, \bibinfo
  {author} {\bibfnamefont {M.}~\bibnamefont {Houssa}}, \ and\ \bibinfo {author}
  {\bibfnamefont {G.}~\bibnamefont {Pourtois}},\ }\href {\doibase
  10.1039/C8CP05665J} {\bibfield  {journal} {\bibinfo  {journal} {Phys. Chem.
  Chem. Phys.}\ }\textbf {\bibinfo {volume} {21}},\ \bibinfo {pages} {1089}
  (\bibinfo {year} {2019})}\BibitemShut {NoStop}%
\bibitem [{\citenamefont {De{\'{a}}k}\ \emph {et~al.}(2020)\citenamefont
  {De{\'{a}}k}, \citenamefont {Han}, \citenamefont {Lorke}, \citenamefont
  {Tabriz},\ and\ \citenamefont {Frauenheim}}]{De_k_2020}%
  \BibitemOpen
  \bibfield  {author} {\bibinfo {author} {\bibfnamefont {P.}~\bibnamefont
  {De{\'{a}}k}}, \bibinfo {author} {\bibfnamefont {M.}~\bibnamefont {Han}},
  \bibinfo {author} {\bibfnamefont {M.}~\bibnamefont {Lorke}}, \bibinfo
  {author} {\bibfnamefont {M.~F.}\ \bibnamefont {Tabriz}}, \ and\ \bibinfo
  {author} {\bibfnamefont {T.}~\bibnamefont {Frauenheim}},\ }\href {\doibase
  10.1088/1361-648x/ab7fdb} {\bibfield  {journal} {\bibinfo  {journal} {Journal
  of Physics: Condensed Matter}\ }\textbf {\bibinfo {volume} {32}},\ \bibinfo
  {pages} {285503} (\bibinfo {year} {2020})}\BibitemShut {NoStop}%
\bibitem [{\citenamefont {Komsa}\ and\ \citenamefont
  {Krasheninnikov}(2015)}]{Komsa2015}%
  \BibitemOpen
  \bibfield  {author} {\bibinfo {author} {\bibfnamefont {H.-P.}\ \bibnamefont
  {Komsa}}\ and\ \bibinfo {author} {\bibfnamefont {A.~V.}\ \bibnamefont
  {Krasheninnikov}},\ }\href {\doibase 10.1103/PhysRevB.91.125304} {\bibfield
  {journal} {\bibinfo  {journal} {Phys. Rev. B}\ }\textbf {\bibinfo {volume}
  {91}},\ \bibinfo {pages} {125304} (\bibinfo {year} {2015})}\BibitemShut
  {NoStop}%
\bibitem [{\citenamefont {Kresse}\ and\ \citenamefont
  {Furthm\"uller}(1996)}]{PhysRevB.54.11169}%
  \BibitemOpen
  \bibfield  {author} {\bibinfo {author} {\bibfnamefont {G.}~\bibnamefont
  {Kresse}}\ and\ \bibinfo {author} {\bibfnamefont {J.}~\bibnamefont
  {Furthm\"uller}},\ }\href {\doibase 10.1103/PhysRevB.54.11169} {\bibfield
  {journal} {\bibinfo  {journal} {Phys. Rev. B}\ }\textbf {\bibinfo {volume}
  {54}},\ \bibinfo {pages} {11169} (\bibinfo {year} {1996})}\BibitemShut
  {NoStop}%
\bibitem [{\citenamefont {Bj\"orkman}\ \emph {et~al.}(2012)\citenamefont
  {Bj\"orkman}, \citenamefont {Gulans}, \citenamefont {Krasheninnikov},\ and\
  \citenamefont {Nieminen}}]{PhysRevLett.108.235502}%
  \BibitemOpen
  \bibfield  {author} {\bibinfo {author} {\bibfnamefont {T.}~\bibnamefont
  {Bj\"orkman}}, \bibinfo {author} {\bibfnamefont {A.}~\bibnamefont {Gulans}},
  \bibinfo {author} {\bibfnamefont {A.~V.}\ \bibnamefont {Krasheninnikov}}, \
  and\ \bibinfo {author} {\bibfnamefont {R.~M.}\ \bibnamefont {Nieminen}},\
  }\href {\doibase 10.1103/PhysRevLett.108.235502} {\bibfield  {journal}
  {\bibinfo  {journal} {Phys. Rev. Lett.}\ }\textbf {\bibinfo {volume} {108}},\
  \bibinfo {pages} {235502} (\bibinfo {year} {2012})}\BibitemShut {NoStop}%
\bibitem [{\citenamefont {Björkman}\ \emph {et~al.}(2012)\citenamefont
  {Björkman}, \citenamefont {Gulans}, \citenamefont {Krasheninnikov},\ and\
  \citenamefont {Nieminen}}]{Bj_rkman_2012}%
  \BibitemOpen
  \bibfield  {author} {\bibinfo {author} {\bibfnamefont {T.}~\bibnamefont
  {Björkman}}, \bibinfo {author} {\bibfnamefont {A.}~\bibnamefont {Gulans}},
  \bibinfo {author} {\bibfnamefont {A.~V.}\ \bibnamefont {Krasheninnikov}}, \
  and\ \bibinfo {author} {\bibfnamefont {R.~M.}\ \bibnamefont {Nieminen}},\
  }\href {\doibase 10.1088/0953-8984/24/42/424218} {\bibfield  {journal}
  {\bibinfo  {journal} {Journal of Physics: Condensed Matter}\ }\textbf
  {\bibinfo {volume} {24}},\ \bibinfo {pages} {424218} (\bibinfo {year}
  {2012})}\BibitemShut {NoStop}%
\bibitem [{\citenamefont {Liu}\ \emph {et~al.}(2020)\citenamefont {Liu},
  \citenamefont {Yang},\ and\ \citenamefont {Guo}}]{PhysRevB.101.045428}%
  \BibitemOpen
  \bibfield  {author} {\bibinfo {author} {\bibfnamefont {X.}~\bibnamefont
  {Liu}}, \bibinfo {author} {\bibfnamefont {J.}~\bibnamefont {Yang}}, \ and\
  \bibinfo {author} {\bibfnamefont {W.}~\bibnamefont {Guo}},\ }\href {\doibase
  10.1103/PhysRevB.101.045428} {\bibfield  {journal} {\bibinfo  {journal}
  {Phys. Rev. B}\ }\textbf {\bibinfo {volume} {101}},\ \bibinfo {pages}
  {045428} (\bibinfo {year} {2020})}\BibitemShut {NoStop}%
\bibitem [{\citenamefont {Freysoldt}\ \emph {et~al.}(2014)\citenamefont
  {Freysoldt}, \citenamefont {Grabowski}, \citenamefont {Hickel}, \citenamefont
  {Neugebauer}, \citenamefont {Kresse}, \citenamefont {Janotti},\ and\
  \citenamefont {Van~de Walle}}]{RevModPhys.86.253}%
  \BibitemOpen
  \bibfield  {author} {\bibinfo {author} {\bibfnamefont {C.}~\bibnamefont
  {Freysoldt}}, \bibinfo {author} {\bibfnamefont {B.}~\bibnamefont
  {Grabowski}}, \bibinfo {author} {\bibfnamefont {T.}~\bibnamefont {Hickel}},
  \bibinfo {author} {\bibfnamefont {J.}~\bibnamefont {Neugebauer}}, \bibinfo
  {author} {\bibfnamefont {G.}~\bibnamefont {Kresse}}, \bibinfo {author}
  {\bibfnamefont {A.}~\bibnamefont {Janotti}}, \ and\ \bibinfo {author}
  {\bibfnamefont {C.~G.}\ \bibnamefont {Van~de Walle}},\ }\href {\doibase
  10.1103/RevModPhys.86.253} {\bibfield  {journal} {\bibinfo  {journal} {Rev.
  Mod. Phys.}\ }\textbf {\bibinfo {volume} {86}},\ \bibinfo {pages} {253}
  (\bibinfo {year} {2014})}\BibitemShut {NoStop}%
\bibitem [{\citenamefont {Alkauskas}\ \emph {et~al.}(2011)\citenamefont
  {Alkauskas}, \citenamefont {De{\'a}k}, \citenamefont {Neugebauer},
  \citenamefont {Pasquarello},\ and\ \citenamefont {Van~de
  Walle}}]{alkauskas2011advanced}%
  \BibitemOpen
  \bibfield  {author} {\bibinfo {author} {\bibfnamefont {A.}~\bibnamefont
  {Alkauskas}}, \bibinfo {author} {\bibfnamefont {P.}~\bibnamefont {De{\'a}k}},
  \bibinfo {author} {\bibfnamefont {J.}~\bibnamefont {Neugebauer}}, \bibinfo
  {author} {\bibfnamefont {A.}~\bibnamefont {Pasquarello}}, \ and\ \bibinfo
  {author} {\bibfnamefont {C.}~\bibnamefont {Van~de Walle}},\ }\href
  {https://books.google.de/books?id=0clIxe9q2asC} {\emph {\bibinfo {title}
  {Advanced Calculations for Defects in Materials: Electronic Structure
  Methods}}}\ (\bibinfo  {publisher} {Wiley},\ \bibinfo {year}
  {2011})\BibitemShut {NoStop}%
\bibitem [{\citenamefont {Chen}\ and\ \citenamefont
  {Pasquarello}(2015)}]{Chen_2015}%
  \BibitemOpen
  \bibfield  {author} {\bibinfo {author} {\bibfnamefont {W.}~\bibnamefont
  {Chen}}\ and\ \bibinfo {author} {\bibfnamefont {A.}~\bibnamefont
  {Pasquarello}},\ }\href {\doibase 10.1088/0953-8984/27/13/133202} {\bibfield
  {journal} {\bibinfo  {journal} {Journal of Physics: Condensed Matter}\
  }\textbf {\bibinfo {volume} {27}},\ \bibinfo {pages} {133202} (\bibinfo
  {year} {2015})}\BibitemShut {NoStop}%
\bibitem [{\citenamefont {Oba}\ and\ \citenamefont {Kumagai}(2018)}]{Oba_2018}%
  \BibitemOpen
  \bibfield  {author} {\bibinfo {author} {\bibfnamefont {F.}~\bibnamefont
  {Oba}}\ and\ \bibinfo {author} {\bibfnamefont {Y.}~\bibnamefont {Kumagai}},\
  }\href {\doibase 10.7567/apex.11.060101} {\bibfield  {journal} {\bibinfo
  {journal} {Applied Physics Express}\ }\textbf {\bibinfo {volume} {11}},\
  \bibinfo {pages} {060101} (\bibinfo {year} {2018})}\BibitemShut {NoStop}%
\bibitem [{\citenamefont {Jung}\ \emph {et~al.}(2015)\citenamefont {Jung},
  \citenamefont {Shojaei}, \citenamefont {Park}, \citenamefont {Oh},
  \citenamefont {Im}, \citenamefont {Jang}, \citenamefont {Park},\ and\
  \citenamefont {Kang}}]{Jung2015}%
  \BibitemOpen
  \bibfield  {author} {\bibinfo {author} {\bibfnamefont {C.~S.}\ \bibnamefont
  {Jung}}, \bibinfo {author} {\bibfnamefont {F.}~\bibnamefont {Shojaei}},
  \bibinfo {author} {\bibfnamefont {K.}~\bibnamefont {Park}}, \bibinfo {author}
  {\bibfnamefont {J.~Y.}\ \bibnamefont {Oh}}, \bibinfo {author} {\bibfnamefont
  {H.~S.}\ \bibnamefont {Im}}, \bibinfo {author} {\bibfnamefont {D.~M.}\
  \bibnamefont {Jang}}, \bibinfo {author} {\bibfnamefont {J.}~\bibnamefont
  {Park}}, \ and\ \bibinfo {author} {\bibfnamefont {H.~S.}\ \bibnamefont
  {Kang}},\ }\href {\doibase 10.1021/acsnano.5b04876} {\bibfield  {journal}
  {\bibinfo  {journal} {ACS Nano}\ }\textbf {\bibinfo {volume} {9}},\ \bibinfo
  {pages} {9585} (\bibinfo {year} {2015})}\BibitemShut {NoStop}%
\bibitem [{\citenamefont {Rybkovskiy}\ \emph {et~al.}(2011)\citenamefont
  {Rybkovskiy}, \citenamefont {Arutyunyan}, \citenamefont {Orekhov},
  \citenamefont {Gromchenko}, \citenamefont {Vorobiev}, \citenamefont
  {Osadchy}, \citenamefont {Salaev}, \citenamefont {Baykara}, \citenamefont
  {Allakhverdiev},\ and\ \citenamefont {Obraztsova}}]{PhysRevB.84.085314}%
  \BibitemOpen
  \bibfield  {author} {\bibinfo {author} {\bibfnamefont {D.~V.}\ \bibnamefont
  {Rybkovskiy}}, \bibinfo {author} {\bibfnamefont {N.~R.}\ \bibnamefont
  {Arutyunyan}}, \bibinfo {author} {\bibfnamefont {A.~S.}\ \bibnamefont
  {Orekhov}}, \bibinfo {author} {\bibfnamefont {I.~A.}\ \bibnamefont
  {Gromchenko}}, \bibinfo {author} {\bibfnamefont {I.~V.}\ \bibnamefont
  {Vorobiev}}, \bibinfo {author} {\bibfnamefont {A.~V.}\ \bibnamefont
  {Osadchy}}, \bibinfo {author} {\bibfnamefont {E.~Y.}\ \bibnamefont {Salaev}},
  \bibinfo {author} {\bibfnamefont {T.~K.}\ \bibnamefont {Baykara}}, \bibinfo
  {author} {\bibfnamefont {K.~R.}\ \bibnamefont {Allakhverdiev}}, \ and\
  \bibinfo {author} {\bibfnamefont {E.~D.}\ \bibnamefont {Obraztsova}},\ }\href
  {\doibase 10.1103/PhysRevB.84.085314} {\bibfield  {journal} {\bibinfo
  {journal} {Phys. Rev. B}\ }\textbf {\bibinfo {volume} {84}},\ \bibinfo
  {pages} {085314} (\bibinfo {year} {2011})}\BibitemShut {NoStop}%
\bibitem [{\citenamefont {De\'ak}\ \emph {et~al.}(2019)\citenamefont {De\'ak},
  \citenamefont {Khorasani}, \citenamefont {Lorke}, \citenamefont
  {Farzalipour-Tabriz}, \citenamefont {Aradi},\ and\ \citenamefont
  {Frauenheim}}]{PhysRevB.100.235304}%
  \BibitemOpen
  \bibfield  {author} {\bibinfo {author} {\bibfnamefont {P.}~\bibnamefont
  {De\'ak}}, \bibinfo {author} {\bibfnamefont {E.}~\bibnamefont {Khorasani}},
  \bibinfo {author} {\bibfnamefont {M.}~\bibnamefont {Lorke}}, \bibinfo
  {author} {\bibfnamefont {M.}~\bibnamefont {Farzalipour-Tabriz}}, \bibinfo
  {author} {\bibfnamefont {B.}~\bibnamefont {Aradi}}, \ and\ \bibinfo {author}
  {\bibfnamefont {T.}~\bibnamefont {Frauenheim}},\ }\href {\doibase
  10.1103/PhysRevB.100.235304} {\bibfield  {journal} {\bibinfo  {journal}
  {Phys. Rev. B}\ }\textbf {\bibinfo {volume} {100}},\ \bibinfo {pages}
  {235304} (\bibinfo {year} {2019})}\BibitemShut {NoStop}%
\bibitem [{\citenamefont {Noh}\ \emph {et~al.}(2014)\citenamefont {Noh},
  \citenamefont {Kim},\ and\ \citenamefont {Kim}}]{Noh2014}%
  \BibitemOpen
  \bibfield  {author} {\bibinfo {author} {\bibfnamefont {J.-Y.}\ \bibnamefont
  {Noh}}, \bibinfo {author} {\bibfnamefont {H.}~\bibnamefont {Kim}}, \ and\
  \bibinfo {author} {\bibfnamefont {Y.-S.}\ \bibnamefont {Kim}},\ }\href
  {\doibase 10.1103/PhysRevB.89.205417} {\bibfield  {journal} {\bibinfo
  {journal} {Phys. Rev. B}\ }\textbf {\bibinfo {volume} {89}},\ \bibinfo
  {pages} {205417} (\bibinfo {year} {2014})}\BibitemShut {NoStop}%
\bibitem [{\citenamefont {Barin}\ \emph {et~al.}(1977)\citenamefont {Barin},
  \citenamefont {Knacke},\ and\ \citenamefont {Kubaschewski}}]{Barin1977}%
  \BibitemOpen
  \bibfield  {author} {\bibinfo {author} {\bibfnamefont {I.}~\bibnamefont
  {Barin}}, \bibinfo {author} {\bibfnamefont {O.}~\bibnamefont {Knacke}}, \
  and\ \bibinfo {author} {\bibfnamefont {O.}~\bibnamefont {Kubaschewski}},\
  }\href {\doibase 10.1007/978-3-662-02293-1_1} {\emph {\bibinfo {title}
  {Thermochemical properties of inorganic substances: Supplement}}}\ (\bibinfo
  {publisher} {Springer Berlin Heidelberg},\ \bibinfo {address} {Berlin,
  Heidelberg},\ \bibinfo {year} {1977})\ pp.\ \bibinfo {pages}
  {1--861}\BibitemShut {NoStop}%
\bibitem [{\citenamefont {Chase}(1998)}]{219851}%
  \BibitemOpen
  \bibfield  {author} {\bibinfo {author} {\bibfnamefont {M.}~\bibnamefont
  {Chase}},\ }\href@noop {} {\emph {\bibinfo {title} {NIST-JANAF Thermochemical
  Tables, 4th Edition}}}\ (\bibinfo  {publisher} {American Institute of
  Physics, -1},\ \bibinfo {year} {1998})\BibitemShut {NoStop}%
\bibitem [{200(2005)}]{2005chemical}%
  \BibitemOpen
  \href {https://books.google.lu/books?id=WwhtlAEACAAJ} {\emph {\bibinfo
  {title} {Chemical Thermodynamics of Selenium}}},\ Chemical Thermodynamics\
  (\bibinfo  {publisher} {Elsevier Science},\ \bibinfo {year}
  {2005})\BibitemShut {NoStop}%
\bibitem [{\citenamefont {Zhang}\ and\ \citenamefont
  {Northrup}(1991)}]{PhysRevLett.67.2339}%
  \BibitemOpen
  \bibfield  {author} {\bibinfo {author} {\bibfnamefont {S.~B.}\ \bibnamefont
  {Zhang}}\ and\ \bibinfo {author} {\bibfnamefont {J.~E.}\ \bibnamefont
  {Northrup}},\ }\href {\doibase 10.1103/PhysRevLett.67.2339} {\bibfield
  {journal} {\bibinfo  {journal} {Phys. Rev. Lett.}\ }\textbf {\bibinfo
  {volume} {67}},\ \bibinfo {pages} {2339} (\bibinfo {year}
  {1991})}\BibitemShut {NoStop}%
\bibitem [{\citenamefont {Stukowski}(2010)}]{0965-0393-18-1-015012}%
  \BibitemOpen
  \bibfield  {author} {\bibinfo {author} {\bibfnamefont {A.}~\bibnamefont
  {Stukowski}},\ }\href {http://stacks.iop.org/0965-0393/18/i=1/a=015012}
  {\bibfield  {journal} {\bibinfo  {journal} {Modelling and Simul. Mater. Sci.
  Eng.}\ }\textbf {\bibinfo {volume} {18}},\ \bibinfo {pages} {015012}
  (\bibinfo {year} {2010})}\BibitemShut {NoStop}%
\bibitem [{\citenamefont {Leslie}\ and\ \citenamefont
  {Gillan}(1985)}]{Leslie_1985}%
  \BibitemOpen
  \bibfield  {author} {\bibinfo {author} {\bibfnamefont {M.}~\bibnamefont
  {Leslie}}\ and\ \bibinfo {author} {\bibfnamefont {N.~J.}\ \bibnamefont
  {Gillan}},\ }\href {\doibase 10.1088/0022-3719/18/5/005} {\bibfield
  {journal} {\bibinfo  {journal} {Journal of Physics C: Solid State Physics}\
  }\textbf {\bibinfo {volume} {18}},\ \bibinfo {pages} {973} (\bibinfo {year}
  {1985})}\BibitemShut {NoStop}%
\bibitem [{\citenamefont {Makov}\ and\ \citenamefont
  {Payne}(1995)}]{PhysRevB.51.4014}%
  \BibitemOpen
  \bibfield  {author} {\bibinfo {author} {\bibfnamefont {G.}~\bibnamefont
  {Makov}}\ and\ \bibinfo {author} {\bibfnamefont {M.~C.}\ \bibnamefont
  {Payne}},\ }\href {\doibase 10.1103/PhysRevB.51.4014} {\bibfield  {journal}
  {\bibinfo  {journal} {Phys. Rev. B}\ }\textbf {\bibinfo {volume} {51}},\
  \bibinfo {pages} {4014} (\bibinfo {year} {1995})}\BibitemShut {NoStop}%
\bibitem [{\citenamefont {Castleton}\ \emph {et~al.}(2006)\citenamefont
  {Castleton}, \citenamefont {H\"oglund},\ and\ \citenamefont
  {Mirbt}}]{PhysRevB.73.035215}%
  \BibitemOpen
  \bibfield  {author} {\bibinfo {author} {\bibfnamefont {C.~W.~M.}\
  \bibnamefont {Castleton}}, \bibinfo {author} {\bibfnamefont {A.}~\bibnamefont
  {H\"oglund}}, \ and\ \bibinfo {author} {\bibfnamefont {S.}~\bibnamefont
  {Mirbt}},\ }\href {\doibase 10.1103/PhysRevB.73.035215} {\bibfield  {journal}
  {\bibinfo  {journal} {Phys. Rev. B}\ }\textbf {\bibinfo {volume} {73}},\
  \bibinfo {pages} {035215} (\bibinfo {year} {2006})}\BibitemShut {NoStop}%
\bibitem [{\citenamefont {Lany}\ and\ \citenamefont
  {Zunger}(2008)}]{PhysRevB.78.235104}%
  \BibitemOpen
  \bibfield  {author} {\bibinfo {author} {\bibfnamefont {S.}~\bibnamefont
  {Lany}}\ and\ \bibinfo {author} {\bibfnamefont {A.}~\bibnamefont {Zunger}},\
  }\href {\doibase 10.1103/PhysRevB.78.235104} {\bibfield  {journal} {\bibinfo
  {journal} {Phys. Rev. B}\ }\textbf {\bibinfo {volume} {78}},\ \bibinfo
  {pages} {235104} (\bibinfo {year} {2008})}\BibitemShut {NoStop}%
\bibitem [{\citenamefont {Freysoldt}\ \emph {et~al.}(2009)\citenamefont
  {Freysoldt}, \citenamefont {Neugebauer},\ and\ \citenamefont {Van~de
  Walle}}]{PhysRevLett.102.016402}%
  \BibitemOpen
  \bibfield  {author} {\bibinfo {author} {\bibfnamefont {C.}~\bibnamefont
  {Freysoldt}}, \bibinfo {author} {\bibfnamefont {J.}~\bibnamefont
  {Neugebauer}}, \ and\ \bibinfo {author} {\bibfnamefont {C.~G.}\ \bibnamefont
  {Van~de Walle}},\ }\href {\doibase 10.1103/PhysRevLett.102.016402} {\bibfield
   {journal} {\bibinfo  {journal} {Phys. Rev. Lett.}\ }\textbf {\bibinfo
  {volume} {102}},\ \bibinfo {pages} {016402} (\bibinfo {year}
  {2009})}\BibitemShut {NoStop}%
\bibitem [{\citenamefont {Komsa}\ \emph {et~al.}(2012)\citenamefont {Komsa},
  \citenamefont {Rantala},\ and\ \citenamefont
  {Pasquarello}}]{PhysRevB.86.045112}%
  \BibitemOpen
  \bibfield  {author} {\bibinfo {author} {\bibfnamefont {H.-P.}\ \bibnamefont
  {Komsa}}, \bibinfo {author} {\bibfnamefont {T.~T.}\ \bibnamefont {Rantala}},
  \ and\ \bibinfo {author} {\bibfnamefont {A.}~\bibnamefont {Pasquarello}},\
  }\href {\doibase 10.1103/PhysRevB.86.045112} {\bibfield  {journal} {\bibinfo
  {journal} {Phys. Rev. B}\ }\textbf {\bibinfo {volume} {86}},\ \bibinfo
  {pages} {045112} (\bibinfo {year} {2012})}\BibitemShut {NoStop}%
\bibitem [{\citenamefont {Kumagai}\ and\ \citenamefont
  {Oba}(2014)}]{PhysRevB.89.195205}%
  \BibitemOpen
  \bibfield  {author} {\bibinfo {author} {\bibfnamefont {Y.}~\bibnamefont
  {Kumagai}}\ and\ \bibinfo {author} {\bibfnamefont {F.}~\bibnamefont {Oba}},\
  }\href {\doibase 10.1103/PhysRevB.89.195205} {\bibfield  {journal} {\bibinfo
  {journal} {Phys. Rev. B}\ }\textbf {\bibinfo {volume} {89}},\ \bibinfo
  {pages} {195205} (\bibinfo {year} {2014})}\BibitemShut {NoStop}%
\bibitem [{\citenamefont {Komsa}\ and\ \citenamefont
  {Pasquarello}(2013)}]{PhysRevLett.110.095505}%
  \BibitemOpen
  \bibfield  {author} {\bibinfo {author} {\bibfnamefont {H.-P.}\ \bibnamefont
  {Komsa}}\ and\ \bibinfo {author} {\bibfnamefont {A.}~\bibnamefont
  {Pasquarello}},\ }\href {\doibase 10.1103/PhysRevLett.110.095505} {\bibfield
  {journal} {\bibinfo  {journal} {Phys. Rev. Lett.}\ }\textbf {\bibinfo
  {volume} {110}},\ \bibinfo {pages} {095505} (\bibinfo {year}
  {2013})}\BibitemShut {NoStop}%
\bibitem [{\citenamefont {Komsa}\ \emph {et~al.}(2014)\citenamefont {Komsa},
  \citenamefont {Berseneva}, \citenamefont {Krasheninnikov},\ and\
  \citenamefont {Nieminen}}]{PhysRevX.4.031044}%
  \BibitemOpen
  \bibfield  {author} {\bibinfo {author} {\bibfnamefont {H.-P.}\ \bibnamefont
  {Komsa}}, \bibinfo {author} {\bibfnamefont {N.}~\bibnamefont {Berseneva}},
  \bibinfo {author} {\bibfnamefont {A.~V.}\ \bibnamefont {Krasheninnikov}}, \
  and\ \bibinfo {author} {\bibfnamefont {R.~M.}\ \bibnamefont {Nieminen}},\
  }\href {\doibase 10.1103/PhysRevX.4.031044} {\bibfield  {journal} {\bibinfo
  {journal} {Phys. Rev. X}\ }\textbf {\bibinfo {volume} {4}},\ \bibinfo {pages}
  {031044} (\bibinfo {year} {2014})}\BibitemShut {NoStop}%
\bibitem [{\citenamefont {Wang}\ \emph {et~al.}(2015)\citenamefont {Wang},
  \citenamefont {Han}, \citenamefont {Li}, \citenamefont {Xie}, \citenamefont
  {Chen}, \citenamefont {Tian}, \citenamefont {West}, \citenamefont {Sun},\
  and\ \citenamefont {Zhang}}]{PhysRevLett.114.196801}%
  \BibitemOpen
  \bibfield  {author} {\bibinfo {author} {\bibfnamefont {D.}~\bibnamefont
  {Wang}}, \bibinfo {author} {\bibfnamefont {D.}~\bibnamefont {Han}}, \bibinfo
  {author} {\bibfnamefont {X.-B.}\ \bibnamefont {Li}}, \bibinfo {author}
  {\bibfnamefont {S.-Y.}\ \bibnamefont {Xie}}, \bibinfo {author} {\bibfnamefont
  {N.-K.}\ \bibnamefont {Chen}}, \bibinfo {author} {\bibfnamefont {W.~Q.}\
  \bibnamefont {Tian}}, \bibinfo {author} {\bibfnamefont {D.}~\bibnamefont
  {West}}, \bibinfo {author} {\bibfnamefont {H.-B.}\ \bibnamefont {Sun}}, \
  and\ \bibinfo {author} {\bibfnamefont {S.~B.}\ \bibnamefont {Zhang}},\ }\href
  {\doibase 10.1103/PhysRevLett.114.196801} {\bibfield  {journal} {\bibinfo
  {journal} {Phys. Rev. Lett.}\ }\textbf {\bibinfo {volume} {114}},\ \bibinfo
  {pages} {196801} (\bibinfo {year} {2015})}\BibitemShut {NoStop}%
\bibitem [{\citenamefont {Wang}\ \emph {et~al.}(2017)\citenamefont {Wang},
  \citenamefont {Han}, \citenamefont {Li}, \citenamefont {Chen}, \citenamefont
  {West}, \citenamefont {Meunier}, \citenamefont {Zhang},\ and\ \citenamefont
  {Sun}}]{PhysRevB.96.155424}%
  \BibitemOpen
  \bibfield  {author} {\bibinfo {author} {\bibfnamefont {D.}~\bibnamefont
  {Wang}}, \bibinfo {author} {\bibfnamefont {D.}~\bibnamefont {Han}}, \bibinfo
  {author} {\bibfnamefont {X.-B.}\ \bibnamefont {Li}}, \bibinfo {author}
  {\bibfnamefont {N.-K.}\ \bibnamefont {Chen}}, \bibinfo {author}
  {\bibfnamefont {D.}~\bibnamefont {West}}, \bibinfo {author} {\bibfnamefont
  {V.}~\bibnamefont {Meunier}}, \bibinfo {author} {\bibfnamefont
  {S.}~\bibnamefont {Zhang}}, \ and\ \bibinfo {author} {\bibfnamefont {H.-B.}\
  \bibnamefont {Sun}},\ }\href {\doibase 10.1103/PhysRevB.96.155424} {\bibfield
   {journal} {\bibinfo  {journal} {Phys. Rev. B}\ }\textbf {\bibinfo {volume}
  {96}},\ \bibinfo {pages} {155424} (\bibinfo {year} {2017})}\BibitemShut
  {NoStop}%
\bibitem [{\citenamefont {Sundararaman}\ and\ \citenamefont
  {Ping}(2017)}]{doi:10.1063/1.4978238}%
  \BibitemOpen
  \bibfield  {author} {\bibinfo {author} {\bibfnamefont {R.}~\bibnamefont
  {Sundararaman}}\ and\ \bibinfo {author} {\bibfnamefont {Y.}~\bibnamefont
  {Ping}},\ }\href {\doibase 10.1063/1.4978238} {\bibfield  {journal} {\bibinfo
   {journal} {The Journal of Chemical Physics}\ }\textbf {\bibinfo {volume}
  {146}},\ \bibinfo {pages} {104109} (\bibinfo {year} {2017})},\ \Eprint
  {http://arxiv.org/abs/https://doi.org/10.1063/1.4978238}
  {https://doi.org/10.1063/1.4978238} \BibitemShut {NoStop}%
\bibitem [{\citenamefont {Wu}\ \emph {et~al.}(2017)\citenamefont {Wu},
  \citenamefont {Galatas}, \citenamefont {Sundararaman}, \citenamefont
  {Rocca},\ and\ \citenamefont {Ping}}]{PhysRevMaterials.1.071001}%
  \BibitemOpen
  \bibfield  {author} {\bibinfo {author} {\bibfnamefont {F.}~\bibnamefont
  {Wu}}, \bibinfo {author} {\bibfnamefont {A.}~\bibnamefont {Galatas}},
  \bibinfo {author} {\bibfnamefont {R.}~\bibnamefont {Sundararaman}}, \bibinfo
  {author} {\bibfnamefont {D.}~\bibnamefont {Rocca}}, \ and\ \bibinfo {author}
  {\bibfnamefont {Y.}~\bibnamefont {Ping}},\ }\href {\doibase
  10.1103/PhysRevMaterials.1.071001} {\bibfield  {journal} {\bibinfo  {journal}
  {Phys. Rev. Materials}\ }\textbf {\bibinfo {volume} {1}},\ \bibinfo {pages}
  {071001} (\bibinfo {year} {2017})}\BibitemShut {NoStop}%
\bibitem [{\citenamefont {Freysoldt}\ and\ \citenamefont
  {Neugebauer}(2018)}]{PhysRevB.97.205425}%
  \BibitemOpen
  \bibfield  {author} {\bibinfo {author} {\bibfnamefont {C.}~\bibnamefont
  {Freysoldt}}\ and\ \bibinfo {author} {\bibfnamefont {J.}~\bibnamefont
  {Neugebauer}},\ }\href {\doibase 10.1103/PhysRevB.97.205425} {\bibfield
  {journal} {\bibinfo  {journal} {Phys. Rev. B}\ }\textbf {\bibinfo {volume}
  {97}},\ \bibinfo {pages} {205425} (\bibinfo {year} {2018})}\BibitemShut
  {NoStop}%
\bibitem [{\citenamefont {Isik}\ \emph {et~al.}(2017)\citenamefont {Isik},
  \citenamefont {Tugay},\ and\ \citenamefont {Gasanly}}]{temp_GaS}%
  \BibitemOpen
  \bibfield  {author} {\bibinfo {author} {\bibfnamefont {M.}~\bibnamefont
  {Isik}}, \bibinfo {author} {\bibfnamefont {E.}~\bibnamefont {Tugay}}, \ and\
  \bibinfo {author} {\bibfnamefont {N.}~\bibnamefont {Gasanly}},\ }\href@noop
  {} {\bibfield  {journal} {\bibinfo  {journal} {Indian Journal of Pure and
  Applied Physics}\ }\textbf {\bibinfo {volume} {55}},\ \bibinfo {pages} {583}
  (\bibinfo {year} {2017})}\BibitemShut {NoStop}%
\bibitem [{\citenamefont {Isik}\ \emph {et~al.}(2016)\citenamefont {Isik},
  \citenamefont {Tugay},\ and\ \citenamefont {Gasanly}}]{temp_GaSe}%
  \BibitemOpen
  \bibfield  {author} {\bibinfo {author} {\bibfnamefont {M.}~\bibnamefont
  {Isik}}, \bibinfo {author} {\bibfnamefont {E.}~\bibnamefont {Tugay}}, \ and\
  \bibinfo {author} {\bibfnamefont {N.~M.}\ \bibnamefont {Gasanly}},\ }\href
  {\doibase 10.1080/14786435.2016.1209593} {\bibfield  {journal} {\bibinfo
  {journal} {Philosophical Magazine}\ }\textbf {\bibinfo {volume} {96}},\
  \bibinfo {pages} {2564} (\bibinfo {year} {2016})},\ \Eprint
  {http://arxiv.org/abs/https://doi.org/10.1080/14786435.2016.1209593}
  {https://doi.org/10.1080/14786435.2016.1209593} \BibitemShut {NoStop}%
\bibitem [{\citenamefont {Ho}\ \emph {et~al.}(1998)\citenamefont {Ho},
  \citenamefont {Wu}, \citenamefont {Huang}, \citenamefont {Liao},\ and\
  \citenamefont {Tiong}}]{Ho_1998}%
  \BibitemOpen
  \bibfield  {author} {\bibinfo {author} {\bibfnamefont {C.~H.}\ \bibnamefont
  {Ho}}, \bibinfo {author} {\bibfnamefont {C.~S.}\ \bibnamefont {Wu}}, \bibinfo
  {author} {\bibfnamefont {Y.~S.}\ \bibnamefont {Huang}}, \bibinfo {author}
  {\bibfnamefont {P.~C.}\ \bibnamefont {Liao}}, \ and\ \bibinfo {author}
  {\bibfnamefont {K.~K.}\ \bibnamefont {Tiong}},\ }\href {\doibase
  10.1088/0953-8984/10/41/014} {\bibfield  {journal} {\bibinfo  {journal}
  {Journal of Physics: Condensed Matter}\ }\textbf {\bibinfo {volume} {10}},\
  \bibinfo {pages} {9317} (\bibinfo {year} {1998})}\BibitemShut {NoStop}%
\bibitem [{\citenamefont {Wu}\ \emph {et~al.}(2014)\citenamefont {Wu},
  \citenamefont {Wu}, \citenamefont {Jadczak}, \citenamefont {Huang},
  \citenamefont {Ho}, \citenamefont {Hsu},\ and\ \citenamefont
  {Tiong}}]{doi:10.1063/1.4882301}%
  \BibitemOpen
  \bibfield  {author} {\bibinfo {author} {\bibfnamefont {Y.~J.}\ \bibnamefont
  {Wu}}, \bibinfo {author} {\bibfnamefont {P.~H.}\ \bibnamefont {Wu}}, \bibinfo
  {author} {\bibfnamefont {J.}~\bibnamefont {Jadczak}}, \bibinfo {author}
  {\bibfnamefont {Y.~S.}\ \bibnamefont {Huang}}, \bibinfo {author}
  {\bibfnamefont {C.~H.}\ \bibnamefont {Ho}}, \bibinfo {author} {\bibfnamefont
  {H.~P.}\ \bibnamefont {Hsu}}, \ and\ \bibinfo {author} {\bibfnamefont
  {K.~K.}\ \bibnamefont {Tiong}},\ }\href {\doibase 10.1063/1.4882301}
  {\bibfield  {journal} {\bibinfo  {journal} {Journal of Applied Physics}\
  }\textbf {\bibinfo {volume} {115}},\ \bibinfo {pages} {223508} (\bibinfo
  {year} {2014})},\ \Eprint
  {http://arxiv.org/abs/https://doi.org/10.1063/1.4882301}
  {https://doi.org/10.1063/1.4882301} \BibitemShut {NoStop}%
\bibitem [{\citenamefont {Choi}\ \emph {et~al.}(2017)\citenamefont {Choi},
  \citenamefont {Kim}, \citenamefont {Jung}, \citenamefont {Kim}, \citenamefont
  {Yu},\ and\ \citenamefont {Chang}}]{Choi2017}%
  \BibitemOpen
  \bibfield  {author} {\bibinfo {author} {\bibfnamefont {B.~K.}\ \bibnamefont
  {Choi}}, \bibinfo {author} {\bibfnamefont {M.}~\bibnamefont {Kim}}, \bibinfo
  {author} {\bibfnamefont {K.-H.}\ \bibnamefont {Jung}}, \bibinfo {author}
  {\bibfnamefont {J.}~\bibnamefont {Kim}}, \bibinfo {author} {\bibfnamefont
  {K.-S.}\ \bibnamefont {Yu}}, \ and\ \bibinfo {author} {\bibfnamefont {Y.~J.}\
  \bibnamefont {Chang}},\ }\href {\doibase 10.1186/s11671-017-2266-7}
  {\bibfield  {journal} {\bibinfo  {journal} {Nanoscale Research Letters}\
  }\textbf {\bibinfo {volume} {12}},\ \bibinfo {pages} {492} (\bibinfo {year}
  {2017})}\BibitemShut {NoStop}%
\end{thebibliography}%

\end{document}